\documentclass[fleqn,usenatbib]{mnras}

\usepackage[T1]{fontenc}
\usepackage{ae,aecompl}
\usepackage{bm}

\usepackage{graphicx}
\usepackage{amsmath}
\usepackage{amssymb}
\usepackage{newtxtext,newtxmath}

\usepackage{multirow}
\usepackage{array}
\usepackage{xcolor}
\usepackage{ulem}
\usepackage{xspace}
\usepackage[abs]{overpic}
\usepackage{pict2e}
\usepackage{lastpage}
\usepackage{cases}

\newcommand{\ud}{\ensuremath{\mathrm{d}}\xspace}
\newcommand{\tpb}{\ensuremath{t_{\mathrm{pb}}}}

\newcommand{\hydrogen}{\ensuremath{\mathrm{H}}\xspace}
\newcommand{\helium}{\ensuremath{\mathrm{^{4}He}}\xspace}
\newcommand{\carbon}{\ensuremath{\mathrm{^{12}C}}\xspace}
\newcommand{\oxygen}{\ensuremath{\mathrm{^{16}O}}\xspace}
\newcommand{\neon}{\ensuremath{\mathrm{^{20}Ne}}\xspace}
\newcommand{\magnesium}{\ensuremath{\mathrm{^{24}Mg}}\xspace}
\newcommand{\silicon}{\ensuremath{\mathrm{^{28}Si}}\xspace}
\newcommand{\nickel}{\ensuremath{\mathrm{^{56}Ni}}\xspace}
\newcommand{\tracer}{\ensuremath{\mathrm{Tr}}\xspace}

\newcommand{\iron}{\ensuremath{\mathrm{^{56}Fe}}\xspace}
\newcommand{\cobalt}{\ensuremath{\mathrm{^{56}Co}}\xspace}

\newcommand{\solm}{\ensuremath{\rm{M_{\odot}}}\xspace}

\newcommand{\kms}{\ensuremath{\mathrm{km\, s^{-1}}}}
\newcommand{\km}{\ensuremath{\mathrm{km}}}
\newcommand{\erg}{\ensuremath{\mathrm{erg}}}
\newcommand{\s}{\ensuremath{\text{s}}}
\newcommand{\ms}{\ensuremath{\text{ms}}}
\newcommand{\gcc}{\text{g}\, \text{cm}^{-3}}

\newcommand{\prom}{\textsc{Prometheus-HOTB}\xspace}
\newcommand{\vertexprom}{\textsc{Vertex-Prometheus}\xspace}
\newcommand{\vertex}{\textsc{Vertex}\xspace}

\newcommand{\rns}{\ensuremath{R_{\mathrm{NS}}}\xspace}
\newcommand{\mns}{\ensuremath{M_{\mathrm{b}}}\xspace}

\newcommand{\onemg}{\ensuremath{\mathrm{e8.8}}\xspace}
\newcommand{\snine}{\ensuremath{\mathrm{s9.0}}\xspace}
\newcommand{\znine}{\ensuremath{\mathrm{z9.6}}\xspace}
\newcommand{\vect}[1]{\boldsymbol{#1}}

\def\prl{Phys. Rev. Lett.}
\def\prd{Phys. Rev. D}

\def\apj{ApJ}
\def\apjs{ApJS}
\def\apjl{ApJL}
\def\aap{A\&A}

\def\nphysa{Nucl. Phys. A.}
\def\mnras{MNRAS}

\title[3D Models of Supernovae From Low-mass Progenitors]{Three-dimensional Models of Core-collapse Supernovae From
Low-mass Progenitors With Implications for Crab}

\author[G. Stockinger et al.]{G.~Stockinger$^{1,2}$,
 H.-T.~Janka$^1$\thanks{Email: thj@mpa-garching.mpg.de},
 D.~Kresse$^{1,2}$,
 T.~Melson$^1$,
 T.~Ertl$^1$,
 M.~Gabler$^3$,  \and
 A.~Gessner$^{4,5}$,  
 A.~Wongwathanarat$^1$,
 A.~Tolstov$^6$,
 S.-C.~Leung$^7$, 
 K.~Nomoto$^8$, \and
 A.~Heger$^{9-13}$ 
 \\
$^1$Max Planck Institute for Astrophysics, Karl-Schwarzschild-Str.~1, 85748 Garching, Germany\\
$^2$Physik-Department, Technische Universit\"at M\"unchen, James-Franck-Str.~1, 85748 Garching, Germany\\
$^3$LATO/DCET, Universidade Estadual de Santa Cruz, Rod.~Jorge Amado, km 16,
Ilh{\'e}us, BA, CEP 45662-900, Brazil\\
$^4$University of T\"ubingen, Maria-von-Linden-Str. 6, 72076 T\"ubingen, Germany\\
$^5$Max Planck Institute for Intelligent Systems, Max-Planck-Ring~4, 72076 T\"ubingen, Germany \\
$^6$The Open University of Japan, 2--11, Wakaba, Mihama-ku, Chiba, Chiba 261--8586, Japan\\
$^7$TAPIR, Walter Burke Institute for Theoretical Physics, Mailcode 350-17, Caltech, Pasadena, CA 91125, USA\\
$^8$Kavli Institute for the Physics and Mathematics of the Universe (WPI), The University of Tokyo Institute for Advanced Study,\\  \hspace{0.08cm} The University of Tokyo, Kashiwa, Chiba 277-8583, Japan\\
$^9$School of Physics \& Astronomy, Monash University, Clayton 3800,
Victoria, Australia\\
$^{10}$Joint Institute for Nuclear Astrophysics, 1 Cyclotron Laboratory,
National Superconducting Cyclotron Laboratory,\\ Michigan State University,
East Lansing, MI 48824-1321, USA\\
$^{11}$Tsung-Dao Lee Institute, Shanghai 200240, China\\
$^{12}$Center of Excellence for Astrophysics in Three Dimensions
(ASTRO-3D), Australia\\
$^{13}$Australian Research Council Centre of Excellence for Gravitational
Wave Discovery, Clayton, VIC 3800, Australia\\
}

\date{Accepted 2020 June 10. Received 2020 June 04; in original form 2020 May 06}

\pubyear{2020}

\begin{document}
\label{firstpage}
\pagerange{\pageref{firstpage}--\pageref{lastpage}}
\maketitle
\pubyear{2020}

\begin{abstract}
We present 3D full-sphere supernova simulations of non-rotating
low-mass ($\sim$\,9\,M$_\odot$) progenitors, covering the entire 
evolution from core collapse
through bounce and shock revival, through shock breakout from the
stellar surface, until fallback is completed several days later.  We
obtain low-energy explosions ($\sim$0.5--$1.0\times10^{50}$\,erg) of
iron-core progenitors at the low-mass end of the core-collapse
supernova (LMCCSN) domain and compare to a super-AGB (sAGB) progenitor
with an oxygen-neon-magnesium core that collapses and explodes as
electron-capture supernova (ECSN).  The onset of the explosion in the
LMCCSN models is modelled self-consistently using the \vertexprom
code, whereas the ECSN explosion is modelled using parametric neutrino
transport in the \textsc{Prometheus-HOTB} code, choosing different
explosion energies in the range of previous self-consistent models.
The sAGB and LMCCSN progenitors that share structural similarities
have almost spherical explosions with little metal mixing into the
hydrogen envelope.  A LMCCSN with less 2nd dredge-up results in a
highly asymmetric explosion.  It shows efficient mixing and dramatic
shock deceleration in the extended hydrogen envelope.  Both properties
allow fast nickel plumes to catch up with the shock, leading to
extreme shock deformation and aspherical shock breakout.  Fallback
masses of $\mathord{\lesssim}\,5\,\mathord{\times}\,10^{-3}$\,M$_\odot$
have no significant effects on the neutron star (NS) masses and kicks.
The anisotropic fallback carries considerable angular momentum, however,
and determines the spin of the newly-born NS.  The LMCCSN model with
less 2nd dredge-up results in a hydrodynamic and neutrino-induced NS
kick of $>$40\,km\,s$^{-1}$ and a NS spin period of $\sim$30\,ms, both
not largely different from those of the Crab pulsar at birth.
\end{abstract}


\noindent

\section{Introduction}

According to current understanding stars with initial masses of
$\mathord{\gtrsim}$\,8--9\,\solm end their lives in a core-collapse 
supernova (CCSN). The explosion is powered by gravitational energy, 
which is released when the core of the star collapses to a compact 
remnant (a neutron star or black hole), and a fraction of which is
transferred to the ejecta by neutrino-energy
deposition \citep{Bethe1985,Colgate1966}. In the past six decades, 
numerous studies have focused on the collapse phase and subsequent 
evolution using 1D simulations. Over the past three decades, 
multi-dimensional simulations with successively improved treatment 
of the microphysics have driven our understanding of the explosion 
mechanism (see e.g. \citealt{Janka2012,Janka2012a,Burrows2013,Janka2016,Mueller2016}).
A close analysis of these models led to the discovery of new hydrodynamic 
instabilities such as the standing accretion shock instability (SASI) 
(\citealt{Blondin2003,Blondin2007,Ohnishi2006,Foglizzo2007,Fernandez2010}). 
The increase of computational capabilities in the recent years along 
with new developments for neutrino transport methods (see, e.g. 
\citealt{Buras2006,Takiwaki2012,OConnor2018b,Skinner2019,Glas2019} and references therein)
have enabled full 3D simulations of the early explosion phase (see e.g. 
\citealt{Takiwaki2014,Couch2014a,Melson2015a,Melson2015,Lentz2015, Summa2016,Roberts2016,Mueller2017,Mueller2018,Vartanyan2018,Ott2018,OConnor2018,Melson2019,Vartanyan2019,Burrows2019}).

Motivated by the historical SN1987A and its progenitor detection, a 
variety of studies in two and three dimensions also investigated the 
propagation of the shock 
wave from its initiation to its breakout from the stellar surface 
in 15--20\,\solm blue supergiant (BSG) models, 
which are suitable as progenitors of SN1987A. In first studies, the blast
wave was launched through artificial energy deposition near the center (see e.g. \citealt{Nomoto1987b,Nomoto1988,Arnett1989b,Mueller1991a,Fryxell1991}),
later the explosion was initiated with the neutrino-driven mechanism (see \citealt{Kifonidis2003,Hammer2010,Wongwathanarat2013,Wongwathanarat2015}).
\citet{Mueller2018} followed the long-time evolution of the explosion of 
ultra-stripped progenitors by 3D simulations, motivated by the importance 
of such stars in understanding the progenitor systems of the recent detections  
of NS-NS mergers by LIGO/Virgo (GW170817, \citealt{Abbott2017}, and
GW190425, \citealt{Ligo2019b}). 

These theoretical works showed that supernova (SN) explosions are by far 
not spherical events as previously thought. Three-dimensional 
instabilities facilitate the explosion
\citep{Herant1994a,Burrows1995,Janka1996} 
and are a necessary ingredient to explain the clumpiness and 
mixing found in photospheric emission \citep{Utrobin2015,Utrobin2017} 
and spectral analyses of the nebular phase of core-collapse events \citep{Jerkstrand2017}. 
\citet{Wongwathanarat2015} showed that the final ejecta 
distribution carries imprints of the 
asphericities produced during the onset of the explosion 
($t\,\mathord{\sim}\, 1\, \rm s$), which are further modified during 
later phases. Depending on the detailed progenitor structure, hydrodynamic 
instabilities arising at the composition interfaces, such as the Rayleigh-Taylor (RT) 
instability and the Richtmeyer-Meshkov (RM) instability, shape 
the final spatial and velocity distributions of nucleosynthetic products. 
The resulting ejecta morphology ranges from quasi-spherical ejecta \citep{Mueller2018} to
strongly pronounced RT-fingers including cases that resemble the geometry found in Cas~A 
\citep{Wongwathanarat2017,Grefenstette2017}. Due to the highly nonlinear and 
stochastic behaviour of non-radial instabilities and turbulence during the 
onset of the explosion phase and the subsequent evolution of RT instabilities,
which depend on the progenitor structure,
a clear connection between the asymmetries, and thus the degree of mixing, and 
the progenitor properties has still to be worked out.

In this paper, we consider CCSN progenitors with initial masses near 
the low-mass end of about 9--10~\solm, where around 20\% of all CCSNe 
are thought to occur (assuming a Salpeter initial mass function  
\citealt{Salpeter1955} and an upper mass limit of $\mathord{\sim}20\,\solm$ 
for CCSNe), to study the differences in their development of mixing 
instabilities during the explosion. To this end, we compare 
ECSNe from a super-AGB progenitor with an ONeMg core and
CCSNe from red supergiant (RSG) progenitors with iron cores, 
all in a zero-age main sequence (ZAMS) mass range around 9~\solm.

The evolution of stars with masses $\mathord{\lesssim}\,12\, \solm$ 
is very sensitive to the initial stellar mass, various 
pulsational instabilities, and mass-loss phenomena \citep{Woosley2015}.
Iron-core CCSN progenitors with initial masses around 9--10~\solm
ignite oxygen burning off-center in contrast to their more massive 
($M\,\mathord{>}\,15\, \solm$) counterparts. After oxygen burning, silicon 
ignites in a degenerate flash which might, in some cases, lead to 
additional mass loss in the last decade of evolution or is speculated to even
eject parts of the hydrogen envelope \citep{Woosley2015}.

The $8.8\, \solm$ progenitor of \cite{Nomoto1984} is even more peculiar. 
It experiences  several thermal pulses and off-center ignition of 
fusion material. In the end, it has a degenerate ONeMg-core 
surrounded by a dilute and extended hydrogen envelope.
 
When the core approaches its Chandrasekhar mass, electron captures 
on $^{24}\mathrm{Mg}$ and $^{20}\mathrm{Ne}$ via the reaction chains 
$^{24}\mathrm{Mg}(e^-,\nu_e)^{24}\mathrm{Na}(e^-,\nu_e)^{24}\mathrm{Ne}$ 
and $^{20}\mathrm{Ne}(e^-,\nu_e)^{20}\mathrm{F}(e^-,\nu_e)^{20}\mathrm{O}$
\citep{Miyaji1980} 
destabilize the core due to a reduction of the effective 
adiabatic index of the electron-degeneracy dominated gas pressure. 
Continuous electron capture on $\mathrm{^{20}Ne}$, which further 
reduces the pressure support, works against the now beginning 
oxygen-burning as temperatures increase during the collapse. 
Simulations in 1D (see \citealt{Kitaura2006,Huedepohl2010,Fischer2010})
and 2D (see \citealt{Janka2008,Radice2017})
suggest that the collapse proceeds despite the oxygen burning. 
\cite{Jones2016} simulated the deflagration of oxygen in ONeMg 
cores with different core densities. 
At $\log_{10}(\rho_c/\mathrm{g\,cm^{-3}})\,\mathord{=}\,9.95$
and lower densities
their cores do not collapse but get partly unbound due to the inefficient 
semi-convective mixing during the electron-capture phase and the 
resulting strong thermonuclear runaway. Only when the central densities 
are higher than this threshold value of $\rho_c$, the core is 
found to collapse to a proto-neutron star (PNS).
Recently, \cite{Kirsebom2019} investigated the influence of a newly 
measured strong transition between the ground states of 
$^{20}\mathrm{Ne}$ and $^{20}\mathrm{F}$ on the electron-capture 
rate and thus on the evolution of ONeMg 
cores. Adding the new transition increases the likelihood that the 
star is (partially) disrupted by a thermonuclear explosion (termed tECSN) 
rather than collapsing to form a PNS. However, \cite{Zha2019}, using
state-of-the-art electron-capture rates including the latest rate for the 
second forbidden transition of $^{20}\mathrm{Ne}(e^-,\nu_e)^{20}\mathrm{F}$
from \cite{Suzuki2019},
found that the oxygen deflagration starting from $\log_{10}(\rho_c/\mathrm{g\,cm^{-3}})\,\mathord{>}\,10.01$  $(\mathord{<}\,10.01)$
leads to collapse (thermonuclear explosion). Their estimate of the central
density when the oxygen deflagration is initiated in an evolving ONeMg core
exceeds this critical value. Therefore they conclude 
that ONeMg cores are likely to collapse.

For this reason, in our study we assume that the ONeMg core collapses to a 
PNS, leading to a ``collapse ECSN'' (cECSN). This assumption
receives additional motivation by the fact that recent studies 
considering the galactic chemical evolution of the Milky Way stress 
the importance of cECSNe to reproduce the solar abundances of several 
important and problematic isotopes including, e.g., 
$\mathrm{^{48}Ca}$, $\mathrm{^{50}Ti}$, and several of the 
isotopes from $\mathrm{Zn}$ to $\mathrm{Zr}$ \citep{Jones2019}.

Despite the narrow range of central densities for which 
cECSNe are expected to occur \citep{Leung2020}, and despite the 
open questions associated with a variety of competing processes that
decide about collapse or thermonuclear explosion and that
depend strongly on many uncertain aspects of the employed physics, 
connections to cECSNe have been made for observations of SN1994N, 1997D, 
1999br, 1999eu, 2011dc and 2005cs \citep{Stevenson2014}. However,
comparisons of the nebular spectra of some of these cases with 
1D neutrino-driven SN models are ambiguous or disfavor the link to cECSNe
\citep{Jerkstrand2018}. Also SN1054 (the Crab) has been speculated 
to be a cECSN \citep{Nomoto1982,Hillebrandt1982,Tominaga2013,Smith2013},
although such an interpretation is 
in conflict with results by \cite{Gessner2018} for the maximum 
kick velocity of PNSs produced by cECSNe.

With the help of full-sphere three-dimensional simulations
we aim at investigating the following questions:

\begin{itemize}
\item What are the differences in the early stages (first seconds) of the 
explosion in low-mass Fe-core and ONeMg-core progenitors?

\item What is the influence of the different progenitor structures on 
the long-time evolution of the explosion? In particular, what 
is the influence on the formation of reverse shocks and the 
efficiency of outward mixing of neutrino-heated material?

\item Are CCSNe of low-mass progenitors able to produce highly 
asymmetric ejecta and strong radial mixing of metals similar 
to findings for more massive RSG and BSG stars in previous studies?

\item How do the properties of the compact remnants change on long time-scales
due to the fallback of matter? Are there significant changes to the remnants' mass, 
kick, and angular momentum?
\end{itemize}

The structure and contents of our paper are the following:
In Section~\ref{sec:progenitors}, the basic properties of the considered 
models of non-rotating, low-mass (super-AGB and RSG) progenitors are 
introduced. Section~\ref{sec:Numerical and physical Setup}
provides a brief description of the numerical methods and input
physics used in our simulations. 
Section~\ref{sec:Evolution during the first second} contains our
results for the first second(s) of the explosion, focusing on shock
dynamics, explosion energies, neutrino emission, PNS properties and
the chemical composition of the ejecta. 
In Section~\ref{sec:Evolution until Shock-Breakout}, for the
first time in 3D explosion modeling, the SN evolution of low-mass 
super-AGB and RSG progenitors is described until and beyond shock 
breakout concerning the development of mixing instabilities and ejecta
asymmetries, the spatial distribution of chemical elements, and
the effects of fallback on the properties of the newly formed NSs.
In Section~\ref{sec:Comparison to previous studies}, we briefly 
compare our results with previous studies, and in Section~\ref{sec:summary},
we conclude with a summary and 
discussion. Several appendices contain basic information on 
more technical aspects concerning the simulation inputs and 
the analysis methods.

\begin{figure*} 
 \centering
 \includegraphics[width=0.96\textwidth,trim=0cm 0.0cm 0cm 0cm,clip]{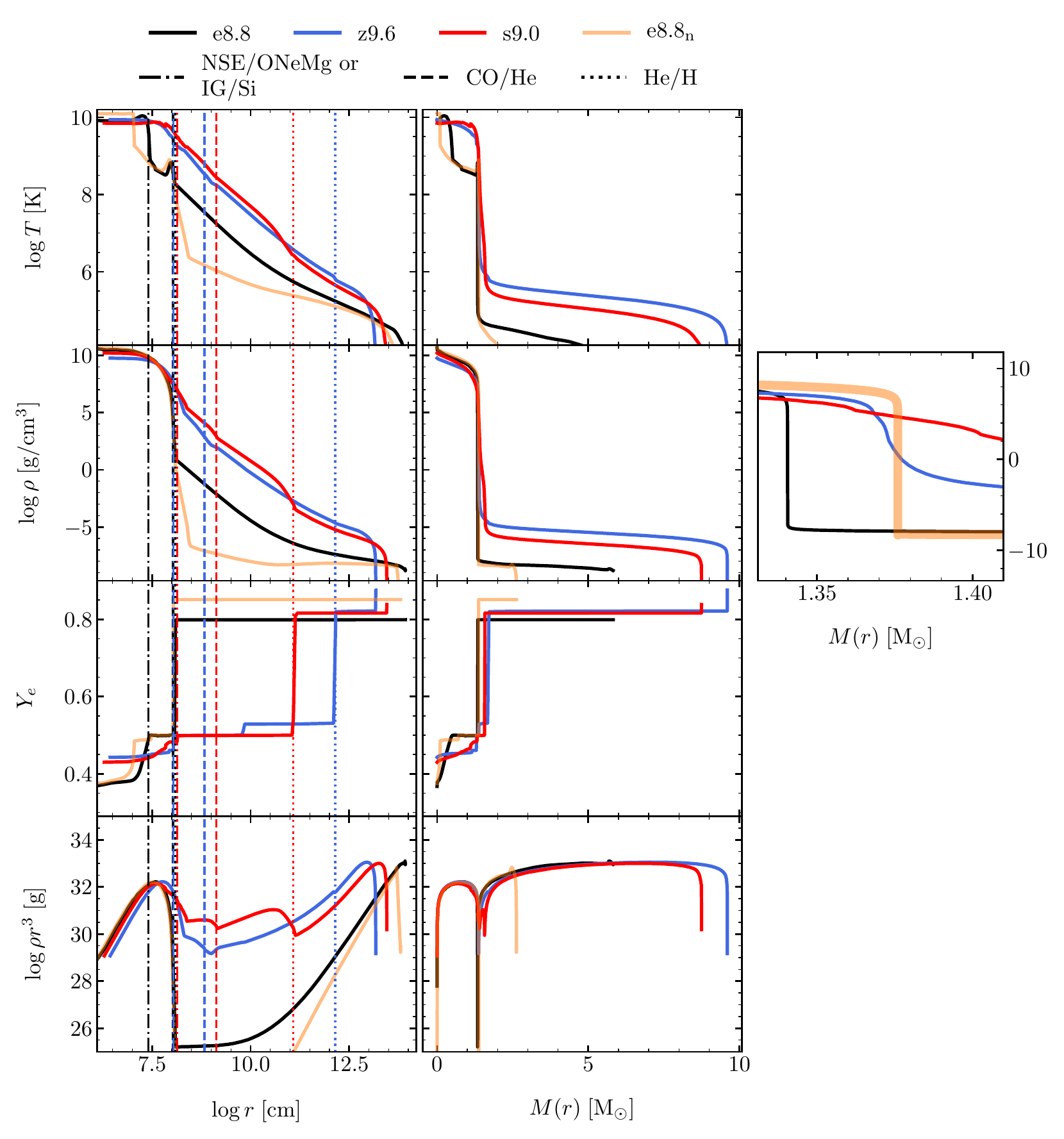}
 \caption{Profiles of the temperature ($T$), density ($\rho$), electron 
 fraction ($Y_e$) and $\rho r^3$ for the pre-collapse progenitor 
 models as functions of radial coordinate (left panels) and mass 
 coordinate (right panels). Indicated by dash-dotted, dashed, and 
 dotted lines are the outer boundaries of the degenerate 
 (iron or NSE), CO, and He 
 cores, respectively. Note the huge differences in the density 
 and $\rho r^3$-profiles between the progenitors with iron 
 and ONeMg-cores, in particular just outside of the CO core. 
 We show the difference in the 
 core structures of our ONeMg-core models in a zoom of the $\rho$ vs.\ $M(r)$
 profiles in the rightmost panel.}
 \label{fig:prog_tem_rho_ye_rhor}
\end{figure*}

\begin{figure*} 
 \centering
 \includegraphics[width=0.9\textwidth,trim=0cm 0.0cm 0cm 0cm,clip]{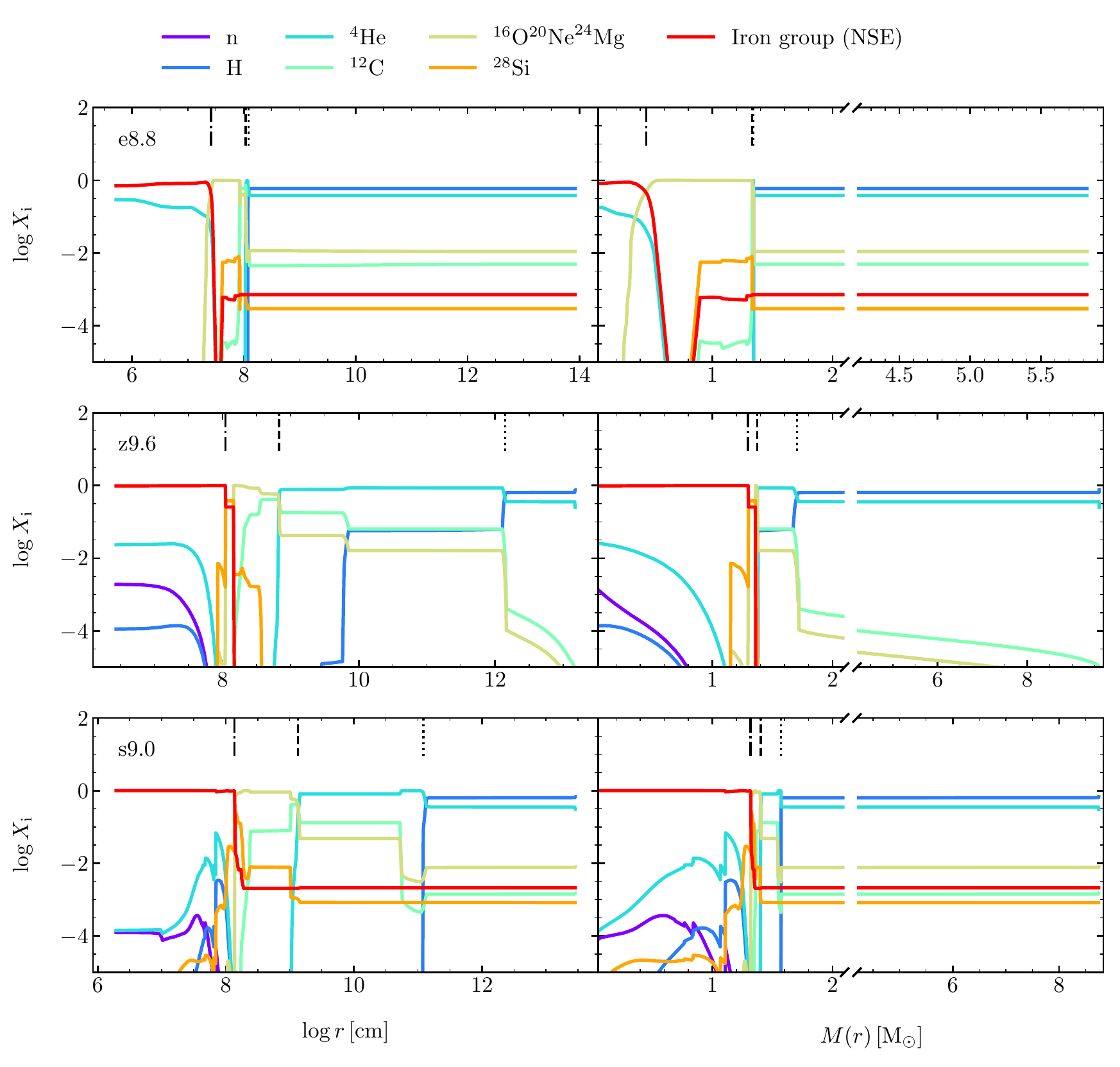}
 \caption{Pre-collapse composition of models 
 \onemg (top), \znine (middle) and \snine (bottom) as a function 
 of radius (left) and enclosed mass (right). Note the broken horizontal 
 axis of the right panels. We combine all chemical elements 
 with mass numbers greater than 28 into the ``iron-group'' (IG). The 
 dash-dotted, dashed, and dotted lines indicate the outer 
 boundaries of degenerate (iron or NSE), CO, and He cores, respectively.}
 \label{fig:composition_all}
\end{figure*}

\section{Progenitors}
\label{sec:progenitors}
In this paper, we study an ECSN of a non-rotating 8.8 \solm 
super-AGB star (\onemg), which is constructed from the envelope
model of \citet{Jones2013} and a collapsing core model 
(\citealt{Leung2020}; A.~Tolstov, S.-C.~Leung, and K.~Nomoto, 
2017, private communication), and two CCSNe resulting from 
non-rotating low-mass RSGs with iron cores 
(\znine, \snine), evolved to the onset of collapse by 
A.~Heger \citetext{2015, private communication} 
and by \cite{Woosley2015}, respectively. The considered ECSN 
progenitor is explored here for the first time, whereas the 
explosions of the iron core progenitors were simulated in 3D 
with the \vertexprom code and some results were published in 
\cite{Melson2015a} and \cite{Melson2019}.

\subsection{ONeMg-core progenitor}
\label{sec:onemgcoreprog}

Model \onemg is a solar-metallicity progenitor with a 
ZAMS mass of $M_{\rm ZAMS}\,\mathord{=}\,8.8\,\solm$.
Its degenerate ONeMg-core undergoes electron capture which ignites the
O-Ne deflagration at the center. The central density ($\rho_\mathrm{c}$) 
and the electron fraction ($Y_e$), and thus the core
mass $M$(ONeMg) at the ignition, depend on the convective stability
criterion, extent of the convective mixing, and the electron capture rate 
\citep{Zha2019}.  \citet{Leung2020} adopted the core with
$\rho_\mathrm{c}\,\mathord{\sim}\,10^{9.975}$\,g\,cm$^{-3}$,
$Y_e\,\mathord{=}\,0.496$, and 
$M(\mathrm{ONeMg})\,\mathord{=}\,1.39$\,M$_\odot$, and calculated the
propagation of the convective deflagration. The deflagration
incinerates ONeMg-core material into nuclear statistical equilibrium
(NSE). In the NSE region, electron capture on iron-group nuclei and
free protons dominates nuclear energy release, thus inducing collapse.
When the central density reaches 
$\rho_\mathrm{c}\,\mathord{\sim}\,10^{10.64}$\,g\,cm$^{-3}$, the
NSE region extends to $\sim$0.45\,M$_\odot$. The SN simulations started
from this progenitor condition. The structure of the core is well
approximated by a spherical model.

When transferring the progenitor model of \onemg to the SN simulations,
the mass of the ONeMg-core was reduced to 1.34\,M$_\odot$ in the course
of providing an easy-to-handle fit to the complex density structure. This
has no relevant influence on the dynamical evolution of the explosion, as
can be concluded from the close similarity of the explosion behavior of
model e8.8 with that of a previous version of the progenitor
(\citealt{Nomoto1984}; \citealt{Nomoto1987}; K.~Nomoto, 2008, private communication),
which is commonly termed \onemg in the literature and $\mathrm{e}8.8_{\mathrm{n}}$ 
in this publication. Model e8.8$_\mathrm{n}$ had a degenerate core of 
1.375\,M$_\odot$, but a smaller hydrogen envelope 
(see Figure~\ref{fig:prog_tem_rho_ye_rhor}). It is a reference case for simulations 
of ECSNe and was used in various studies, focusing on explosion properties 
\citep{Kitaura2006,Fischer2010}, nucleosynthesis \citep{Janka2008,Wanajo2011,Wanajo2018}, 
effects of microphysics \citep{Huedepohl2010,Groote2014,Radice2017}, and PNS 
kicks \citep{Gessner2018}. In order to compensate for possible uncertainties in
our ECSN simulations with the new progenitor model of e8.8,
we vary its explosion energy in a set of 2D simulations, 
denoted by e8.8$_3$, e8.8$_6$, e8.8$_{10}$, and e8.8$_{15}$ 
for $E_\mathrm{exp}\,\mathord{=}\,(3,\,6,\,10,\,15)\times 10^{49}$\,erg.

Progenitor model \onemg has the very sharp density gradient at $1.34\,\solm$
near the edge of its compact
ONeMg-core ($\xi_{2.5}\,\mathord{=}\,5.7\times10^{-6}$, 
$\xi_{1.5}\,\mathord{=}\,8.0\times10^{-6}$)\footnote{Because the core 
structure of higher-mass progenitors can be characterized by the compactness 
parameter
{$\xi_{M}\,\mathord{\equiv}\,\frac{M/\solm}{R(M_{\rm bary}\mathord{=}M)/10^{3}\, \rm km}$}
\citep{OConnor2011}, we provide values for this quantity here.
However, due to the steep density gradient outside of the core in the 
considered progenitors, 
the value of this parameter in the context of our work is limited.} that
is characteristic of such progenitors prior to the onset of core collapse.
This steep gradient
is a prominent feature of the density ($\rho$) profile in Figure~\ref{fig:prog_tem_rho_ye_rhor}.
In the same same figure
we also show the $\rho r^3$-profile as well as the electron fraction 
$Y_e$ and the temperature $T$. All of these quantities are displayed
as functions of enclosed mass and radius. The inner $\sim$0.45\,M$_\odot$
of the degenerate core contain iron-group nuclei and $\alpha$-particles
in NSE. The dash-dotted, dashed, and dotted lines indicate the positions 
of the NSE/ONeMg, CO/He, and He/H composition interfaces, respectively. 
We define the locations of the composition interfaces, similar to 
\cite{Wongwathanarat2015}, as those positions at the bottom of the respective 
layers of the star where the mass fractions $X_{i}$ drop below half of their 
maximum values in the layer. The radial positions of the composition 
interfaces are summarized in Table~\ref{tab:progenitors}.
In the top panel of Figure~\ref{fig:composition_all} we present the 
composition of model \onemg, where we combine all elements with mass 
numbers greater than 28 into ``iron-group'' (IG) material or nuclei
in NSE.
The ONeMg-core is surrounded by very thin carbon and helium shells 
($M_{\rm C}\,\mathord{\approx}\,8.1\mathord{\times}10^{-3}\, \solm$, 
$M_{\rm He}\,\mathord{\sim}\,2.1\mathord{\times}10^{-6}\, \solm$) and a 
hydrogen ($\mathrm{H}\mathrm{+}\mathrm{He}$) 
envelope ($M_{\rm H}\,\mathord{\approx}\, 4.49\,\solm$).  
The total masses of the different nuclei present in the entire pre-SN model
are listed in Table~\ref{tab:composition tmap}.

For collapse and post-bounce evolution of an ONeMg-core progenitor we
employ in all of this paper
the new progenitor model e8.8 in 1D, 2D, and 3D simulations with the 
\prom code as detailed in Section~\ref{sec:Collapse and post-bounce setup in prom}.
The profile of the old progenitor e8.8$_\mathrm{n}$ 
is shown in Figure~\ref{fig:prog_tem_rho_ye_rhor} 
merely for illustration and reference.

\begin{table*}
   \caption{Shell structure of the pre-collapse progenitor models.}
   \label{tab:progenitors}
    \renewcommand{\arraystretch}{1.11}
   \begin{tabular}{l l l l l c} 
   \hline
     Model      & 
     Interface & 
     $R_\mathrm{shell}\,$ &
     $M_\mathrm{shell}$ &
     $t_{\mathrm{sh,max}}^{\mathrm{3D}}$ &
     $E_{\mathrm{bind}}$ \\ [0.5ex] 
                &          
                & 
           [cm] & 
      $[\solm]$ & 
      $[\s]$    &
       $[10^{49}\,\erg]$ \\ [0.5ex] 
   \hline
   \multirow{5}{*}{\onemg} & NSE/ONeMg & $2.57\mathord{\times} 10^{7}$  & 0.45 & - & \\ 
                           & ONeMg/C  & $8.48\mathord{\times} 10^{7}$  & 1.33 & 0.19 & $E_{\mathrm{bind}}(m>M_{\mathrm{map}})=-5.99$ \\
                           & C/He     & $1.09\mathord{\times} 10^{8}$  & 1.34 & 0.20 & \\
                           & He/H     & $1.21\mathord{\times} 10^{8}$  & 1.34 & 0.21 & \\
                           & Surface  & $8.43\mathord{\times} 10^{13}$ & 5.83 & $4.1\mathord{\times}10^5$ & \\
   \hline
   \multirow{6}{*}{\znine} & IG/Si    & $1.10\mathord{\times} 10^{8}$  & 1.30 & 0.09 &  \\ 
                           & Si/CO    & $1.45\mathord{\times} 10^{8}$  & 1.36 & 0.11 &  $E_{\mathrm{bind}}(m>M_{\mathrm{map}})=-5.82$ \\ 
                           & CO/He    & $6.48\mathord{\times} 10^{8}$  & 1.37 & 0.40 &\\
                           & He/H (H$<$1\%)     & $6.24\mathord{\times} 10^{9}$  & 1.38 & 2.64 & \\
                           & He/H (H$<$10\%)    & $1.40\mathord{\times} 10^{12}$ & 1.70 & $1.8\mathord{\times}10^{3}$ & \\
                           & Surface  & $1.50\mathord{\times} 10^{13}$ & 9.60 & $1.1\mathord{\times}10^5$ & \\
   \hline
   \multirow{5}{*}{\snine} & IG/Si    & $1.24\mathord{\times} 10^{8}$  & 1.30 & 0.08 &   \\ 
                           & Si/CO    & $1.55\mathord{\times} 10^{8}$  & 1.33 & 0.30 & $E_{\mathrm{bind}}(m>M_{\mathrm{map}}) =-2.63$\\ 
                           & CO/He    & $1.34\mathord{\times} 10^{9}$  & 1.40 & 1.30 & \\
                           & He/H     & $1.22\mathord{\times} 10^{11}$ & 1.57 & 124.0 & \\
                           & Surface  & $2.86\mathord{\times} 10^{13}$ & 8.75 & $1.8\mathord{\times} 10^5$ & \\
  \hline
\end{tabular}
\flushleft
\textit{Notes}: The radii of the composition interfaces, $R_{\mathrm{shell}}$,
are defined as those positions at the bottom of the stellar layers (e.g. CO)
where the mass fractions (e.g. C+O) drop below half of their maximum values 
in the respective layer. 
Progenitor model z9.6 is in the process of deep 2nd
dredge-up with hydrogen reaching basically to the bottom of the former helium
layer; hence two values are provided for the He/H interface. In all
Figures we refer to the ``high'' value, since it is often used also in other context, 
e.g., for the ``$\alpha$-parameter'' in common-envelope (CE) studies.
We also show the mass $M_{\mathrm{shell}}$ contained within the corresponding 
radius and the post-bounce time when the outermost radius of the forward 
shock of our 3D models reaches the interface. $E_{\mathrm{bind}}$ is the binding
 (i.e., internal + kinetic + potential) energy in the progenitor star outside of
$M_\mathrm{map}$, which is the location of the final mass cut. For values see 
Table~\ref{tab:long term boundaries}.  
\end{table*}

\subsection{Fe-core progenitors}

As a second progenitor we employ a zero-metallicity 
$M_{\text{ZAMS}}\,\mathord{=}\,9.6\,\solm$ star, termed \znine.
It was first used by \citet{Janka2012} and was also considered in other 
studies such as \citet{Mueller2013,Radice2017,Mueller2019}. This iron core 
progenitor is structurally similar to the ECSN model. It also shows a 
sharp decline of the density at the edge of its iron core, enabling low-energy 
explosions in 1D \citep{Melson2015a,Radice2017}. 
Evolved by A.~Heger (2012, private communication) as an extension 
to \cite{Heger2010}, the pre-SN model develops an 
iron core of about 1.30\,\solm. The iron core is surrounded by a $0.061\,\solm$ 
Si-layer, a $0.016\,\solm$ CO-layer, and has a hydrogen-free helium layer of 
about 0.004\,M$_\odot$ below a 0.33\,M$_\odot$ convective H/He-layer with a 
hydrogen mass fraction of $\sim$6\%, which is surrounded by a massive hydrogen 
envelope of nearly $8\,\mathrm{M}_\odot$
($\xi_{2.5}\,\mathord{=}\,7.66 \times 10^{-5}$, $\xi_{1.5}\,\mathord{=}\,2.38 \times 10^{-4}$).  
The H/He-layer is in the process of extended,
but incomplete, 2nd dredge-up of the He-layer that is typical for
AGB/sAGB stars and those at the low-mass end of the CCSN domain. The
star is basically an iron-core AGB star, maybe should be called a
hyper-AGB (hAGB) star.

As the envelope is not rich of metals, mass loss is expected to play only a negligible 
role during the star's evolution. This leaves the total mass of the star 
almost unchanged (see Table~\ref{tab:progenitors}). 
Due to its structure it was one of the first iron-core progenitors that exploded 
in fully self-consistent 3D simulations by \cite{Melson2015a} with \vertexprom and 
was also investigated by \cite{Radice2017} and \cite{Mueller2019}. 
The result of \cite{Melson2015a} provides the initial state for our investigation. 

Moreover, we investigate a solar-metallicity $M_{\text{ZAMS}}\,\mathord{=}\,9.0\,\solm$ 
star, termed \snine, of \citet{Sukhbold2016}. Its 1.30\,\solm iron core is 
surrounded by a silicon shell of 0.03\,\solm, a carbon-oxygen layer of 0.068\,\solm, 
and a helium shell of 0.169\,\solm (see Table~\ref{tab:progenitors}). 
The hydrogen envelope extends from 1.57\,\solm up to 8.75\,\solm 
($\xi_{2.5}\,\mathord{=}\,3.83 \times 10^{-5}$, 
$\xi_{1.5}\,\mathord{=}\,5.24 \times 10^{-3}$). In comparison to model \znine,
model \snine is just slightly less evolved on its track to 2nd dredge-up of 
the He core, however, the convection is better driven by the iron-peak
opacity from the outside-in.
The \snine progenitor was chosen to be representative for low-mass CCSNe 
by \cite{Jerkstrand2018}, who studied the late-time nebular spectra 
of the supernova (SN), by \cite{Glas2019} focusing on the neutrino 
emission during the explosion, and by \cite{Burrows2019} in a large set of 
3D simulations. The three-dimensional exploding model for 
our investigation is provided by \citet{Melson2019} and has also been 
modeled with \vertexprom.

Although the progenitors considered in this study have very 
similar ZAMS masses, 
their pre-collapse core structures differ significantly 
(see Figure~\ref{fig:prog_tem_rho_ye_rhor}).
We stress that the $\rho r^3$-profiles of the progenitors are 
decisive for the long-time evolution of the explosion 
\citep{Kifonidis2003,Wongwathanarat2015}. 
The behavior of the $\rho r^3$-profile yields important information on 
the propagation of the shock through the stellar structure, because, 
according to \cite{Sedov1961}, positive gradients of $\rho r^3$ cause 
shock deceleration, whereas negative gradients cause the opposite. 
Additionally, variations of the shock velocity produce crossing 
pressure and density gradients behind the shock front near the 
composition-shell interfaces \citep{Chevalier1978}. 
Such conditions are essential for RT instabilities as detailed in 
Section~\ref{sec:Linear stability analysis}, assigning them a 
crucial role for explaining high-velocity metal-rich ejecta 
and radial mixing of heavy elements 
\citep[e.g.,][]{Arnett1989,Nomoto1990,Wongwathanarat2015}.

The ECSN progenitor exhibits an extremely sharp drop of the 
$\rho r^3$-profile just outside of the ONeMg-core. It is this drop 
in density that enables fast explosions due to an early, rapid decline
of the mass accretion rate and of the corresponding ram pressure at the 
SN shock \citep{Kitaura2006}. Outside of the core, the $\rho r^3$-profile 
grows monotonically as no other composition interfaces are encountered.
The \znine progenitor falls into the class of ECSN-like progenitors also 
in this respect: Similar to the electron-capture progenitor, the \znine
model shows a monotonic growth of the $\rho r^3$-profile exterior to the CO 
core, where only a small step in the density profile can be seen at the 
He/H interface.
Model \snine on the other hand exhibits strong variations in its 
$\rho r^3$-profile. Each interface of different composition layers
is accompanied
by a negative $\rho r^3$-gradient close to the interface. Of 
particular interest are the CO/He and He/H interfaces. These interfaces
have an impact on the long-time evolution of the explosion as will be 
discussed in Section~\ref{sec:Linear stability analysis}.

For this paper we performed spherically symmetric (1D), 
axisymmetric (2D) and fully three-dimensional (3D) 
simulations for all progenitors beyond the moment when the shock reaches 
the surface of the star. The different setups and approaches 
for these simulations will be described in the following section.

\clearpage

\section{Numerical Methods and Physical Setup}
\label{sec:Numerical and physical Setup}

In order to cover collapse, shock revival, 
and shock propagation through the envelope and 
circumstellar material we employ a step-wise approach.
Core collapse, bounce, and post-bounce evolution until the 
explosion is well on its way, i.e., the neutrino-dominated
phases of the first second(s) around and after core bounce, of the 
iron-core progenitors are simulated with the \vertexprom code.
The corresponding long-time simulations covering the shock propagation 
through the star, after the onset of the explosion until and beyond shock 
breakout, are conducted with the \prom code. The explosion of the 
ECSN progenitor is simulated entirely with \prom.\footnote{The reason for these different treatments is
mainly technical: The \vertexprom version used for ONeMg core collapse by
\citet{Kitaura2006,Janka2008}, and \citet{Huedepohl2010} has not yet been updated
with the 3D developments and parallelisation optimization applied to the 
version used for Fe-core collapse. However, for suitable choices of parameter 
values, the explosion dynamics of ECSNe 
computed with \prom is very similar to the fully self-consistent 
2D \vertexprom and {\textsc{CoCoNuT-Vertex}} explosion models 
discussed by \citet{Janka2008} and \citet{Janka2012}, respectively.}

In the following sections we describe the numerical and physical 
features of the codes and the setups applied during the different 
stages of the evolution.

\subsection{\vertexprom code}
\vertexprom is a hydrodynamics code based on an implementation of the 
Piecewise Parabolic Method (PPM) of \cite{Colella1984}, coupled 
with a three-flavor, energy-dependent, ray-by-ray-plus (RbR+) 
neutrino transport scheme that iteratively solves the neutrino energy and momentum 
equations with a closure determined from a tangent-ray Boltzmann solver
\citep{Rampp2002}. It employs the full set of neutrino reactions and 
microphysics presented in \cite{Buras2006} and the high-density equation 
of state (EoS) of \cite{Lattimer1991} with a nuclear incompressibility of 
$\mathrm{K}\,\mathord{=}\,220\,\mathrm{MeV}$.
At low densities ($\rho\,\mathord{\leq}\,\rho_{\mathrm{HD}}\,\mathord{=}\,10^{11}\,\mathrm{g/cm^3}$) 
\vertexprom uses the EoS of \cite{Janka1995}, which includes the contributions of 
photons, arbitrarily degenerate and arbitrarily relativistic $e^+/e^-$, 
and non-degenerate, non-relativistic nucleons and nuclei. 
The relative abundances of 23 nuclear species (including some neutron-rich
nuclei) are determined by an NSE solver in 
regions with temperatures above $T_{\mathrm{NSE}}$. Below $T_{\mathrm{NSE}}$ a 
flashing scheme is used to approximately treat nuclear burning 
(see \citealt{Rampp2002}). For unshocked, collapsing stellar matter we choose 
$T_{\mathrm{NSE}}\,\mathord{=}\,0.5\,\mathrm{MeV}$, and for neutrino-heated
postshock matter we take $T_{\mathrm{NSE}}\,\mathord{=}\,0.5\,\mathrm{MeV}$ for
the simulation of model \znine and $T_{\mathrm{NSE}}\,\mathord{=}\,0.34\,\mathrm{MeV}$ for model \snine.\footnote{\vertexprom does not apply a nuclear network
for $T\,\mathord{<}\,T_{\mathrm{NSE}}$. The choice of a lower value of $T_{\mathrm{NSE}}$ permits us to follow the ejection of mass through 
neutrino heating for a longer 
time period in model \snine, where the expansion velocity of the expelled 
matter is smaller than in \znine, thus facilitating
nucleon recombination to $\alpha$ particles and heavy nuclei.}

The simulations presented here are performed with a 1D gravitational potential 
including general relativistic corrections, Case~A of \citet{Marek2006}.
The neutrino transport solver contains corrections for general relativistic
redshift and time dilation effects. \vertexprom
makes use of the axis-free Yin-Yang grid \citep{Kageyama2004} based on 
the implementation of \cite{Melson2016}.

\subsection{\prom code}
\label{sec:PHOTBcode}

\prom is based on the same hydrodynamics module as 
\vertexprom and uses the implementation of the Yin-Yang grid 
presented in \cite{Wongwathanarat2010}. It employs the EoS 
of \cite{Lattimer1991} for high densities above a threshold value 
$\rho_{\mathrm{HD}}$ (usually $10^{11}\,\mathrm{g/cm^3}$) and the 
``Helmholtz'' EoS of \cite{Timmes1999} for densities below 
$\rho_{\mathrm{HD}}$, which takes into account arbitrarily degenerate 
and relativistic electrons and positrons, photons, and a set of 
non-degenerate, non-relativistic nuclei.
The set of nuclei consists of neutrons $n$, protons $p$, 
13 $\alpha$-nuclei ($^4{\mathrm{He}}$, $^{12}\mathrm{C}$, $^{16}\mathrm{O}$, 
$^{20}\mathrm{Ne}$, $^{24}\mathrm{Mg}$, $^{28}\mathrm{Si}$, $^{32}\mathrm{S}$, 
$^{36}\mathrm{Ar}$, $^{40}\mathrm{Ca}$, $^{44}\mathrm{Ti}$, $^{48}\mathrm{Cr}$, 
$^{52}\mathrm{Fe}$, $^{56}\mathrm{Ni}$), and an additional tracer nucleus \tracer,
which tracks the production of neutron-rich nuclei and replaces \nickel
in environments with low electron fraction, $Y_{e}\mathord{<}0.49$. 
These nuclear species 
are described as non-relativistic Boltzmann 
gases. The advection of the species is treated with the Consistent Multi-fluid 
Advection (CMA) scheme of \cite{Plewa1999}.
NSE is assumed above $T_{\mathrm{NSE}}\,\mathord{=}\,9\times 10^9$\,K
and accounted for by an NSE table including the nuclei listed above 
\citetext{Kifonidis 2004, private communication}.
Nuclear burning is considered at temperatures below $T_{\mathrm{NSE}}$ with a 
13-species $\alpha$-network, which is consistently coupled to the hydrodynamic 
modeling. At the boundary between network and NSE we assume that all free 
neutrons and protons recombine to yield \helium. We thus add the mass fractions 
of $p$ and $n$ onto the mass fraction of \helium, 
accounting for the corresponding energy 
release\footnote{In a newer version of the code we allow only paired 
free neutrons and protons to recombine to \helium, thus also satisfying 
charge conservation.}.
The \prom code uses a 3D gravitational potential with the general-relativistic 
(GR) monopole correction of \cite{Marek2006} as discussed by \cite{Arcones2007},
while higher multipoles are obtained from a solution of Poisson's equation as 
described in \cite{Mueller1995}.

Different from \vertexprom, \prom uses a three-flavor grey 
neutrino transport scheme as presented in \cite{Scheck2006},\footnote{Some 
improvements to the neutrino transport module are described 
in Appendix~\ref{Appendix:Neutrino}.} which is applicable at low and
moderate optical depths. Therefore, the high-density core of the PNS, with 
a mass of $M_{\mathrm{c}}\,\mathord{=}\,1.1\,\solm$ and densities well above 
those of the neutrinospheric layer, is replaced by a closed
(Lagrangian) inner grid boundary at radius $R_{\mathrm{ib}}$. The 
excised 1.1\,\solm PNS core is taken into account in the gravitational 
potential as a central point mass. The contraction of 
the PNS is mimicked by an inward movement of the boundary radius 
$R_{\mathrm{ib}}$, whose motion is followed by all grid points
in the computational domain. We use the contraction of $R_{\mathrm{ib}}$ 
as prescribed by \cite{Ertl2016}.
The time-dependent neutrino luminosities at $R_{\mathrm{ib}}$ are imposed as 
boundary conditions as provided by an analytic one-zone cooling model following
\cite{Ugliano2012}, \cite{Sukhbold2016}, \cite{Ertl2016}, and \cite{Ertl2020}.
This time-dependent treatment of the central-core region employs five 
parameters ($p$, $a$, $R_{\mathrm{c,f}}$, $t_{0}$, $R_{\mathrm{ib,f}}$;
see Appendix~\ref{Appendix:prom inner boundary} for definitions), 
which can be calibrated to yield explosions that fulfill the constraints 
set by observed SNe or by fully self-consistent 3D simulations of CCSNe.
The reader is referred to 
Appendix~\ref{Appendix:prom inner boundary} for a more detailed 
description of the parametric approach and to Table~\ref{table:e8param}
for the parameters of the core model employed in our work.

\subsection{Collapse and post-bounce setup in \vertexprom}
\label{sec:Collapse and post-bounce setup in vertexprom}

The collapse of the iron core progenitors is computed in 1D using the 
full set of neutrino interactions until 10 ms after bounce. Thereafter, 
the simulations are mapped onto the three-dimensional Yin-Yang 
grid and random cell-to-cell density perturbations are imposed 
with an amplitude of 0.1\%. The simulations employ a 
non-equidistant radial grid with initially 400 zones extending 
to $10^9\,\text{cm}$, which is refined in steps to more than 600 zones. 
This guarantees a resolution $\Delta r/r$ of better than $1\%$ at 
the gain radius. The innermost 1.6 km are calculated in spherical 
symmetry to avoid time stepping constraints at the grid center.
The angular resolution of the \znine model is $2^{\circ}$. 
The post-bounce evolution of model \snine is computed 
with a newly implemented static mesh refinement (SMR) scheme presented in 
\cite{Melson2019}, which increases the angular
resolution to $1^{\circ}$ outside of
the gain radius and to $0.5^{\circ}$ exterior to a radius of 160 km.

The simulations with full neutrino transport are
too expensive to continue them to late post-bounce times. 
At $\tpb\,\mathord{\gtrsim}\,0.5\,\s$ the neutrino transport is therefore 
switched off and replaced by a simplified scheme for neutrino heating 
and cooling, which ensures an essentially seamless continuation 
with a minimum of transient artifacts. Details of this scheme are 
given in Appendix~\ref{appendix:scaling relations}.
During this phase of simplified neutrino treatment both model
\znine and \snine are simulated with uniform angular resolution
of 2$^\circ$.

\subsection{Collapse and post-bounce setup in \prom}
\label{sec:Collapse and post-bounce setup in prom}

During the spherically symmetric simulation of the collapse up 
to core bounce, \prom uses the parametrized deleptonization scheme 
described in \citet{Liebendorfer2005}. The necessary 
$Y_{e}(\rho)$-trajectory was provided by H\"udepohl \citetext{2018, private communication} from 
his core-collapse simulations of the ONeMg-core progenitor 
$\mathrm{e8.8_{n}}$ with \vertexprom.

For the simulation of the ECSNe progenitor we take
$\rho_{\mathrm{HD}}\,\mathord{=}\,10^{11}\,\mathrm{g\,cm^{-3}}$ and 
assume NSE in regions where the temperature exceeds
$T_{\mathrm{NSE}}\,\mathord{=}\,9\mathord{\times}10^9\,\mathrm{K}$
and apply the $\alpha$-network for temperatures lower than this
value. After bounce \prom employs the grey neutrino transport scheme
and modeling approach as presented in \cite{Scheck2006}. Thus,  
the neutrino-opaque central core of the PNS is excised from the
computational domain and replaced by the analytic core model of
\citet{Ugliano2012}.
In Table~\ref{table:e8param} we list the parameter values of the 
PNS core model used for a set of simulations of model \onemg. 
We perform 1D and 2D simulations for all four sets of parameter 
values and 
choose the $\rm e8.8_{10}$ calibration as our reference case for
a 3D simulation. The 1D and 2D simulations possess a 
non-equidistant radial grid with 2000 zones up to a radius of 
$R_{\mathrm{ob}}\,\mathord{=}\,2\mathord{\times}10^{10}\,\text{cm}$. 
The 3D run has only 1400 radial zones for computational efficiency. 
The multi-dimensional simulations are conducted with an angular 
resolution of $2^{\circ}$, and the 3D simulation makes use of
the Yin-Yang grid.
We restrict ourselves to a 1D gravitational potential with GR
corrections \citep{Arcones2007} because the explosions
of model \onemg are nearly spherical.

\begin{table}
\centering
\caption{Summary of the PNS core parameter values used for the 
\onemg model and resulting explosion energies from 1D simulations. 
Definitions of these parameters in the context of our 
modeling approach are given in Appendix~\ref{Appendix:prom inner boundary}.
The explosion energy is essentially independent of dimensionality 
(1D, 2D, 3D).}
  \label{table:e8param}
    \renewcommand{\arraystretch}{1.11}
   \begin{tabular}{l  c   c   c   c   c   c}
  \hline
  Model &
  $E_{\mathrm{exp}}$ &
  $p$ & 
  $a$ & 
  $R_{\mathrm{c,f}}$ &
  $t_0$ & 
  $R_{\mathrm{ib,f}}$ \\
                &
  [foe] &
  [index] &
  [factor] &
  [$\km$]  &
  [$\s$] &
  [$\km$] \\
  \hline
  $\mathrm{e}8.8_{3}$  & 0.03 & $-3$ & $1.0\times 10^{-2}$ & 27 & 0.1 &  40 \\
  $\mathrm{e}8.8_{6}$  & 0.06 & $-3$ & $1.2\times 10^{-2}$ & 22 & 0.1 &  40 \\
  $\mathrm{e}8.8_{10}$ & 0.10 & $-3$ & $4.0\times 10^{-1}$ & 20 & 0.1 &  40 \\
  $\mathrm{e}8.8_{15}$ & 0.15 & $-3$ & $5.8\times 10^{-1}$ & 18 & 0.1 &  40 \\
  \hline
  \end{tabular}
\end{table}

\subsection{Setup for the long-time simulations}
\label{Setup during the long-term evolution}

The simulations of the long-time evolution of all models are 
computed with \prom. For this we map the final state of a post-bounce 
simulation at time $t_\mathrm{map}$ onto a 
new computational grid within \prom, similar to the procedure 
described in \cite{Wongwathanarat2015}. We also add the low-density 
extensions to the Helmholtz EoS described therein. In 
Table~\ref{tab:long term boundaries} we list the times of mapping 
and the inner and outer radii of the new computational domain,
$R_{\mathrm{ib}}$ and $R_{\mathrm{ob}}$, respectively.
The mass contained within $R_{\mathrm{ib}}$ 
is treated as a point mass\footnote{We ensure that matter at radii 
smaller than $R_{\mathrm{ib}}$ has velocities smaller than the 
local escape velocity and will thus eventually contribute the to final NS mass.}
and is called $M_{\mathrm{map}}$.

The time $t_\mathrm{map}$ is chosen such that the explosion
energy has effectively converged to its asymptotic value and a
neutrino-driven wind region has developed around the PNS, where
the outflow is essentially spherical and reaches supersonic velocity.

All long-time simulations are computed with an angular 
resolution of $2^{\circ}$. We use a non-equidistant (geometrically 
increasing) radial grid from the inner to the outer boundary. In 
order to guarantee sufficient resolution where needed, the radial 
grid is allowed to move with the ejecta starting from $\tpb\mathord{\sim}10\,\s$. 
Gravity is accounted for by a 1D GR-corrected potential and nuclear reactions 
are still considered. 
When mapping the iron core models, z9.6 and e9.0, into \prom, we recombine free 
$n$ and $p$ from the freeze-out of NSE into \helium under the 
condition of charge conservation and account for the energy release. Moreover, we combine
all neutron-rich nuclei formed in neutrino-heated ejecta into tracer ($\tracer$) material.

\begin{figure} 
 \centering
 \includegraphics[width=0.50\textwidth,trim=0.2cm 0.2cm 0.2cm 0.2cm,clip]{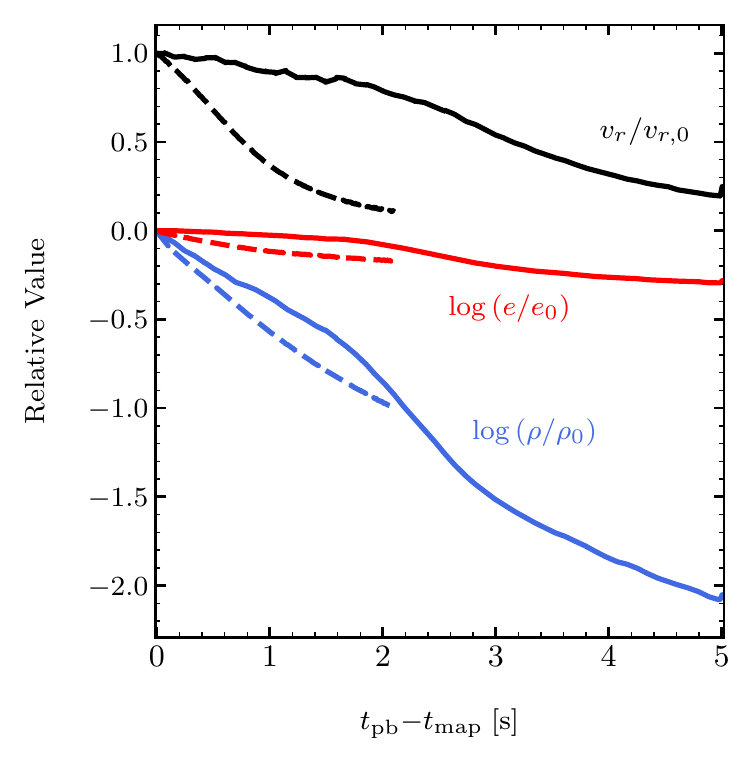}
 \caption{Time-dependent behavior of neutrino-wind density $\rho$, radial velocity $v_r$, 
 and total energy density $e$, normalized to their initial values at 
 $t_{\mathrm{map}}$ (see Table~\ref{tab:long term boundaries}),
 which defines the start of the long-time simulations 
 of model \onemg and \znine. The data for model \znine are extracted from a 1D simulation of the explosion of the 
 $9.6\,\solm$ progenitor with \vertexprom (see \citealt{Mirizzi2016}), evaluated at a radius of 600 km (solid lines). For model \onemg we use data of the respective 1D
 simulation with \prom (dashed lines).}
\protect{\label{fig:wind}}
\end{figure}

When mapping from the simulations of the onset of the explosion to the 
follow-up simulations, the central region interior to $R_{\mathrm{ib}}$ 
(Table~\ref{tab:long term boundaries}) is removed from the computational 
domain. Similar to \cite{Wongwathanarat2015} we prescribe an inflow boundary 
condition at $R_{\mathrm{ib}}$, which corresponds to the neutrino-driven 
baryonic mass-loss (``neutrino-wind''; e.g \citealt{Qian1996}) generated 
by ongoing neutrino-energy deposition in the surface layers of the cooling 
PNS. In contrast to \cite{Wongwathanarat2015}, we employ neutrino-wind results 
adopted from 1D simulations of the explosions, seamlessly connected to the 
fully multi-dimensional explosion simulations by choosing $R_{\mathrm{ib}}$ 
to be in the supersonic wind region (ensuring that perturbations cannot 
propagate back to the inner boundary creating artifacts) and by applying 
the wind data at times when the outflow properties match closely between 
the 1D and the (angle-averaged) multi-dimensional models. 
This is possible because the PNSs in 1D and 
multi-dimensional models are extremely similar and the neutrino-emission and 
thus the neutrino-driven winds also have very similar properties. 
For the long-time run of model \znine, we therefore employ neutrino-wind 
conditions of a 1D simulation of this model with the \vertexprom code. This 
model treats PNS convection with an approach based on mixing-length theory
and exhibits neutrino-emission properties that are hardly distinguishable  
from the multi-dimensional calculation \citep[see]{Mirizzi2016}. 
The time dependence of the radial velocity 
$v_r$, density $\rho$, and total (i.e. kinetic + internal) energy density 
$e$, which are needed for setting the boundary condition, are shown in 
Figure~\ref{fig:wind} (solid lines) as extracted from the 1D explosion 
simulation of the \znine model at a radius of 600 km (in the supersonic 
wind domain). For the 3D simulation of model \onemg we use the neutrino-wind 
results from the corresponding 1D run with the same explosion energy (see 
dashed lines in Figure~\ref{fig:wind}), whereas we do not impose a wind 
boundary condition in the long-time simulation of model \snine, since the 
neutrino-driven wind in this model is already very weak at the time of 
mapping. Using time dependences of the boundary conditions normalized by 
the initial value at the mapping time $t_{\mathrm{map}}$ guarantees a 
smooth, seamless transition from the earlier evolution to the long-time 
evolution of the explosion. Transient artifacts are thus kept minimal.

\begin{table} 
 \centering
 \caption{Initial positions of inner ($R_\mathrm{ib}$) and 
 outer ($R_\mathrm{ob}$) grid boundaries, baryonic mass contained within
 the inner boundary ($M_{\mathrm{map}}$)
 and mapping times $t_\mathrm{map}$ in seconds 
 after bounce for our long-time simulations. }
\label{tab:long term boundaries}
\renewcommand{\arraystretch}{1.07}
\begin{tabular}{cccccc}
  \hline
 Model & Dim. & $R_\mathrm{ib}$ & $R_\mathrm{ob}\,$ & $M_{\mathrm{map}}$ &  $t_{\mathrm{map}}\,$   \\
       & & [km] & [km] & $[\solm]$ & [s]  \\
  \hline
 $\mathrm{e}8.8_{3}$  & 2D & 1000 & $8.4\times10^{8}$ & 1.334 & 2.515 \\
 $\mathrm{e}8.8_{6}$  & 2D & 1000 & $8.4\times10^{8}$ & 1.327 & 2.515 \\
 $\mathrm{e}8.8_{10}$ & 2D & 1000 & $8.4\times10^{8}$ & 1.319 & 2.515 \\
 $\mathrm{e}8.8_{15}$ & 2D & 1000 & $8.4\times10^{8}$ & 1.309 & 2.515 \\
 $\mathrm{e}8.8$      & 3D & 500  & $8.4\times10^{8}$ & 1.326 & 0.470 \\
 $\mathrm{z}9.6$      & 3D & 600  & $1.5\times10^{8}$ & 1.353 & 1.440 \\
 $\mathrm{s}9.0$      & 3D & 1000  & $2.9\times10^{8}$ & 1.351 & 3.140 \\
  \hline
\end{tabular}
\end{table} 

Additionally, for the long-time simulations 
we include the decay of radioactive nickel, which 
becomes a relevant source of energy during late phases of the explosion.
Radioactive \nickel 
(half-life $t_{1/2}\,\mathord{=}\,  6.077\, \mathrm{days}$) 
decays to \cobalt via electron capture (EC) decay. The resulting 
\cobalt nucleus is unstable ($t_{1/2}\,\mathord{=}\,77.23\,\mathrm{days}$) 
and decays to \iron by means of electron capture
and via positron decay ($\beta^+$). We thus add \cobalt 
and \iron to our set of nuclei.
The respective decay reactions are given by
\begin{align*}
   \mathrm{EC:}&\qquad e^- + _{28}^{56}\mathrm{Ni} \rightarrow _{27}^{56}\mathrm{Co} + \nu_e + \gamma\,, \\
   \mathrm{EC:}&\qquad e^- + _{27}^{56}\mathrm{Co} \rightarrow _{26}^{56}\mathrm{Fe} + \nu_e + \gamma \quad\ (81\%)\,, \\
   \mathrm{\beta^+ :}&\qquad _{27}^{56}\mathrm{Co} \rightarrow _{26}^{56}\mathrm{Fe} + e^+ + \nu_e + \gamma \quad\ (19\%)\,.
\end{align*}
The above reactions provide an energy source for the 
surrounding plasma in the form of gamma radiation ($E_{\gamma}$) 
and kinetic energy ($E_{\mathrm{kin,e^{+}}}$) of the positrons 
that are produced in the $\beta^+$ decays. We include the 
annihilation energy of the positrons with electrons ($E_{\mathrm{ann}}$)
in $E_{\gamma}$. The produced neutrinos escape freely.
The average energy available (including the kinetic energy 
of the positron in the cobalt decay) per decay is 
$E_{\gamma,\mathrm{Ni}}\,\mathord{=}\,1.72\,\mathrm{MeV}$, 
$E_{\gamma,\mathrm{Co}}\,\mathord{=}\,3.735\,\mathrm{MeV}$ \citep{Nadyozhin1994}.
A fraction of the $\gamma$'s may escape depending on the (radial)
optical depth $\tau(r)$ of the gas up to the stellar surface at radius
$R_*$. This optical depth is defined as
\begin{equation}
    \tau(r) = -\int_{R_*}^{r} \kappa_{\gamma} Y_e(r') \rho(r')\, \ud r',
\label{eq:tau}
\end{equation}
where $Y_e$ 
the electron fraction and $\kappa_{\gamma}$ the optical opacity.
In the practical application the integral boundary $r$ in
Equation~(\ref{eq:tau}) is the radial location of a considered grid cell, 
and we assume that the trapped fraction of the locally produced
$\gamma$ radiation, $(1\,\mathord{-}\,\exp[-\tau(r)])$, deposits 
its energy locally in the same cell of the computational grid.
Assuming Compton-scattering is the dominant opacity source, we 
adopt a constant value of 
$\kappa_{\gamma}\,\mathord{=}\,6.0\,\mathord{\times}\,10^{-2}\,\mathrm{cm^2\,g^{-1}}$ \citep{Swartz1995}.
Therefore, the energy per mass $\Delta E_i/\Delta M$
deposited by each species $i$, with mass fraction 
$X_i$ and 
nuclear mass $m_i$, into the surrounding plasma during a 
time step $\Delta t$ is given by
\begin{equation}
        \frac{\Delta E_i}{\Delta M} =  \frac{\Delta X_i}{m_i}
        \left[ E_{\gamma} \left( 1 - \mathrm{e}^{-\tau(r)} \right) + E_{\mathrm{kin,e^{+}}}\right],
\end{equation}
where $\Delta X_i\,\mathord{=}\,\left(1-\mathrm{e}^{-\Delta t/t_{0,i}}\right)$ 
is the change of $X_i$ during $\Delta t$ with 
$t_{0,i}\,\mathord{=}\,t_{1/2,i}(\ln(2))^{-1}$ being the 
life-time of species $i$, and $E_{\mathrm{kin,e^{+}}}$ is the kinetic 
energy of the positron in the cobalt decay. The energy is assumed to be
deposited locally, and thermodynamic quantities are self-consistently updated.

\section{Evolution during the first seconds}
\label{sec:Evolution during the first second}

\subsection{Shock propagation and explosion energetics}
\label{sec:Shock propagation and explosion energetics}

\begin{figure*}
 \centering
 \includegraphics[width=\textwidth,trim=0.1cm 2.1cm 0cm 2.3cm,clip]{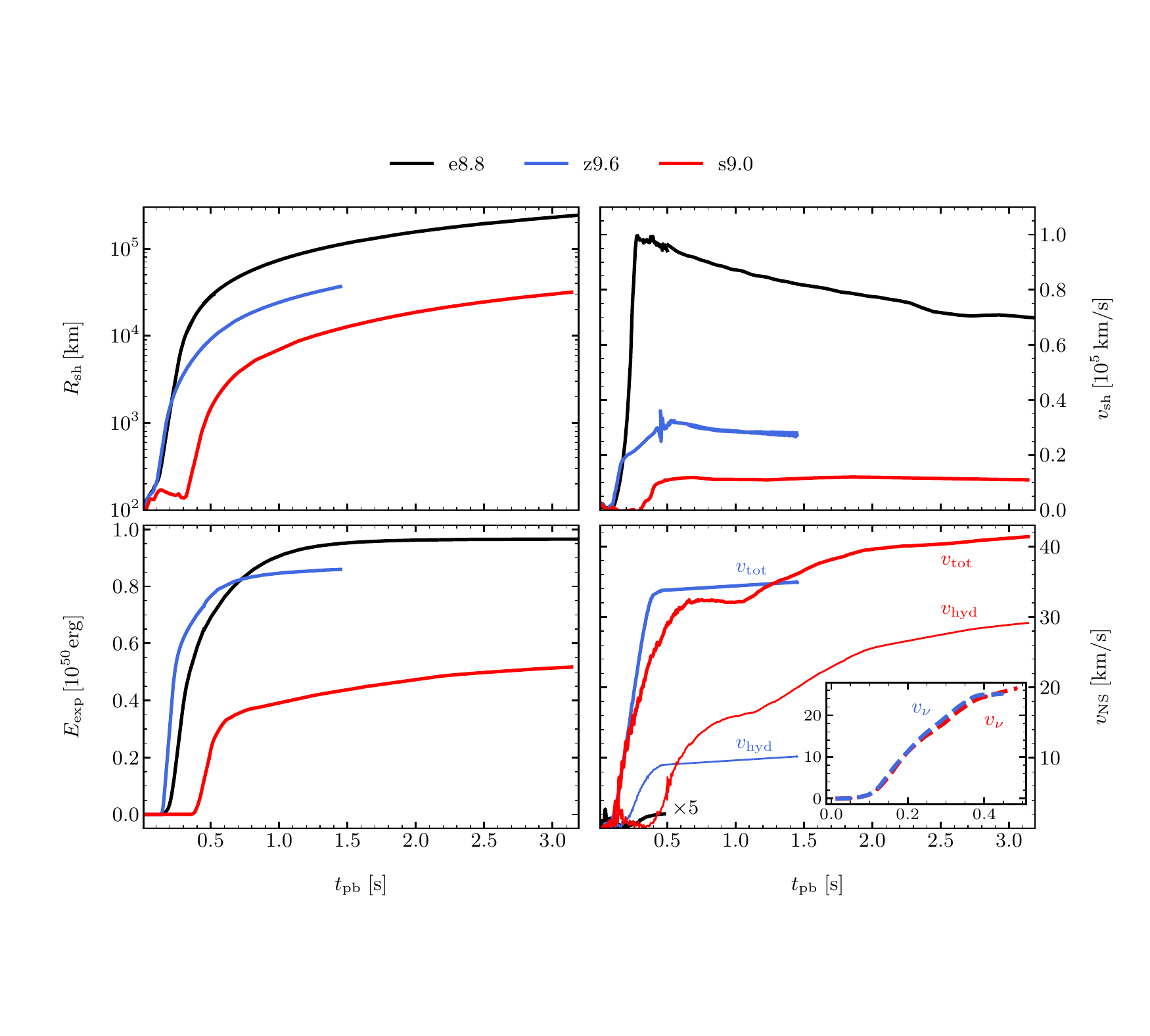}
 \caption{Shock radius (upper left panel), shock velocity (upper right panel) and diagnostic 
 explosion energy (lower left panel) versus post-bounce time for all of our 3D models. The wiggles in the shock velocity of model \znine at $\sim$0.45\,s are a 
 consequence of small-amplitude neutron-star vibrations when the \textsc{Vertex}
 neutrino transport is switched off and the heating/cooling scheme of 
 Appendix~\ref{appendix:scaling relations} is switched on. 
 Owing to an improved treatment this numerical transient is much reduced in model
 \snine. We also show the total PNS kick velocities (thick lines) and the
 hydrodynamically induced PNS kick velocities (thin lines) in the lower right panel.
 For better visibility, the inset of this panel
 displays the PNS kick velocities caused by asymmetric neutrino emission due to the LESA 
 phenomenon (see text for details). The total velocities are the vector sum of the hydrodynamic 
 PNS kick and the neutrino-induced kick. Despite nearly equal neutrino-induced kicks in \znine and
 \snine and lower hydrodynamic kick in \znine, this latter model has a higher total kick velocity
 for some time, because its hydrodynamic and neutrino-induced kick directions are essentially parallel,
 in contrast to the situation in model \snine. For the iron-core progenitors we can track only hydrodynamic 
 contributions to the kick after the transport module of \vertexprom is switched off at 
 $t_\mathrm{neut}\,\mathord{=}\,0.45\,\s$ and $t_\mathrm{neut}\,\mathord{=}\,0.49\,\s$ 
 for models \znine and \snine, respectively.  
 For the ONeMg case our simplified neutrino treatment with the excised core of the PNS 
 does not allow us to monitor the LESA induced kick. The kick of model \onemg is scaled by a
 factor of 5 for better visibility. }
\label{fig:eexp all}
\end{figure*}

In the following we provide a brief overview of the most important features
of the early post-bounce evolution in the 3D simulations of models e8.8, z9.6, 
and s9.0. The reader
is referred to \cite{Melson2015a} and \cite{Melson2019} for a detailed analysis 
of the post-bounce phase of the iron-core progenitors in \vertexprom simulations. 
Generic properties of the 
explosion of ECSNe are given in \cite{Kitaura2006,Janka2008,Huedepohl2010,Fischer2010,Radice2017,Gessner2018}. 
The dynamics of our simulations for the 8.8\,M$_\odot$ progenitor
closely resemble these previous findings.
In Figure~\ref{fig:eexp all} we show the evolution of the angle-averaged radius of the SN 
shock and the diagnostic explosion energy of our three-dimensional simulations during the first
three seconds after bounce. 
The angle-averaged shock radius is calculated as
\begin{equation}
    \langle R_{\mathrm{sh}} \rangle =  \frac{1}{4\pi}\int R_{\mathrm{sh}}(\theta,\phi)\ud\Omega\,,
    \label{equ:avg rsh}
\end{equation}
where $\ud \Omega\mathord{=}\sin{\theta}\ud\theta\ud\phi$.
The diagnostic explosion energy at all times is given by the integral
of the total (i.e., kinetic plus internal plus gravitational) energy density
in the postshock region, defined as 
$e_{\text{b}} = e_{\mathrm{int}} + e_{\mathrm{kin}} + e_{\text{grav}}$, 
over volume elements where it has a positive value 
\begin{equation}
    E_{\mathrm{exp}} = \int_{V_{\mathrm{postshock}}(e_{\mathrm{b}} > 0)}
    \mathrm{dV}\, e_{\mathrm{b}}  \,.
    \label{equ:ene exp}
\end{equation}

The sharp drop in density outside of the ONeMg-core in model \onemg 
leads to an early and strong decrease of the accretion rate and hence 
ram pressure at the shock.
Consequently, the SN shock expands rapidly and reaches the 
core/envelope boundary ($R_{\mathrm{He/H}}\,\mathord{=}\,1210\,\text{km}$), 
at $0.21\,\s$ after core bounce.
This is in stark contrast to the typical CCSN, where the 
high ram pressure stalls the shock expansion at a small radius for several 
$100\,\text{ms}$. The explosion energy starts rising steeply, fuelled by
the onset of a neutrino-driven wind, as soon 
as the shock leaves the ONeMg-core.

The acceleration of the shock at the core/envelope boundary is followed by
a drastic switch to deceleration after the shock passes the lower boundary of 
the H-envelope (see upper right panel of Figure~\ref{fig:eexp all}). This is caused 
by a sudden change of the density gradient at the core/envelope transition.  
Consequently, the neutrino-heated ejecta pile up in a dense, compressed and
decelerated shell behind the SN shock.

\begin{table*}
\caption{Overview of PNS properties in our multi-dimensional models at $t_{\mathrm{map}}$.}
\label{tab:neutron star}

\renewcommand{\arraystretch}{1.20}
\begin{tabular}{ccccccccccccc||ccc}
    \hline 
            & Dim. & 
            $t_{\mathrm{map}}$ & 
            $E_{\mathrm{exp}}$ & 
            $v_{\mathrm{NS}}^\mathrm{tot}$ & 
            $v_{\mathrm{NS}}^{\mathrm{hyd}}$ & 
            $v_{\mathrm{NS}}^{\nu}$ &
            $J_{\mathrm{NS}}/10^{45}$    & 
            $\theta_{vJ}$    & 
            $\alpha_{\mathrm{ej}}$&
            $\alpha_{\nu}$&
            \mns &
            \rns &
            $M_{\mathrm{map}}$ &
            $M_{\mathrm{g}}$ &
            $P_{\mathrm{NS}}$ 
            \\
    Model &
     &
    [s]   &
    $[10^{50}\, \erg]$   &
    [km/s] &
    [km/s] &
    [km/s] &
    $[\mathrm{cm^2 g/s}]$ &
    $[^\circ]$ &
    [\%]              &  
    [\%]              &  
    [\solm]        &
    [km] &
    [\solm]         &
    [\solm]    &
    [s]     \\
    
    \hline
    $\rm e8.8_{3}$ & 2D & 2.515 & 0.3   & 1.55  & 1.55  & -     &  1.78  & -     & 1.154 &  -     & 1.323 &  49.85 &  1.334 &  1.216 &  4.18  \\
    $\rm e8.8_{6}$ & 2D & 2.515 & 0.6   & 0.94  & 0.94  & -     &  2.75  & -     & 0.041 &  -     & 1.316 &  50.22 &  1.327 &  1.210 &  2.69  \\
    $\rm e8.8_{10}$& 2D & 2.515 & 1.0   & 0.13  & 0.13  & -     &  4.19  & -     & 0.004 &  -     & 1.308 &  50.57 &  1.319 &  1.203 &  1.75  \\
    $\rm e8.8_{15}$& 2D & 2.515 & 1.5   & 0.59  & 0.59  & -     &  1.67  & -     & 0.011 &  -     & 1.299 &  50.86 &  1.309 &  1.195 &  4.37  \\
    \onemg         & 3D & 0.470 & 1.0   & 0.44  & 0.44  & -     &  0.70  & 90.0  & 0.004 &  -     & 1.307 &  50.57 &  1.326 &  1.210 &  10.58  \\
    \znine         & 3D & 1.440 & 0.86  & 34.90 & 10.16 & 24.89 &  2.55  & 45.4  & 4.623 & 1.354  & 1.340 &  20.98 &  1.353 &  1.231 &  2.96  \\
    \snine         & 3D & 3.140 & 0.48  & 40.87 & 28.45 & 26.46 &  8.05  & 31.3  & 10.02 & 1.178  & 1.350 &  19.58 &  1.351 &  1.230 &  0.94  \\

    \hline
\end{tabular}
\flushleft
\textit{Notes}: All values are given at the end of our explosion simulations with
neutrino treatment ($t_{\mathrm{map}}$). $E_{\mathrm{exp}}$ is the diagnostic 
explosion energy, which is essentially identical to the final explosion energy
because of the small envelope binding energy (see Table~\ref{tab:composition tmap}).
$v_{\mathrm{NS}}^{\mathrm{tot}}$ is the total NS kick velocity resulting from the 
hydrodynamic ($v_{\mathrm{NS}}^{\mathrm{hyd}}$)
plus the neutrino-induced ($v_{\mathrm{NS}}^{\mathrm{\nu}}$) contributions.
We measure the contribution of neutrinos
to the total kick until the neutrino transport is switched of at 
$t_{\mathrm{neut}}\,\mathord{=}\,0.45\,\s$
and $t_{\mathrm{neut}}\,\mathord{=}\,0.49\,\s$ after core
bounce for models \znine and \snine, respectively. Hydrodynamic
contributions and total kick velocities are monitored until $t_{\mathrm{map}}$.
$J_{\mathrm{NS}}$ is the total angular momentum transported to the PNS through 
a radius of 100 km until $t_{\mathrm{map}}$. $\theta_{vJ}$ is the angle between
the direction of the total kick velocity $\pmb{v}$ and the direction of $\pmb{J}$. $\alpha_{\mathrm{ej}}$ and 
$\alpha_{\nu}$ are the final hydrodynamic and neutrino anisotropy-parameters, 
respectively (see Appendix~\ref{Appendix:Neutron Star Properties}). For the 
iron-core progenitors we average $\alpha_{\nu}$ over the time when the emission dipole
is largest until the end of the simulation. $M_{\mathrm{b}}$ is the baryonic PNS 
mass, which is defined as the enclosed mass within the radius $R_{\mathrm{NS}}$, at which the density drops 
below $10^{11}\,\mathrm{g\,cm}^{-3}$. Note that for the simulations of the \onemg progenitor,
\rns is determined by the chosen parameters of the inner grid boundary (see Table~\ref{table:e8param}).
$M_{\mathrm{map}}$ (see also Table~\ref{tab:long term boundaries}) is the 
central mass contained within the excised region, from which we calculate the 
gravitational mass $M_{\mathrm{g}}$ for a PNS radius of 12~km (see Appendix~\ref{Appendix:Neutron Star Properties}). 
$P_{\mathrm{NS}}$ is the PNS spin period at the end of our post-bounce 
simulations, assuming a final NS radius $R_{\mathrm{NS}}$ of 12 km, angular 
momentum conservation, and a gravitational mass of $M_{\mathrm{g}}$. 
\end{table*}

The ECSN-like structure of model \znine is reflected in the 
evolution of the SN shock front and in the growth of the diagnostic explosion energy. 
The shock radii remain almost perfectly spherical in both the \onemg and \znine models. 
The acceleration of the blast wave outside of the iron core, however, ends earlier in 
model \znine than in model \onemg because of the more gradual changes in the density profile. 
In both cases the explosion energies also start to rise early and saturate just after 
$\tpb\mathord{\sim}1\,\s$.

In contrast to the \znine and \onemg models, model \snine lacks
the very steep density gradient at the edge of the Fe-core, as can be 
seen in Figure~\ref{fig:prog_tem_rho_ye_rhor}. The shock can expand initially up 
to $180\,\km$ at $\tpb\mathord{\sim}130\,\ms$, but then 
it enters a phase of recession. The arrival of the Si/O interface at the shock and 
the decreasing mass-accretion rate within the oxygen shell eventually lead to shock 
expansion at $\mathord{\sim}0.32\,\s$ after bounce. Shock expansion is aided by strong 
convection behind the shock front \citep[see also][]{Melson2019}. 
Similar to the results presented in \cite{Glas2019}, 
who used the same progenitor, the model remains convection-dominated and does not 
exhibit any sign of the oscillatory growth of the SASI \citep{Blondin2003,Foglizzo2007}.
However, although SASI does not develop in the simulation, the forward shock experiences 
large-scale deformation with a dipole amplitude of $\mathord{\sim}10\%$ 
(compared to the angle-averaged shock radius). This deformation is driven by 
big plumes that form in the post-shock layer. 
Contrary to the ECSN-like models, strong anisotropic, non-radial mass flows 
persist around the PNS during several seconds after bounce. Continuous mass accretion 
onto the PNS through narrow funnels delays the emergence of the spherical 
neutrino-driven wind, which is why we needed to continue the simulations 
including PNS and neutrino treatment for more than three seconds after bounce.
Explosion energy and shock velocity in model \snine remain considerably lower
than in the other two progenitors (see Figure \ref{fig:eexp all}).

\subsection{Neutrino emission properties}
\label{sec:neutrinos}

\begin{figure*}
\includegraphics[width=2.0\columnwidth]{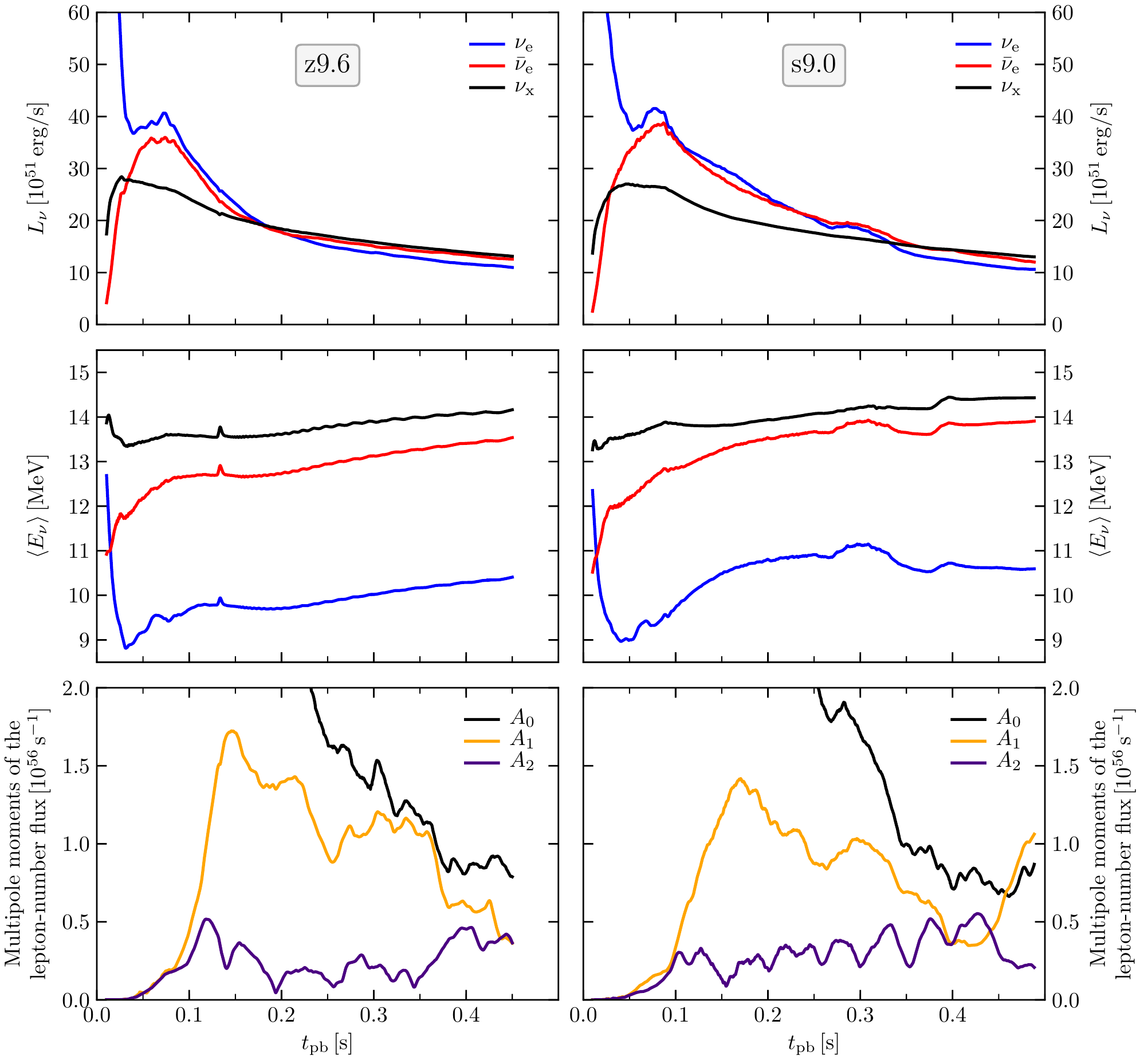}
\caption{Neutrino luminosities (top), mean energies (middle), all
averaged over all viewing directions, and lowest-order multipole moments 
($A_0$, $A_1$, $A_2$ for monopole, dipole, and quadrupole)
of the electron-neutrino lepton-number flux (bottom) as
functions of time for models \znine (left) and \snine (right).
All quantities are 
evaluated at 400\,km, transformed to an observer frame at rest at infinity.
With $\nu_x$ we denote one species of the heavy-lepton neutrinos.
Note that the dipole moment, $A_1$, plotted here is one third
of the dipole amplitude of the lepton-number flux defined by 
\protect\cite{Tamborra2014a} 
(see Appendix~\ref{app:neutrinomultipoles} for details).}
\label{fig:nu-flux+multipoles}
\end{figure*}

In the following we present the neutrino emission properties of our
two 3D simulations performed with the \textsc{Vertex-Prometheus} code.
Figure~\ref{fig:nu-flux+multipoles} displays the neutrino luminosities
(defined as $4\pi$-integrated energy fluxes) and mean neutrino energies 
(defined as ratio of angle-averaged energy density to number density) 
for $\nu_e$, $\bar\nu_e$ and heavy-lepton neutrinos $\nu_x$, as well as the
three lowest-order multipoles (monopole, dipole, and quadrupole)
of the electron-neutrino lepton-number flux for models \znine and \snine 
as functions of post-bounce time (\znine left column, \snine right column). 
All quantities are evaluated at 400\,km
(transformed to an observer frame at rest at infinity). The formulas for the 
spherical harmonics decomposition are provided in 
Appendix~\ref{app:neutrinomultipoles}.

In the case of \znine the luminosities of all three species 
become very similar after only $\sim$180\,ms, signalling the end
of PNS accretion caused by the quick onset of the explosion and
the rapid shock expansion. In contrast, PNS accretion continues at
a significant rate until roughly 350\,ms in model \snine.
Only afterwards the luminosities in this model converge to nearly
the same level, mirroring the characteristic trend when the 
cooling emission of the newly formed NS begins. Overall, the
time evolution and the values of the neutrino luminosities and
mean energies of both models are very similar to each other,  
consistent with the nearly equal masses of the PNSs in both cases
(see Table~\ref{tab:neutron star}). Model \snine exhibits additional
accretion emission between $\sim$150\,ms and $\sim$350\,ms, 
which enhances the $\nu_e$ and $\bar\nu_e$ luminosities slightly 
and drives a continuous increase of the mean energies of
$\nu_e$ and $\bar\nu_e$. However, the trend does not persist for
long enough (and the PNS does not gain enough mass) to reach
a crossing of the electron antineutrino energy with the 
heavy-lepton neutrino energy as reported by \cite{Marek2009}
for more massive progenitors with more massive PNSs.

\begin{figure*}
\includegraphics[width=2.0\columnwidth]{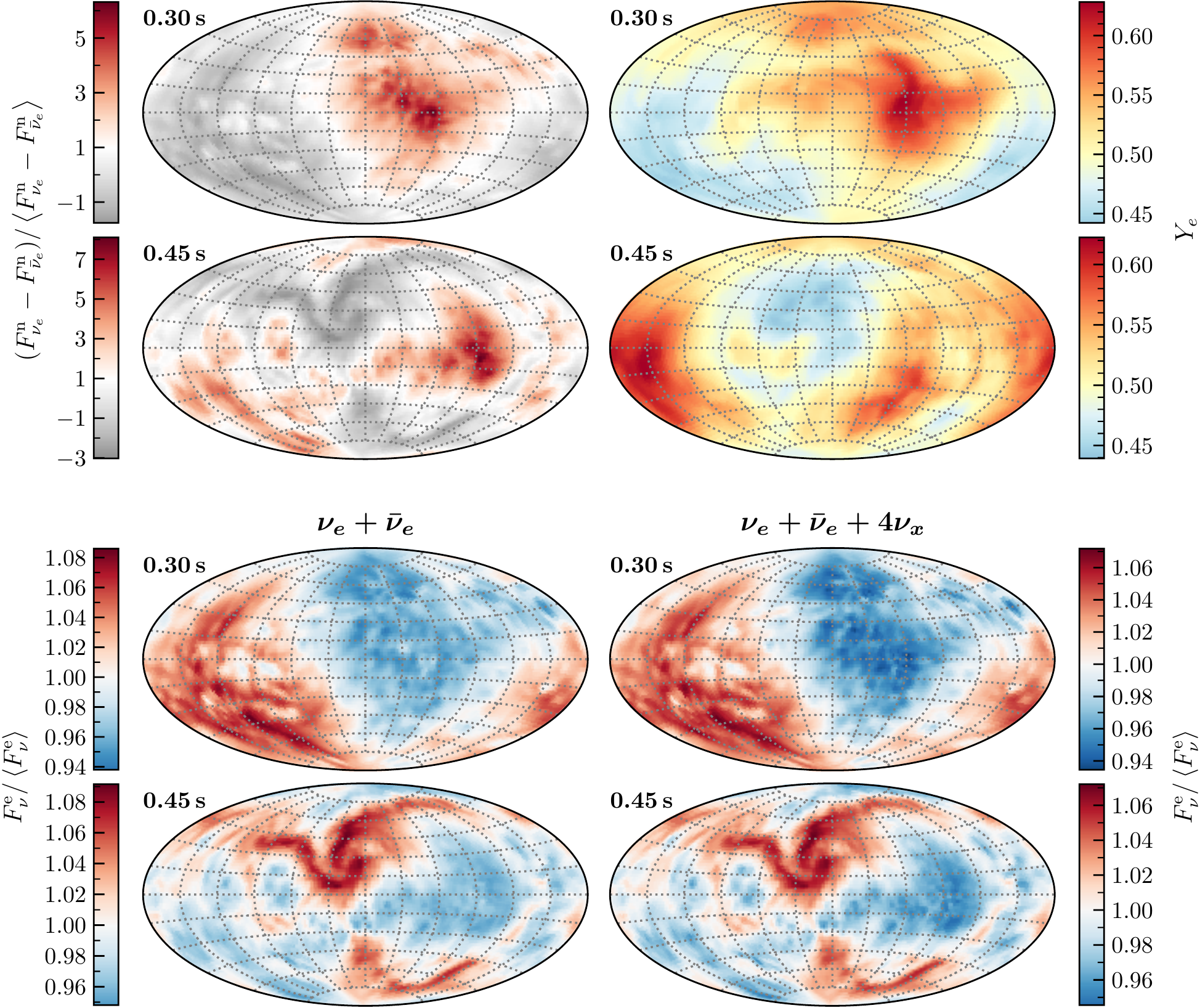}
\caption{Full-sphere Aitoff projections of various LESA related
quantities for model \znine at 0.30\,s and 0.45\,s after bounce.
The two upper left panels display the relative variation of the 
electron-neutrino lepton-number flux, 
$(F^\mathrm{n}_{\nu_e}-F^\mathrm{n}_{\bar{\nu}_e})/
\left< F^\mathrm{n}_{\nu_e} - F^\mathrm{n}_{\bar{\nu}_e}\right>$ 
(normalized by the angular average)
at a fixed radius of 400\,km as function of polar angles 
$\theta$ and $\phi$. The two upper right panels show the variation of 
the electron fraction $Y_e$ in the neutrino-heated ejecta at 250\,km.
The two lower left panels visualize the relative variation of the 
energy flux of $\nu_e$ plus $\bar\nu_e$, and the two lower right panels
the corresponding relative variation of the total energy flux as sum
of all contributions of $\nu_e$, $\bar\nu_e$ and $\nu_x$, again
evaluated for lab-frame quantities at 400\,km. Note that the amplitude
of the energy-flux variation is considerably lower than that of the
lepton-number flux, and the dipole directions of both fluxes possess 
opposite orientations.}
\label{fig:LESA-Aitoff}
\end{figure*}

The bottom panels of Figure~\ref{fig:nu-flux+multipoles} 
demonstrate that both models develop a lepton-number emission
dipole that dominates the higher-order multipoles and can 
reach and even exceed the monopole for longer periods
of time (i.e., hundreds of milliseconds, see also \citealt{Tamborra2014}). 
Note that the dipole amplitude displayed in
Figure~\ref{fig:nu-flux+multipoles} is, by normalization, one third
of the dipole amplitude considered by \cite{Tamborra2014}
(see also Appendix~\ref{app:neutrinomultipoles}).
A long-lasting lepton-number emission dipole, whose
direction is nearly stable or migrates only very slowly,
is a characteristic feature of the LESA (Lepton-Emission
Self-sustained Asymmetry) phenomenon that was first witnessed in 3D 
\textsc{Vertex-Prometheus} simulations with neutrino transport 
by \cite{Tamborra2014} and \cite{Janka2016}.
This striking phenomenon has meanwhile been confirmed with 
fully multi-dimensional instead of ray-by-ray neutrino transport by \cite{OConnor2018}, 
\cite{Glas2019b} and \cite{Vartanyan2019}.

Neither in model \znine nor in \snine does SASI play a role and, in
addition, model \znine explodes after a very brief period of post-bounce
accretion. In both models the PNS emission is therefore not masked by
asymmetric accretion due to SASI shock sloshing or spiral motions.
Because accretion in particular in model \znine does not contribute to
the neutrino emission at any significant level after the onset of the
explosion, its lepton-number flux asymmetry is merely determined by 
asymmetric convection inside of the PNS, where outward transport of
lepton number is strongly suppressed in the anti-LESA direction. 
Therefore the lepton-number flux can even be negative in the anti-LESA
direction (see Fig.~\ref{fig:LESA-Aitoff}, two upper left panels). 
This means that one hemisphere of the PNS radiates a
greater number of $\bar\nu_e$ than $\nu_e$, whereas there is the usual
excess of $\nu_e$ number loss from the other hemisphere.
The models, in particular \znine with no long-lasting accretion,
therefore confirm that LESA is a phenomenon primarily
generated by hemispherically asymmetric convection in the interior
of the PNS, well below the neutrinosphere; the reader is referred to the
more detailed discussion by \cite{Glas2019b}, where also a
3D simulation with successful explosion of the 9.0\,$M_\odot$ progenitor  
is evaluated.\footnote{In constrast to the models of \cite{Glas2019b},
in which a 1D core of 10\,km was used, the 3D simulations discussed
here were computed with a very small central 1D core of only 
1.6\,km radius. Both sets of simulations show very similar LESA
features, which means that the 10\,km core had no relevant influence
on the previous results.}
It is also interesting to note that the lepton-emission multipole 
($\ell\,\mathord{\neq}\,0$) that grows fastest and thus develops high amplitudes
of asymmetry first is the one of order $\ell\,\mathord{=}\,4$, which is also
in line with the results reported by \cite{Glas2019b}.

\begin{figure*}
 \centering
 \includegraphics[width=0.75\textwidth,trim=0cm 0.1cm 0.0cm 0.0cm,clip]{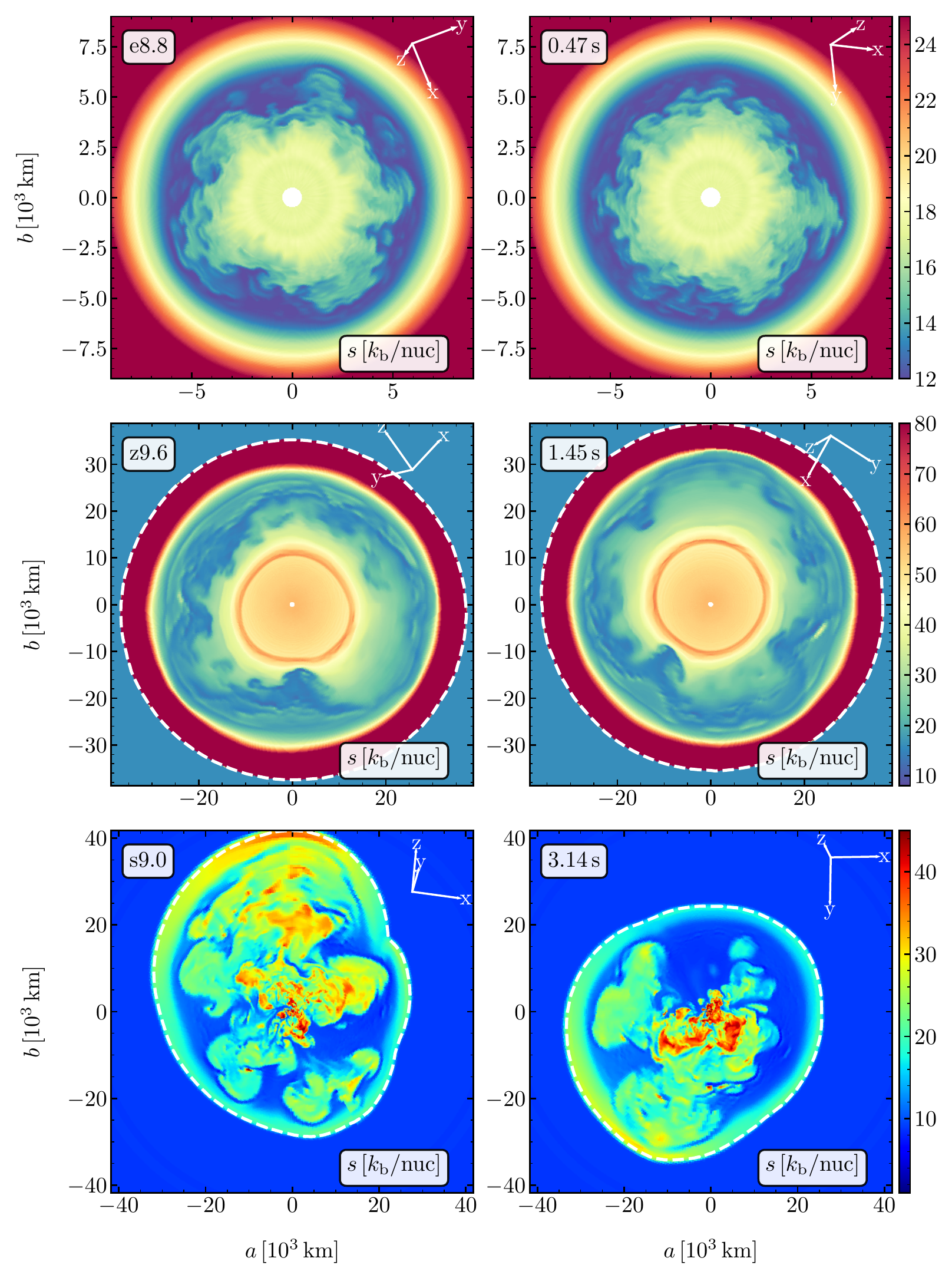}
 \caption{Planar slices of our 3D models showing the entropy color-coded at 
 $t_{\mathrm{map}}$. The left panels display the plane of largest shock deformation,
 whereas the right panels present the plane of smallest shock expansion. The 
 coordinate directions of the plots (indicated by the tripods in the top
 right corners) have no association with the coordinates of the computational 
 grid. Note the almost spherical morphology of model \onemg and the deformed ejecta 
 morphology of models \snine and \znine. For better visibility of the small-scale 
 structures of model \snine we choose a different color representation in this case.
 The white dashed line marks the shock surface. This line is missing in the top two panels because
 in model \onemg the shock is at more than 20,000~km at this time already, far ahead of all
 explosion asymmetries.}
 \label{fig:slices first mapping}
\end{figure*}

\begin{figure*}
 \centering
 \includegraphics[width=0.80\textwidth,trim=0cm 0.0cm 0cm 0.0cm,clip]{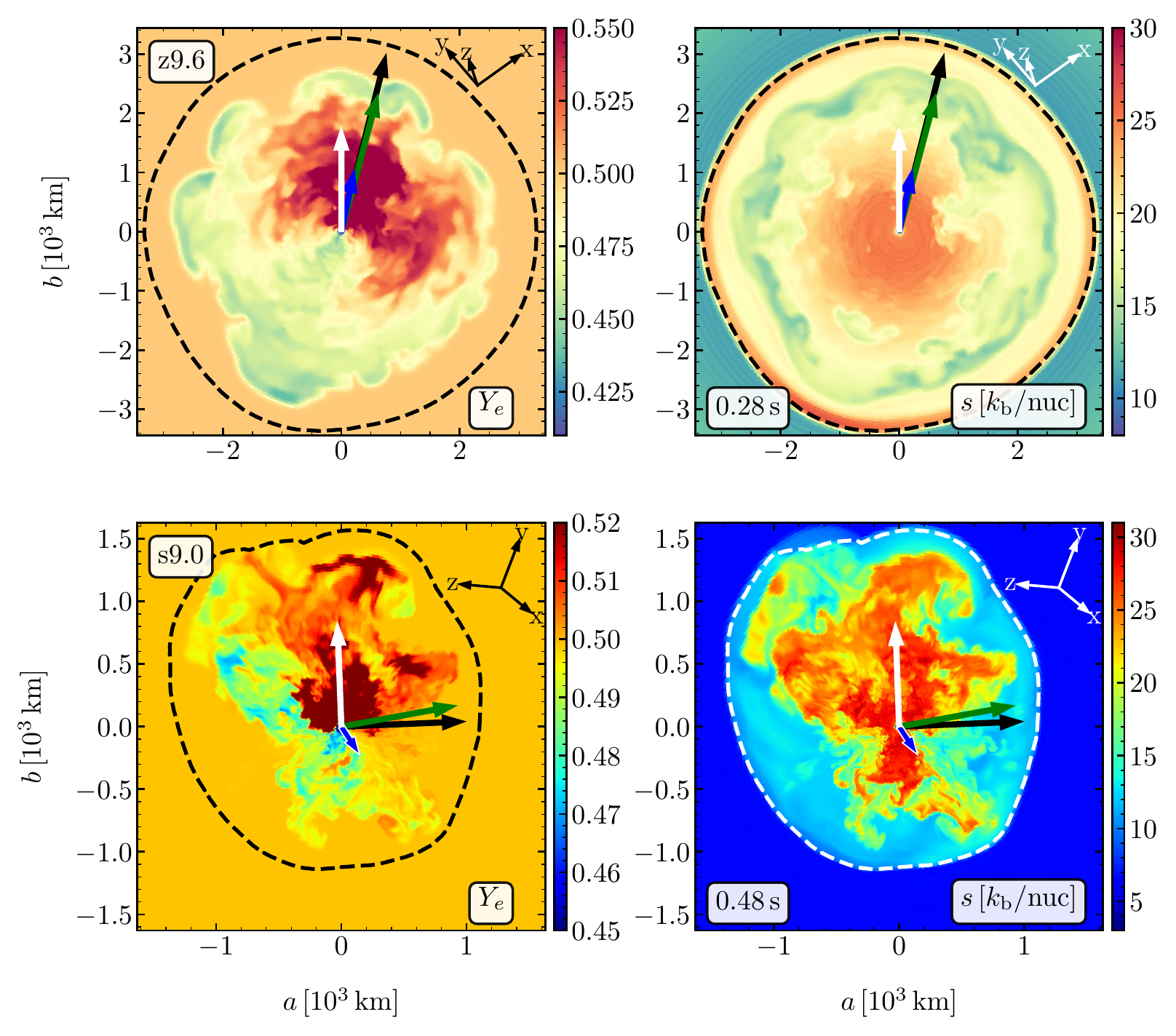}
 \caption{Planar slices of the electron fraction $Y_{e}$ (left panels) 
 and entropy per nucleon (right panels) of models \znine and \snine at 0.28~s and 0.48~s after bounce, 
 respectively. We choose times where the LESA dipole is strong in both models. The planes are spanned
 by the LESA vector (white arrow) and the vector of the total PNS kick (black arrow) and oriented 
 such that the LESA dipole direction points to the north. The positioning of the plane in the global 
 coordinate system of the simulation output is shown by the tripod in the upper right corner of each panel. 
 The arrows indicate the directions of the total PNS kick velocity (black), 
 the hydrodynamic PNS kick velocity (blue), and the neutrino-induced kick velocity of the PNS (green),
 all computed as integrals of the accelerations until the displayed times. The black or white dashed 
 lines mark the shock radius. Note that in model \znine the blue and green arrows are nearly aligned, 
 adding up to the black arrow.}
 \label{fig:sto ye s9 z9 kick}
\end{figure*}

Interestingly, there are phases in both of our models when the dipole
and quadrupole amplitudes become similar (Figure~\ref{fig:nu-flux+multipoles}). 
This is also suggested by the
Aitoff projections for time $t_\mathrm{pb} \mathord{=} 0.45$\,s near the end of 
our simulation of \znine (Fig.~\ref{fig:LESA-Aitoff}). 
At this late time a quadrupole
pattern is superimposed on a hemispheric dipole asymmetry, whereas at
$t_\mathrm{pb}\,\mathord{=}\,0.30$\,s a much cleaner dipole is present. This reflects
the evolution of the multipole amplitudes visible for \znine 
in the bottom left panel of Figure~\ref{fig:nu-flux+multipoles}.
The late-time rapid growth of the dipole in the lepton-number emission
of model s9.0 at $t_\mathrm{pb}\,\mathord{>}\,0.43$\,s 
(bottom right panel of Figure~\ref{fig:nu-flux+multipoles}) is an apparent
feature caused by a transient phase of a reduced lepton-emission dipole
between $\sim$0.35\,s and $\sim$0.45\,s. This reduction is a consequence
of an accretion asymmetry that channels matter towards the PNS
predominantly in one hemisphere, whereas a mighty outflow in the form 
of a huge, rising bubble develops in the opposite hemisphere (see the 
bottom panels of Figures~\ref{fig:slices first mapping} and 
\ref{fig:sto ye s9 z9 kick}). The main downflow 
direction is misaligned with the LESA dipole direction by roughly
90$^\circ$ and fuels enhanced lepton-number emission by the 
one-sided accretion. This enhanced lepton-number emission combined
with the displaced LESA dipole strengthens the quadrupole component of  
the electron lepton-number emission and at the same time weakens its 
dipole. The remaining dipole level in the interval of 
[0.35\,s, 0.45\,s] signifies that the LESA dipole is stronger than the
emission of lepton number by accretion. The dipole amplitude recovers
to the full LESA strength after $t_\mathrm{pb}\mathord{\sim}0.45$\,s because 
the mass-accretion rate onto the PNS declines continuously.

The Aitoff projections show that not only the electron-neutrino
lepton-number flux exhibits large-scale (low-order multipolar)
asymmetries, but also the $\nu_e$ plus $\bar\nu_e$ energy flux
as well as the total neutrino energy flux (the summed energy fluxes
of $\nu_e$ plus $\bar\nu_e$ plus four times that of $\nu_x$).  
However, while the directional variations of the lepton-number
flux can be several times bigger than the angular average of
this quantity, the $\nu_e$ plus $\bar\nu_e$ energy flux varies only
within roughly 6--8\% and the total neutrino energy flux even less
within about 4--6\% (Figure~\ref{fig:LESA-Aitoff}). 
These directional variations, in particular the
hemispheric asymmetries, have consequences for NS kicks and
the neutron-to-proton ratio in the ejecta. This will be discussed
in the following Sections~\ref{sec:nskicks} and 
\ref{sec:electronfraction}.

\subsection{Neutron star kicks by asymmetric mass ejection and 
neutrino emission}
\label{sec:nskicks}

In order to compare the final state of the post-bounce simulations of our model set, we 
present in Figure~\ref{fig:slices first mapping} planar slices showing the entropy at 
the times $t_{\mathrm{map}}$ when we start our long-time simulations. The left panels are 
chosen such that they align with the plane of the largest shock deformation. The right 
panels align with the plane of the smallest shock deformation.
Note the basically spherical shape of the shock and the mild post-shock asymmetries
in both the \onemg and \znine models.
In strong contrast to the ECSN-like models, model \snine shows a clear 
dipolar shock deformation and ejecta morphology. 

The asymmetries that develop during the explosion also affect the kick of the PNS.
In Table~\ref{tab:neutron star} we provide properties of the PNS resulting 
from our post-bounce simulations.\footnote{We refer the reader to 
Appendix~\ref{Appendix:Neutron Star Properties} for the equations involved in 
the analysis underlying the results 
presented in Table~\ref{tab:neutron star}.} We list the PNS kick velocity 
($v_{\mathrm{NS}}$), ejecta anisotropy parameter ($\alpha_{\mathrm{ej}}$), PNS angular 
momentum ($J_{\mathrm{NS}}$), angle between total PNS kick and angular momentum vector 
$\theta_{vJ}$ and PNS spin period ($P_{\mathrm{NS}}$) along with the PNS 
radius (\rns) and baryonic PNS mass ($M_{\mathrm{b}}$) at the moment in time 
($t_\mathrm{map}$) when we terminate the post-bounce simulations that include neutrino transport
or neutrino heating (see also Table~\ref{tab:long term boundaries}). The PNS radius $\rns$ 
is defined as the radius where the angle-averaged density drops below 
$10^{11}\,\mathrm{g\,cm}^{-3}$. The PNS mass $M_{\mathrm{b}}$ is the mass contained 
interior to $\rns$. $M_{\mathrm{map}}$ is the mass contained within our inner grid boundary,
and thus removed from the hydrodynamic grid at the start of the long-time simulations 
at $t_\mathrm{map}$ (see Table~\ref{tab:long term boundaries}). 
$M_{\mathrm{g}}$ the corresponding gravitational mass.

The acceleration of the PNS is caused by two different mechanisms.
Firstly, aspherical ejection of matter leads to a gravitational tug, which 
accelerates the PNS into the hemisphere opposite to the maximum shock expansion and fastest 
ejecta \citep{Scheck2006,Wongwathanarat2010a,Wongwathanarat2013}. 
Secondly, anisotropic neutrino emission, due to the LESA phenomenon \citep{Tamborra2014}, can 
accelerate the PNS opposite to the direction of the largest total neutrino-energy flux.
LESA manifests itself in a dominant and stable $\ell\,\mathord{=}\,1$ spherical 
harmonics mode of the lepton-number emission and a corresponding energy-emission 
dipole amplitude of several percent compared to the monopole 
(see \citealt{Tamborra2014a,Tamborra2014}, and Section~\ref{sec:neutrinos}).
LESA is observed in both simulations conducted with \vertexprom.
The almost spherical explosions of the ECSN-like progenitor yield very low hydrodynamic 
kick velocities by the ``gravitational tug-boat effect'' \citep{Gessner2018}. 
Anisotropic neutrino emission cannot be evaluated in our simulation 
of model \onemg, because of the spherical treatment of the central PNS region. The 
kick contribution by anisotropic neutrino emission in model \snine is of a magnitude
comparable to the contribution associated with the aspherical ejection of matter and 
even exceeds the hydrodynamic kick contribution in model \znine.

We show in Figure~\ref{fig:sto ye s9 z9 kick} planar slices of the electron fraction 
(left panels) and entropy per nucleon (right panels) of the iron-core progenitors. The black, 
blue, and green arrows indicate the total, hydrodynamic, and neutrino-induced directions of the 
PNS kick. In the case of model \znine, the hydrodynamic and neutrino-induced kicks are nearly 
aligned. This results from the anti-correlation of the LESA lepton-number emission dipole and 
the direction of the maximum $\nu_{\mathrm{e}}\mathord{+}\bar{\nu}_{\mathrm{e}}$ as well as 
total neutrino luminosity (see \citealt{Tamborra2014} and Section~\ref{sec:neutrinos}). Increased 
heating by $\nu_{\mathrm{e}}\mathord{+}\bar{\nu}_{\mathrm{e}}$ absorption in the postshock region
in the hemisphere opposite to the LESA dipole pushes the SN shock to larger 
radii. The induced asymmetry of the post-shock ejecta leads to the hydrodynamic 
acceleration of the PNS in the opposite direction due to the gravitational pull
of the slower ejecta. The neutrino-induced NS kick acts in the same direction, i.e.,
nearly aligned with the LESA dipole, because of the maximum total neutrino luminosity
being in the anti-LESA direction.

Due to the weak postshock convection and only small-scale ejecta asymmetries in model \znine, 
the anisotropic neutrino emission produces the dominant contribution to the PNS kick. 
While the gravitational tug of the ejecta can only account for  
$v_{\mathrm{NS}}^{\mathrm{hyd}}\,\mathord{\sim}\,10\,\kms$, 
the asymmetric emission of neutrinos alone would cause a kick velocity of
$v_{\mathrm{NS}}^{\mathrm{\nu}}\,\mathord{\sim}\,25\,\kms$. 
The total kick sums up to $v_{\mathrm{NS}}^{\mathrm{tot}}\,\mathord{\approx}\,35\,\kms$.
A correlation or even alignment of the neutrino-induced and hydrodynamic kicks
is less clear in model \snine.
The asymmetric neutrino energy emission is also responsible for a sizeable contribution to the PNS
kick in this case, and even for the dominant contribution during the first 0.5\,s after bounce.
However, the neutrino-induced kick in \snine, despite being in the same hemisphere as the LESA dipole 
vector, is not closely aligned with the LESA dipole direction as in model \znine.
The hydrodynamic PNS kick, however, is in the opposite hemisphere, different from the case of model 
\znine. Both, this different orientation of the hydrodynamic kick relative to the LESA dipole, and 
the misalignment of LESA direction and neutrino-induced kick in model \snine, are a consequence of 
the fact that hydrodynamic instabilities in the postshock layer are much stronger and longer-lasting.
Therefore the corresponding asymmetries of the mass distribution around the PNS are more extreme 
in model \snine than in \znine.
The direction of the neutrino-induced kick is thus affected by the neutrino-emission dipole that is 
associated with one-sided PNS accretion. This asymmetric neutrino-emission due to accretion
is transiently superimposed on the LESA dipole during the time interval of [0.35\,s,0.45\,s]
after bounce. It is displaced by roughly 90$^\circ$ from the LESA direction (see Section~\ref{sec:neutrinos}) and thus shifts the neutrino-induced kick away from the LESA
direction (Figure~\ref{fig:sto ye s9 z9 kick}, lower panels).

The neutrino-heating asymmetry associated with LESA, causing more heating in the anti-LESA 
direction (where the $\nu_e$ plus $\bar\nu_e$ luminosity is higher and also the number flux of
the $\bar\nu_e$ with their harder spectra is higher), has a mild influence only in model \znine.
Therefore the explosion in this model is slightly stronger in
the anti-LESA direction, and the hydrodynamic PNS kick is nearly aligned with the LESA vector and
with the direction of the neutrino-induced PNS kick. In contrast, in model \snine the
neutrino-heating asymmetry associated with LESA is not powerful enough to determine
the deformation of the explosion. Convective mass motions, the corresponding accretion asymmetries, 
and the associated anisotropic neutrino emission and absorption between PNS and SN shock play a 
more important role and have a dominant influence on the morphology of the postshock flow.
In model \snine the shock expansion and explosion are weaker in the anti-LESA direction, where convective 
downdrafts towards the PNS are numerous and massive and thus direct the hydrodynamic PNS kick towards this
hemisphere (blue arrows in the southern hemisphere of the lower panels of Figure~\ref{fig:sto ye s9 z9 kick}). 
As explained above, accretion-associated asymmetries of the total neutrino luminosity are also 
the reason why the neutrino-induced kick is not well aligned with the LESA lepton-emission dipole.
The combined hydrodynamic and neutrino-induced kicks in model \snine lead to a PNS velocity of 
$\mathord{\sim}41\,\kms$ at mapping time $t_{\mathrm{map}}$.
It must be emphasized that in both models, \znine and \snine, the \vertexprom neutrino transport 
was switched off after about 0.5\,s of post-bounce evolution, namely at $\sim$0.45\,s and 
$\sim$0.49\,s after bounce, respectively. At this time the neutrino-induced kicks have not reached
their final values, which might be higher than the numbers presented in Table~\ref{tab:neutron star}.

\begin{figure*}
\includegraphics[width=\textwidth]{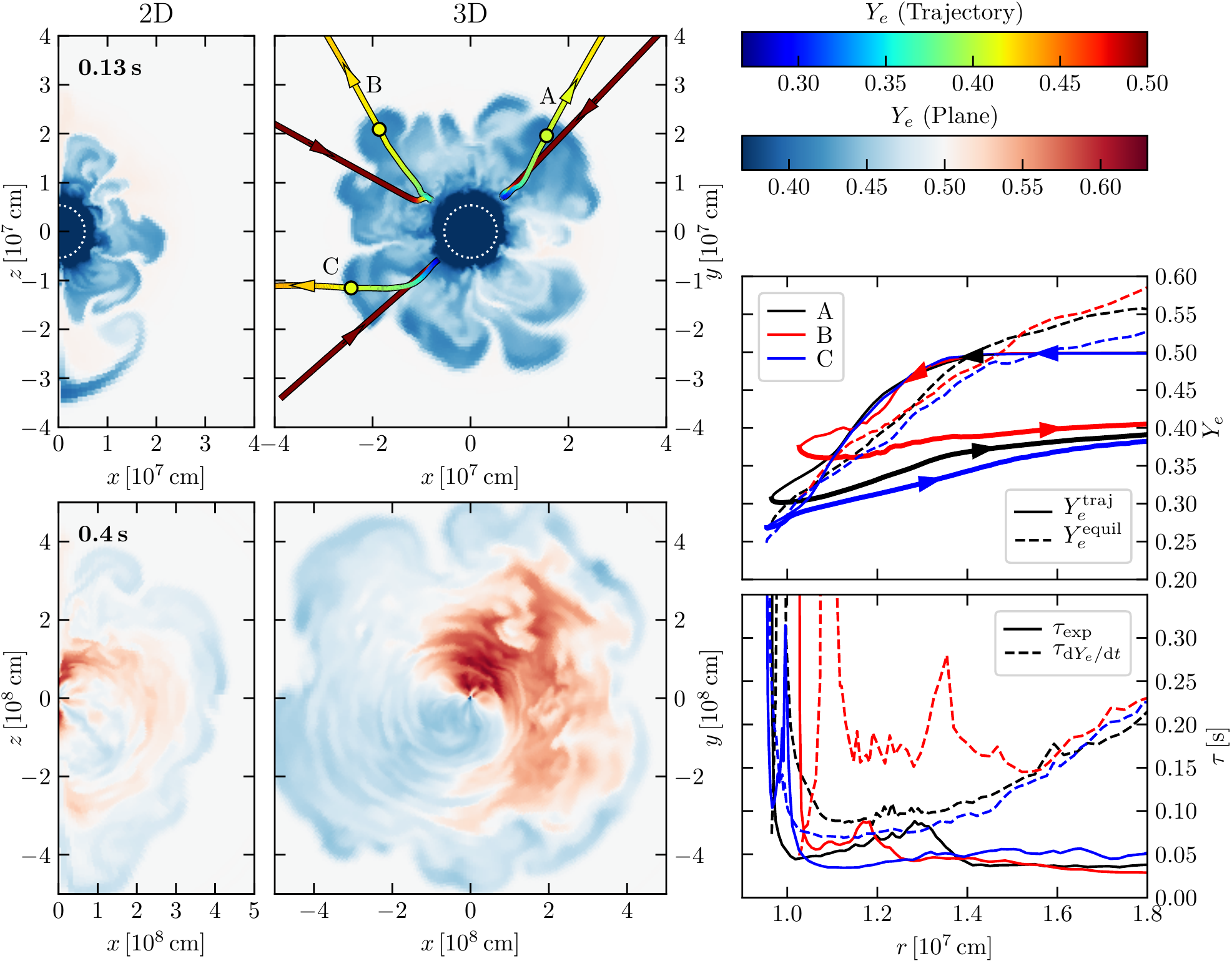}
\caption{\textit{Left panels:}
Distribution of $Y_e$ in cross-sectional slices of 2D (left) and 3D (right) simulations
of model \znine at 0.13\,s (top) and 0.4\,s (bottom) after core bounce.
The ejection of neutron-rich matter 
(with $0.39\,\mathord{\lesssim}\,Y_e\,\mathord{\lesssim}\,0.50$) 
in fast-rising, buoyant plumes is visible. Note that the chosen
color range (lower color bar) does not allow to display the $Y_e$ variation
at radii below $\sim$100\,km. The colored lines visualize the infall and
escape trajectories (projected onto the $x$-$y$-plane of the cross-sectional
slice, the flow direction is marked by arrows) for three selected mass elements 
(trajectories A,B,C), which ultimately get expelled in the neutron-rich mushroom 
heads. The local $Y_e$ value along the trajectories is color
coded according to the upper color bar. The $Y_e$ evolution 
of these fluid elements is more quantitatively represented in the right panels.
The locations of these elements at the time of the plot (0.13\,s) are
indicated by small colored circles. The white dotted circle marks the mean neutrinosphere,
defined by a spectrally averaged optical depth of 2/3. 
The lower left panels, besides showing the farther expanded, neutron-rich plumes,
display neutrino-driven wind ejecta, which develop neutron-rich conditions in one
hemisphere (bluish colors) and proton-richness in the opposite hemisphere (reddish
colors) because of the LESA dipole
asymmetry of the $\nu_e$ and $\bar\nu_e$ emission from the nascent neutron star.
We stress that the increase of $Y_e$ in the
expanding mushroom heads between 0.13\,s and 0.40\,s is not a consequence of neutrino
reactions (which basically cease at $r\,\mathord{\gtrsim}\,250$\,km). Instead, a 
gradual, slow $Y_e$ increase is caused by numerical diffusion connected with the
ejecta mass flowing over the Eulerian grid.
\textit{Upper right panel:} Evolution of the electron fraction for the
three selected mass elements that are ejected in the neutron-rich
mushroom heads. The $Y_e$ evolution during the infall is shown by thin solid
lines, during the re-ejection by thick solid lines (indicated by arrows). 
Note that during the infall
all three lines, A, B, and C, lie on top of each other, starting at 
$Y_e\,\mathord{=}\,0.5$. The dashed lines represent the
local electron fraction, $Y_e^\mathrm{equil}$, for reactive equilibrium. 
\textit{Bottom right panel:} Expansion timescale, $\tau_\mathrm{exp}$ (solid lines),
and timescale of $Y_e$ changes, $\tau_{\mathrm{d}Y_e/\mathrm{d}t}$ (dashed lines),
measured along the outflow parts of the solid curves in the upper right panel.}
\label{fig:yecuts}
\end{figure*}

\subsection{Electron fraction of neutrino-heated ejecta}
\label{sec:electronfraction}

Since the explosions of models \znine and \snine are computed with detailed
neutrino transport, we describe the evolution and distribution of their electron
fraction, $Y_e$, in the following. Corresponding 2D results obtained for ECSN models
in simulations with the \textsc{Vertex-Prometheus} and \textsc{CoCoNuT-Vertex} codes
were presented by \citet{Wanajo2011} and \citet{Wanajo2018}, respectively. 
The electron fraction in the neutrino-heated ejecta of our 3D model \onemg is only
approximate because of the simplified treatment of the neutrino physics and the 
replacement of the central 1.1\,M$_\odot$ PNS core by an inner grid 
boundary in the simulations with the \textsc{Prometheus-HOTB} code 
(see Section~\ref{sec:PHOTBcode}). Therefore we will not further discuss the
neutron-to-proton ratio in the ejecta of this model but refer the reader to
\citet{Wanajo2011} and \citet{Wanajo2018}.

The $Y_e$ distributions of models \znine and \snine in the left panels of
Figure~\ref{fig:sto ye s9 z9 kick} exhibit a clear asymmetry: proton-rich
neutrino-driven outflow (red and orange) in the hemisphere of the 
LESA dipole vector, and neutron-rich neutrino-wind ejecta (blue hues) 
in the opposite hemisphere.
The proton excess is a consequence of the fact that the PNS develops
an enhanced emission of $\nu_e$ relative to $\bar\nu_e$ 
in the hemisphere of the LESA direction, which leads to 
a predominant production of protons by $\nu_e$ absorption on neutrons
($\nu_e + n \to e^- + p$) in the ejecta of the same hemisphere. In the 
anti-LESA direction the emission of $\bar\nu_e$ by the PNS 
is relatively higher, allowing for more efficient creation of
neutrons through the reaction $\bar\nu_e + p \to e^+ + n$ in the 
neutrino-heated outflow.

In addition to this spatial asymmetry in the neutron-to-proton ratio, 
which is present in both explosion models
computed with the \textsc{Vertex-Prometheus} code, model \znine exhibits
neutron-richness in the mushroom-shaped heads of Rayleigh-Taylor plumes 
in all directions, very similar to previous findings in 2D explosion
simulations of the low-mass progenitors of models \onemg and \znine
with self-consistent neutrino transport \citep[see][]{Wanajo2011,Wanajo2018}. 
The plumes contain the earliest neutrino-heated ejecta. Their neutron 
excess is explained by the extremely fast outward propagation of the SN shock 
in these two models, which is enabled when the mass-accretion rate of the
shock plummets because of the steep density decline at the edge of the
degenerate core. The fast shock acceleration allows the neutrino-heated, 
buoyant gas to expand very quickly away from the gain radius. The gas 
rises so rapidly that neutrino absorption is unable to increase $Y_e$ from 
its low values ($\sim$0.25--0.35) near the gain radius to 0.5 or higher. 
Instead, the mushroom plumes stay considerably neutron rich in model \znine,
in contrast to model \snine, where the slower shock acceleration and the
correspondingly slower expansion of the first bubbles of 
neutrino-processed matter lead to $Y_e$ around 0.5 due to the
dominant absorption of $\nu_e$ (Figure~\ref{fig:sto ye s9 z9 kick}).

\begin{figure*}
\includegraphics[width=2.0\columnwidth]{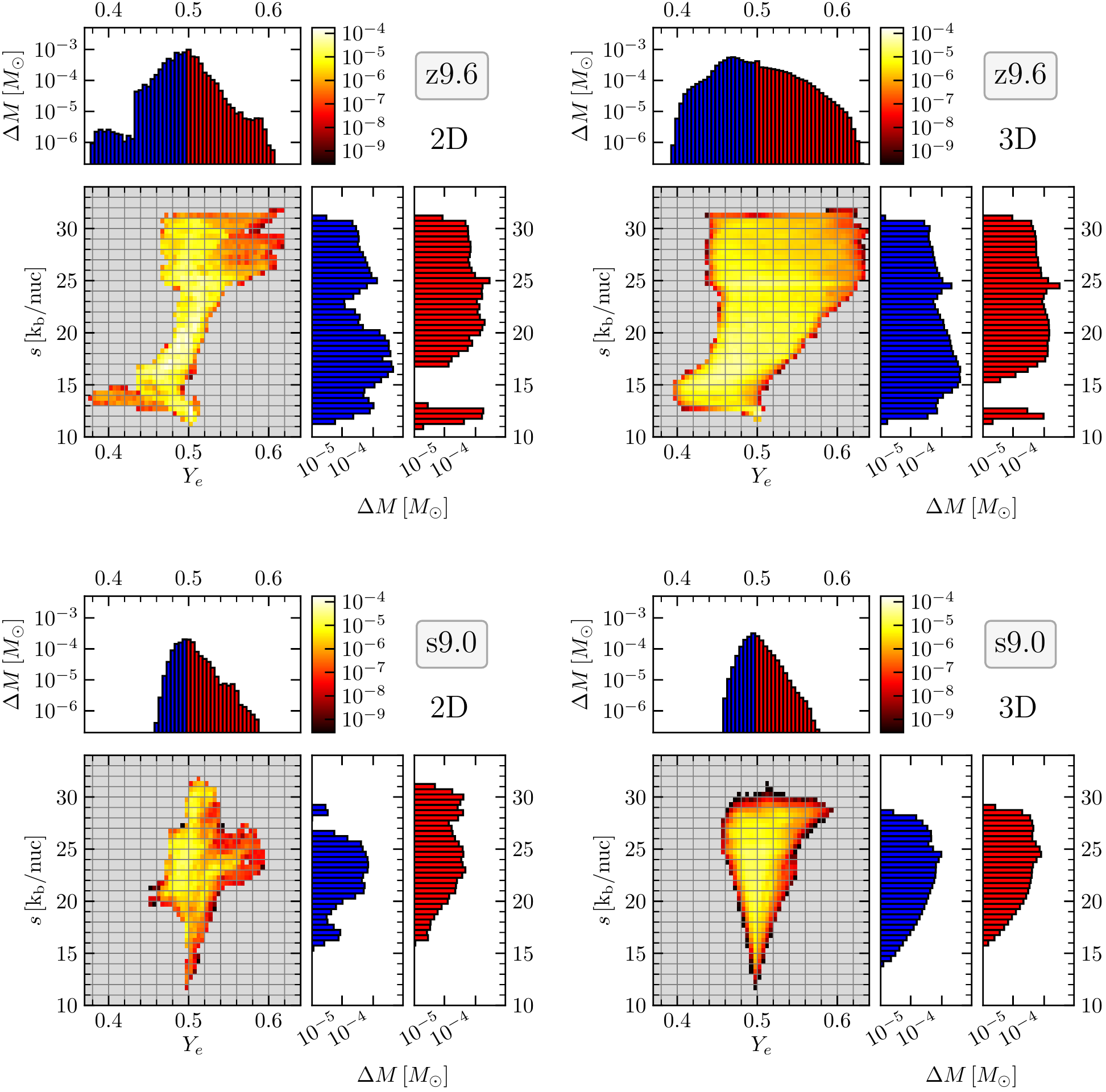}
\caption{Ejecta mass distributions versus $Y_e$ and entropy per nucleon,
$s$ (in units of Boltzmann's constant), for 2D (left) and
3D (right) simulations of model \znine (top) and model \snine (bottom).
The central panels in each case display the mass distribution
in the $Y_e$-$s$-plane (color coding in units of M$_\odot$ according to
the color bar), the corresponding panels above and to the right show
the marginal distributions. Blue and red bars indicate $Y_e\,\mathord{<}\,0.5$ and 
$Y_e\,\mathord{>}\,0.5$, respectively. In order to minimize effects of numerical 
diffusion, the mass distributions for $Y_e$ and $s$ were measured for
outflowing material ($v_r\,\mathord{>}\,0$) at a radius of $r\,\mathord{=}\,250$\,km.}
\label{fig:yedistributions}
\end{figure*}

Figure~\ref{fig:yecuts} permits a closer look at the situation in model \znine.
The upper left panel displays trajectories (projected onto the $x$-$y$-plane)
of three infalling, then neutrino-heated, and finally outgoing mass elements.
The color-coding of these trajectories as well as the upper right panel show that
the infalling matter starts with a value of $Y_e\,\mathord{=}\,0.5$, gets neutronized
by electron captures during infall, reverses direction at a minimum radius around
100\,km close to the gain radius, and experiences an increase of $Y_e$ 
again as it speeds away from the gain radius. The rise of $Y_e$, however,
flattens when the expansion time scale, $\tau_\mathrm{exp}\,\mathord{=}\,r\left|v_r\right|^{-1}$
(with $v_r$ being the local radial component of the velocity), becomes shorter than
the timescale of $Y_e$ changes, 
$\tau_{\mathrm{d}Y_e/\mathrm{d}t}\,\mathord{=}\,Y_e \left|\dot Q_{Y_e}\right|^{-1}$
(with $\dot Q_{Y_e}$ being the local net
rate of $\nu_e$ and $\bar\nu_e$ emission and absorption reactions as computed 
during the hydrodynamic simulation). This can be seen in the upper
and lower right panels. The electron fraction in the expanding ejecta 
stays well below the value for local kinetic equilibrium, $Y_e^\mathrm{equil}$,
which corresponds to the condition when $\nu_e$ and $\bar\nu_e$
absorption and emission reactions are balanced for the
density and temperature of the tracked mass elements at each radius $r$. 
Close to the point of return the radial velocities become small and
$\tau_\mathrm{exp}$ increases steeply, whereas farther out $|v_r|\,\propto\,r$
roughly holds and $\tau_\mathrm{exp}$ becomes nearly constant. The
values of $\tau_{\mathrm{d}Y_e/\mathrm{d}t}$ for trajectories A and C
increase at small radii, because these mass elements approach the NS
so closely that their $Y_e$ at the return point drops to values near
$Y_e^\mathrm{equil}$, for which reason $\dot Q_{Y_e}$, the net rate of 
$Y_e$ changes, becomes low. In contrast, mass element B returns outward at
a considerably larger radius and its $Y_e$ value never gets close to
the local $Y_e^\mathrm{equil}$. Only at $r\,\mathord{\approx}\,110$\,km the condition
$Y_e^\mathrm{traj}\,\mathord{\sim}\,Y_e^\mathrm{equil}$ is incidentally fulfilled
in mass element B, and $\tau_{\mathrm{d}Y_e/\mathrm{d}t}$ exhibits a prominent 
maximum at this position. With growing distance from the NS,
$\tau_{\mathrm{d}Y_e/\mathrm{d}t}$ increases and exceeds the expansion
timescale, because all neutrino rates become slow and ultimately cease
when the mass elements gain distance from the neutrino source and move
to low densities and temperatures.
This explains why the asymptotic values of $Y_e$ along the trajectories 
stay well below the local equilibrium values at large radii.

\begin{table*}
    \centering
     \caption{Total masses of chemical elements $M_\mathrm{i}$ of the 
    pre-collapse progenitor models as shown in Figure~\ref{fig:composition_all} (top),
    behind the shock front in our 3D SN simulations at the time of mapping to the 
    long-time runs, $t_{\mathrm{map}}$, and at the time when the entire shock has broken
    out from the surface of the exploding star, $t_\mathrm{sbo}$. 
    Also given are the binding energies still ahead of the shock at time $t_{\mathrm{map}}$.}
    \label{tab:composition tmap}
    \renewcommand{\arraystretch}{1.14}
    \begin{tabular}{l|ccc||ccc}
          &
         \onemg &
         \znine &
         \snine  &
         \onemg &
         \znine &
         \snine  \\
         & & $M_{\mathrm{i}}\,[\solm] $  & & & $M_{\mathrm{i}}\,[\solm] $  &   \\
         \hline 
          Species &
         \multicolumn{3}{c}{Progenitor} & & &  \\
         \hline 
         $\mathrm{H}$           & $2.68$              & $5.10$              & $4.54$             & & &  \\
         \helium                & $1.79$              & $3.11$              & $2.68$             & & &  \\
         \carbon                & $2.71\times10^{-2}$ & $2.24\times10^{-2}$ & $3.37\times10^{-2}$& & &  \\
         \oxygen\neon\magnesium & $0.94$              & 0.04                & 0.13               & & &  \\
         \silicon               & $3.98\times10^{-3}$ & $2.32\times10^{-2}$ & $1.34\times10^{-2}$& & &  \\
         Iron-group/NSE         & $0.39$              & 1.30                & 1.32               & & &  \\
         \hline 
         Species &
         \multicolumn{3}{c}{Postshock ejecta at $t_\mathrm{map}$} & \multicolumn{3}{c}{Postshock ejecta at $t_\mathrm{sbo}$}  \\
         \hline 
         $n$                            & -                   & $4.25\times10^{-5}$ & $8.22\times10^{-8}$ & -                   & -                   & -                   \\
         $p$                            & -                   & $4.64\times10^{-4}$ & $7.27\times10^{-5}$ & -                   & $5.34\times10^{-4}$ & $5.81\times10^{-5}$ \\
         \hydrogen                      & $1.97\times10^{-7}$ & 0                   & 0                   & $2.68$              & 5.10                & 4.54                 \\
         \helium                        & $4.90\times10^{-3}$ & $8.24\times10^{-3}$ & $5.48\times10^{-5}$ & 1.74                & 3.12                & 2.68                \\
         \carbon                        & $4.08\times10^{-5}$ & $1.52\times10^{-3}$ & $3.25\times10^{-3}$ & $2.19\times10^{-2}$ & $2.26\times10^{-2}$ & $4.04\times10^{-2}$ \\
         \oxygen\neon\magnesium         & $1.00\times10^{-4}$ & $2.20\times10^{-3}$ & $4.13\times10^{-2}$ & $4.95\times10^{-2}$ & $7.76\times10^{-3}$ & 0.11                \\
         \silicon                       & $1.63\times10^{-5}$ & $2.27\times10^{-4}$ & $1.29\times10^{-3}$ & $1.33\times10^{-3}$ & $2.71\times10^{-4}$ & $7.10\times10^{-3}$ \\
         \nickel \ \ or \nickel\cobalt\iron & $1.05\times10^{-3}$ & $6.50\times10^{-4}$ & $4.72\times10^{-3}$ & $3.41\times10^{-3}$ & $3.93\times10^{-3}$ & $6.35\times10^{-3}$ \\
         \tracer                        & $6.69\times10^{-3}$ & $1.05\times10^{-3}$ & $1.06\times10^{-6}$ & $1.30\times10^{-2}$ & $1.05\times10^{-3}$ & $8.62\times10^{-7}$ \\
         \hline
          Binding energy ahead of the &
         \multicolumn{3}{c}{} &  \multicolumn{3}{c}{}\\
          shock at $t_{\mathrm{map}}$ &
         \multicolumn{3}{c}{$[10^{47}\,\erg]$} & & & \\
         \hline
          $E_{\mathrm{bind}}(r>R_{\mathrm{sh}})$ & $-0.56$ & $-6.89$ & $-9.76$ & & &  \\
     \end{tabular}
 \flushleft
 \textit{Notes:} $n$ and $p$ label free neutrons and protons, respectively,
 left unbound after the freeze-out from NSE, whereas \hydrogen labels hydrogen from the envelope. Since 
 the onset of the explosion of model \onemg was simulated with \prom, no free protons remain in the 
 ejecta.
 The high mass of \helium compared to \nickel and \tracer in model \znine at $t_{\mathrm{map}}$ 
 is a consequence of the high value of the temperature when NSE was switched off in this simulation, $T_{\mathrm{NSE}}\,\mathord{\sim}\,5.8\mathord{\times}10^{9}\,\mathrm{K}$,
 instead of $T_{\mathrm{NSE}}\,\mathord{\sim}\,4.0\mathord{\times}10^{9}\,\mathrm{K}$ in model
 \snine. $E_{\mathrm{bind}}(r\,\mathord{>}\,R_{\mathrm{sh}})$ 
 is the total binding (i.e. internal + kinetic + gravitational)
 energy ahead of the shock at the mapping time $t_{\mathrm{map}}$. Free neutrons at $t_{\mathrm{map}}$
 are assumed to decay and are added to free protons at $t_{\mathrm{sbo}}$. Since our network in the long-time
 runs contains only \nickel, \cobalt, and \iron, all other iron-group species (from the explosion models
 or the progenitor) are included in the mass of these nuclei listed for $t_{\mathrm{sbo}}$.
 \end{table*}

In order to unravel similarities and differences between our current 3D models
and the 2D explosion simulations of low-mass progenitors considered previously 
by \citet{Wanajo2018}, we compare the ejecta properties of our 3D models to 
corresponding 2D results in the following three paragraphs. 
Neutron-rich, buoyant, fast plumes  
are not only present in the 3D case but also in the corresponding
2D models (left panels in Figure~\ref{fig:yecuts}). However, in the 2D case
extended regions around the PNS with proton excess or neutron excess exist 
in all directions in the northern as well as southern hemisphere. If this
finding is connected to a 2D phenomenon that corresponds to LESA in 3D, the
LESA direction is less stable in 2D than in 3D. Indeed, an inspection of
the dipole of the electron-neutrino lepton-number flux shows that the 
dipole direction in 2D flips from one hemisphere to the other with a 
full cycle period of $\sim$0.1\,s. Moreover, the grid axis seems to have
a disturbing influence that leads to artificial effects, because both in 
the northern and southern directions proton-rich, collimated outflows 
appear very close to the axis, where they are much stronger than at large
angles away from the axis. Because of the influence of the artificial
symmetry axis, it is not easy to diagnose whether the flipping 2D dipole
has any physical relation to the stable or slowly migrating lepton-emission 
dipole of LESA in 3D.

In Figure~\ref{fig:yedistributions}, we present the ejecta mass distributions 
versus entropy per nucleon, $s$, and electron fraction, $Y_e$,
for 2D and 3D simulations of \znine (top) and \snine (bottom), including 
the fast convective plumes as well as the neutrino-driven wind material.
The mass distributions are constructed by integrating all matter that
flows through a sphere of 250\,km radius with positive radial velocities.
This choice of radius ensures that neutrino interactions have essentially
ceased at this location, but effects due to numerical diffusion and mixing
in the outflowing material, when it moves across the Eulerian (i.e., spatially
fixed) computational grid, are minimized in the mass-versus-$Y_e$ 
distributions extracted from the simulations for nucleosynthesis 
discussions. A comparison of the upper and lower left panels in 
Figure~\ref{fig:yecuts} demonstrates the consequences of this diffusion.
The mushroom heads are very neutron rich at 200--300\,km and 
$t_\mathrm{pb}\,\mathord{=}\,0.13$\,s. However, although the 
neutrino reactions basically cease at $r\,\mathord{\gtrsim}\,250$\,km,
the electron fraction in the nearly self-similarly expanding mushroom heads
still continues to increase slowly to values closer to 0.5 (see lower left
panels for $t_\mathrm{pb}\,\mathord{=}\,0.4$\,s). This is partly 
an unphysical, but unavoidable, consequence of the fact that the mushroom
heads experience gradual numerical mixing (besides some unclear degree
of physical mixing) with the surrounding postshock 
matter, which possesses $Y_e\,\mathord{=}\,0.5$.

The 2D and 3D mass distributions exhibit close similarity in their shapes 
as well as widths, with differences only in details of their substructure. 
These differences originate mainly from the fact that in 2D there are only
a few plumes, which are axially symmetric, massive objects, whereas
in 3D there is a greater number of such bubbles, of which each one
contains less mass than the toroidal 2D objects. The 3D distributions
are therefore smoother and possess less fine structure.
The total masses of neutrino-processed ejecta until about 0.5\,s
after bounce are $10.34\mathord{\times} 10^{-3}$\,M$_\odot$
in model \znine and $2.64\mathord{\times} 10^{-3}$\,M$_\odot$
in model \snine. In both cases about 60\% of these masses are in the
fastest mushroom-shaped plumes, the remaining 40\% are carried by the
neutrino-heated subsequent outflows. Also in both cases roughly 
50\% of this latter component are neutron-rich ($1.90\mathord{\times} 10^{-3}$\,M$_\odot$
in \znine and $0.55\mathord{\times} 10^{-3}$\,M$_\odot$ in \snine), and about
50\% are proton-rich ($2.31\mathord{\times} 10^{-3}$\,M$_\odot$ in \znine and 
$0.51\mathord{\times} 10^{-3}$\,M$_\odot$ in \snine).

The mass-versus-$Y_e$ distributions of model \znine in 2D and 3D
resemble closely the corresponding distributions obtained in 
2D explosion simulations for the ONeMg-core progenitor 
e8.8$_\mathrm{n}$ (mentioned in Section~\ref{sec:onemgcoreprog})
and for our 9.6\,M$_\odot$ Fe-core progenitor and 
another ultra-metal-poor Fe-core progenitor of 8.1\,M$_\odot$, whose results were
presented in \citet{Wanajo2011} and \citet{Wanajo2018} (see figure~2 in both papers).
In all cases the distributions are very wide, stretching from
$Y_e\,\mathord{\sim}\,0.55$--0.6 on the proton-rich side to
$Y_e\,\mathord{\sim}\,0.40$ (or even a bit lower in the 2D models) on the
neutron-rich side. The corresponding nucleosynthetic abundance patterns
are extremely similar and characteristic of this class of ``ECSN-like'' 
explosions, with high
production factors for light trans-Fe elements from Zn to Zr, resulting
from the appreciable ejection of neutron-rich matter in these models
\citep{Wanajo2018}. 
In contrast, model \snine displays a mass-vs-$Y_e$ distribution that
resembles the result for the 11.2\,$M_\odot$ model s11 in \citet{Wanajo2018}.
It is more strongly peaked around 0.5 and has steeper and less extended 
left and right wings. In particular, considerably less matter with neutron
excess is ejected, and the minimal $Y_e$ is around 0.46 instead of 
$\lesssim$0.40. The reason for this difference are the lower velocities 
with which the neutrino-heated ejecta expand outward from the gain radius.
This slower expansion allows $\nu_e$-absorption to lift $Y_e$ from its low initial values at the gain radius 
to values closer to 0.5. Such conditions are less favorable for the 
production of neutron-rich trans-Fe species. The abundance pattern in
the neutrino-processed ejecta of model \snine must be expected to show
similarities with model s11 of \citet{Wanajo2018}. A detailed investigation
of the formation of chemical elements in \snine in comparison to \znine is 
deferred to future work.

\subsection{Elemental distribution shortly after shock revival}
\label{sec:Extent of mixing at shock revival}

\begin{figure*}
 \centering
 \includegraphics[width=\textwidth,trim=0.0cm 0.0cm 0cm 0cm,clip]{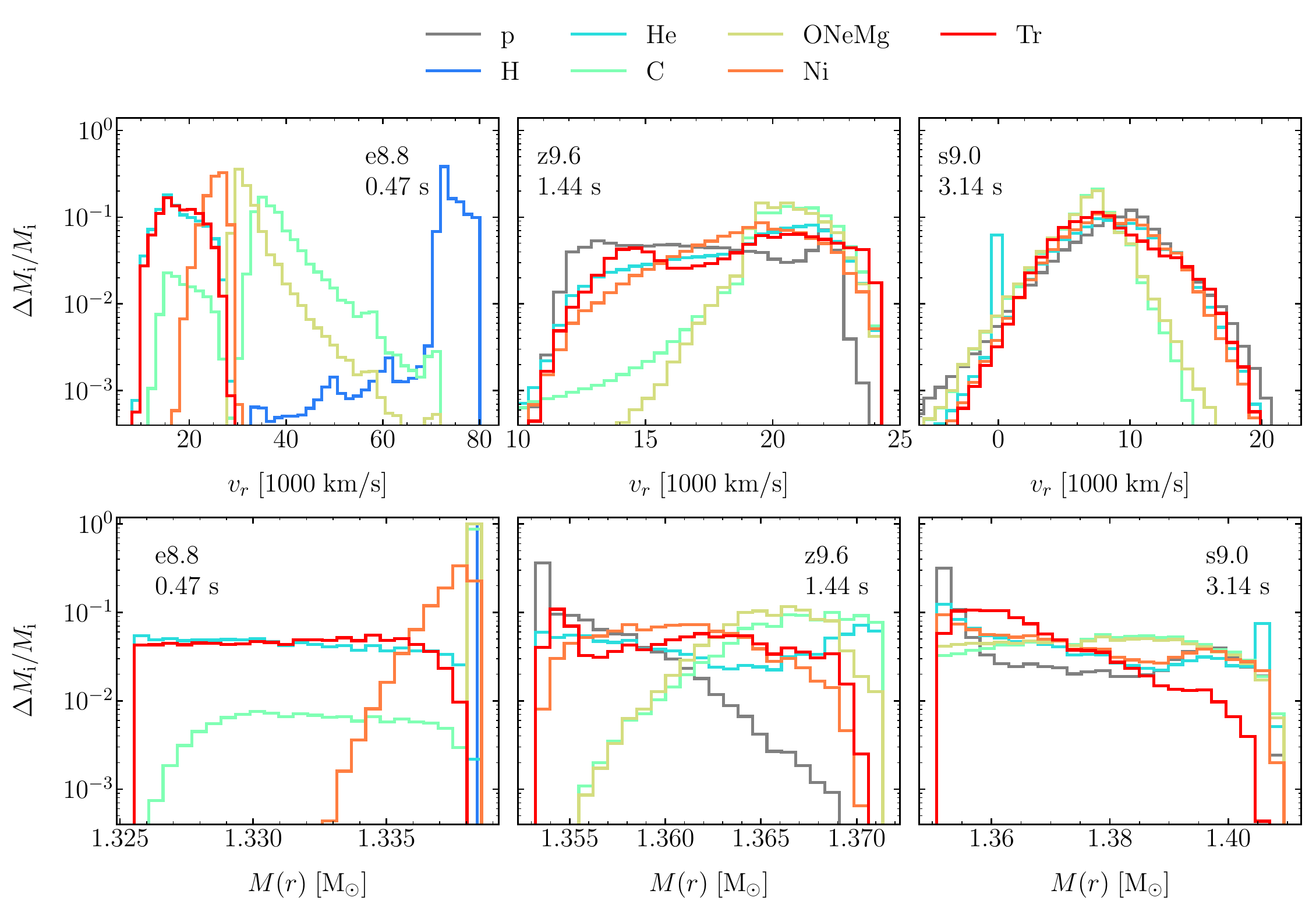}
 \caption{Normalized binned distributions of chemical elements as functions of radial 
 velocity (top panels) and enclosed mass (bottom panels) for all of the
 unbound postshock material in our 3D models at $t_{\mathrm{map}}$ 
 (see Table~\ref{tab:long term boundaries}). We use 50 bins in velocity 
 space and 30 bins in the mass coordinate, and the normalization for 
 each species employs the total unbound mass behind the shock. 
 The mass coordinate in the 3D models
 is defined by the mass enclosed by given radius and starts at the 
 mass value contained by the PNS. Note that according to 
 Table~\ref{tab:composition tmap} only model e8.8 contains a small amount
 of hydrogen from the envelope in the postshock region at $t_{\mathrm{map}}$.}
 \label{fig:mdp first mapping}
\end{figure*}

In order to assess the strength of physical mixing due to multidimensional
hydrodynamical flows during the first seconds in 
our 3D models, we present in Figure~\ref{fig:mdp first mapping} normalized 
mass distributions for selected chemical elements behind the shock radius 
as functions of velocity (top panels) and mass coordinate (lower panels). 
To sample the velocity space, we choose 50 bins between the maximal 
and minimal velocities within the considered region. We use 30 bins to 
sample the distribution in the mass coordinate. The total masses 
in the postshock volume at the time $t_\mathrm{map}$ when we map our 
models onto the new computational grid for the long-time simulations
are listed in Table~\ref{tab:composition tmap}.

The quasi-spherical explosion of model \onemg manifests itself also in the distribution of 
the chemical elements over mass and velocity coordinates, since the initial shell structure 
of the progenitor is essentially preserved. Because of the extreme shock acceleration 
in the steep density gradient at the edge of the degenerate core, most
material of the thin carbon shell of the progenitor travels with $\mathord{\gtrsim} 30,000\mathord{-}70,000\,\kms$ ahead of the outer mass shells of the former ONeMg-core, 
which propagate with up to $\mathord{\sim} 60,000\,\kms$. 
Most of the newly synthesized iron-group and $\alpha$-nuclei expand with velocities below 
$\mathord{\sim} 30,000\,\kms$. Inefficient mixing also confines the neutrino-heated ejecta 
within $M(r)\,\mathord{\le}\,1.338\,\solm$.

Slightly more powerful convective overturn in model \znine leads to more efficient mixing 
in mass and velocity space (see middle panels of Figure~\ref{fig:mdp first mapping}). 
Therefore some neutrino-heated material gets mixed into the carbon and oxygen shells and 
travels at more than $\mathord{\gtrsim} 20,000\,\kms$. 
The bulk of the metal-rich ejecta still expands with slower velocities, however. 
Similar to model \onemg, the deceleration of the SN shock at the CO/He interface in model 
\znine also compresses parts of the ejecta into a dense shell. 
Model \snine, again, differs distinctively from the ECSN-like models. 
Strong and long-lasting convective overturn in the postshock layer completely erases the 
initial onion-shell structure of the progenitor. The chemical elements become nearly 
homogenously mixed over mass and velocity coordinates (see right panels of 
Figure~\ref{fig:mdp first mapping}).
The fastest neutrino-heated ejecta are associated with a big high-entropy plume 
inducing a dipolar deformation of the shock wave, which can be seen in 
Figure~\ref{fig:slices first mapping}.


\begin{figure*}
 \centering
 \includegraphics[width=\textwidth]{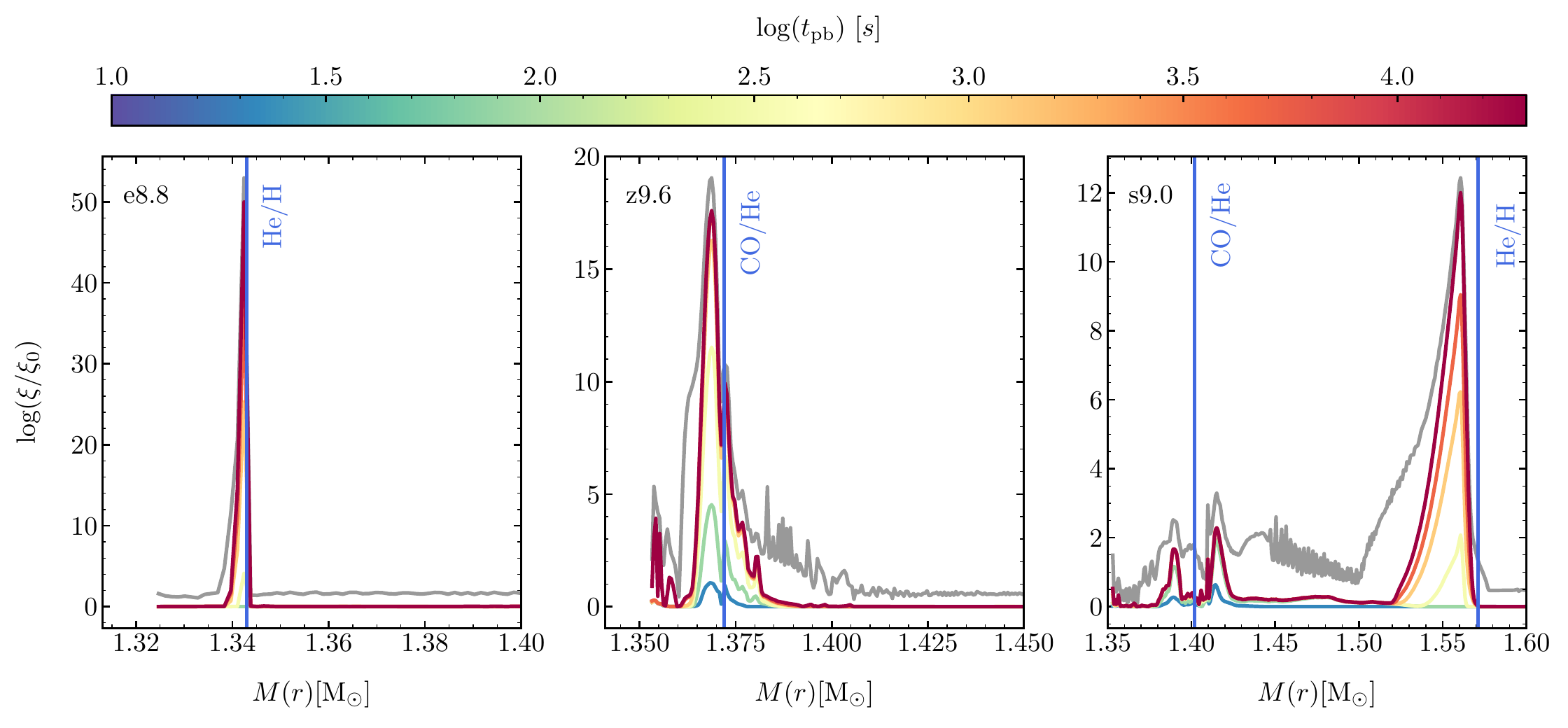}
 \caption{Integrated growth rates for RT instability for all considered models at different times evaluated with 
 Equation~\eqref{equ:growth rates int} for the incompressible case (colored lines). The different
 composition interfaces of the progenitors are indicated by
 vertical blue lines. The gray line denotes the growth factor in the compressible 
 case at $t\mathord{=}10^{4}\,\text{s}$. The different progenitor structures have a large impact on the estimated 
 growth factor as can be seen most prominently by comparing models z9.6 and s9.0. While the ECSN-like progenitors 
 develop only a narrow region of instability around the interface with the steepest density gradient 
 (e.g. He/H for model \onemg and CO/He for model \znine), model \snine shows two distinct regions of high 
 growth factor around the CO/He and the He/H interfaces. }
 \label{fig:growth rates}
\end{figure*}

\section{Evolution beyond Shock Breakout}
\label{sec:Evolution until Shock-Breakout}

\subsection{Linear stability analysis}
\label{sec:Linear stability analysis}

As mentioned already, the velocity of the forward shock depends on the progenitor structure, 
in particular the $\rho r^3$-profile. According to \cite{Sedov1961} one expects an increase/decrease in the shock 
velocity when the gradient of the $\rho r^3$-profile is negative/positive. 
This acceleration and deceleration lead to density and pressure gradients in the post-shock 
region with opposite signs. Thus, perturbations 
in the matter distribution become unstable to the Rayleigh-Taylor (RT) instability \citep{Rayleigh1882,Chevalier1978}. As the 
$\rho r^3 $-profiles, and hence the shock speed, vary significantly between 
the progenitors, the growth of 
RT instabilities will affect the long-time evolution of our models in different ways.

In order to aid us with the interpretation of our three-dimensional simulations, 
we appeal to 1D long-time simulations 
which are started from angle-averaged states of the 3D explosion
models at the same times as given in Table~\ref{tab:long term boundaries}. The numerical and physical setup for the
spherically symmetric simulations remains unchanged compared to the 3D runs. The 1D models 
can teach us about the behavior of the shock, while it propagates through the envelope. 
Additionally, we can compute the linear RT growth rates $\sigma_{\mathrm{RT}}$ of small initial perturbations by 
tracking Lagrangian mass coordinates in our 1D simulations, following \cite{Mueller1991}. 
In the incompressible case the growth rate is given by
\begin{equation}
  \label{equ:growth rates incmp}
  \sigma_{\mathrm{RT,incmp}} = \sqrt{- \frac{p}{\rho}\frac{\partial \ln p}{\partial r}\frac{\partial \ln \rho}{\partial r}},
\end{equation}
where $p$ and $\rho$ are the pressure and density. In the compressible case the growth rate becomes
\begin{equation}
  \sigma_{\mathrm{RT, cmp}} = \frac{c_{s}}{\gamma}\sqrt{\Big(\frac{\partial \ln p}{\partial r}\Big)^ 2 - 
  \gamma \frac{\partial \ln p}{\partial r}\frac{\partial \ln \rho}{\partial r}},
  \label{equ:growth rates cmp}
\end{equation}
where $c_{\mathrm{s}}$ is the speed of sound and $\gamma$ the adiabatic index. 
Equation~\eqref{equ:growth rates cmp} is less restrictive than its incompressible 
counterpart\footnote{Note that we are interested in the locations of maximal growth but not in the actual values. 
Calculating the growth factors from multidimensional models using angle averaged-values gives similar 
overall structure but lower amplitudes \citep{Mueller2018}.}. 
For the time-dependent amplification factor of an initial perturbation $\xi_0$ 
we integrate Equation~\eqref{equ:growth rates cmp} according to
\begin{equation}
  \Sigma_\mathrm{RT}(t) = \frac{\xi}{\xi_0}(t) = \exp \left(\int_0^t \sigma_\mathrm{RT} (t') dt' \right).\label{equ:growth rates int}
\end{equation}
This analysis enables us to estimate the time and locations in mass coordinate
where the fluid becomes unstable to the RT instability, helping us to understand 
the origin of outward/inward mixing of different layers of chemical elements.

In Figure~\ref{fig:growth rates} we show the amplification factors in the incompressible (colored lines) 
and compressible (gray line) cases for our spherically symmetric simulations. Significant differences between the models are evident.

The left panel of Figure~\ref{fig:growth rates} shows the integrated growth rate of the 1D version of model $\mathrm{e}8.8_{10}$.
The extreme deceleration of the forward shock when moving into the hydrogen envelope induces very high 
growth factors at $M(r)\,\mathord{\sim}\,1.34\,\mathrm{M_{\odot}}$.
The ECSN-like structure of model \znine is reflected in its amplification factors.
Strong deceleration of the forward shock at the CO/He interface creates the necessary 
condition for the RT instability to grow there. As can be seen in 
Figure~\ref{fig:growth rates}, the amplification factors at 
$\mathord{\sim}1.37\,\solm$ reach extremely high values as observed in the ECSN case, 
too. Interestingly, no growth is expected at the He/H 
interface, which is a consequence of the tiny step in the $\rho r^3$ profile 
at this interface (see Figure~\ref{fig:prog_tem_rho_ye_rhor}, bottom left 
panel, blue curve, at $r\,\mathord{\approx}\,10^{12}$\,cm).

Model \snine shows striking differences to the models described above. 
Due to its more shallow $\rho r^3$-profile in the core region and the consequently weaker episodes of shock deceleration 
and acceleration, the peak amplitudes of the growth factors are smaller. They are, however, of similar magnitude as in the more massive models investigated by \cite{Wongwathanarat2015}. 
Two distinct regions of instability can be discerned around the CO/He and He/H interfaces. 
Similar to the RSGs presented in \citet{Wongwathanarat2015}, the strongest contribution arises 
from the He/H interface followed by the CO/He interface, where the 
amplification factors are about 10 orders of magnitude lower.

While the compressible and incompressible analyses give similar results for models \onemg and \znine, 
the compressible evaluation of the growth factors in model \snine predicts additional growth within the 
He-shell of the star. This is caused by the passage of the reverse shock at later times 
($\tpb\,\mathord{>}\,3\mathord{\times}10^3\,\s$). Note, however, that at this advanced stage
of evolution the instability is already in a 
strongly non-linear regime, where Equation~\eqref{equ:growth rates cmp} loses its validity.

\begin{figure*}
 \centering
 \includegraphics[width=\textwidth]{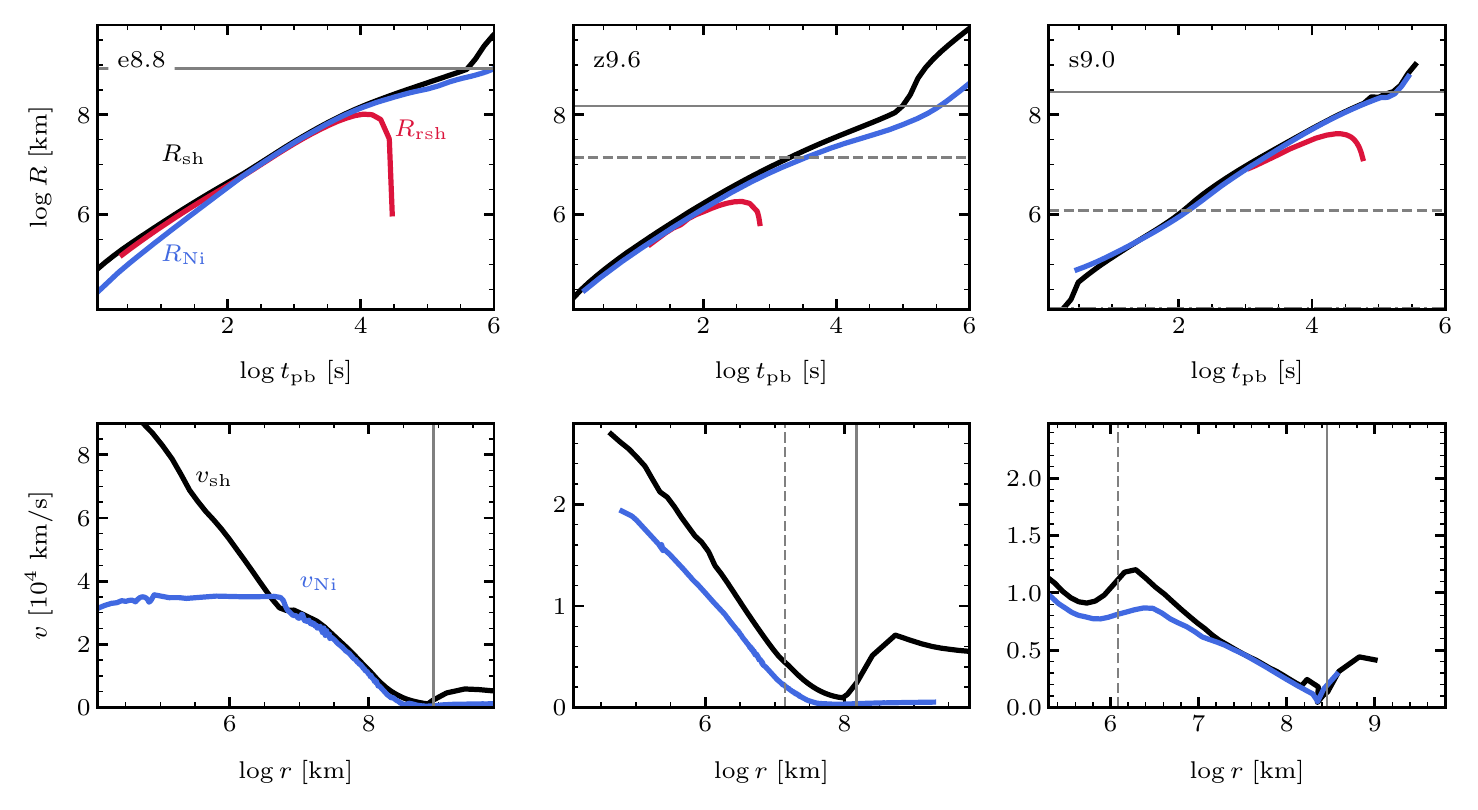}
 \caption{   
  \textit{Upper~panels}: Angle-averaged shock radius $R_{\mathrm{sh}}$ (black),
 angle-averaged reverse shock radius $R_{\mathrm{rsh}}$ (red) and maximum radius of the 
 iso-surface of mass fraction $X_{\nickel\mathord{+}\tracer}\,\mathord{=}\,0.03$ (blue). 
 \textit{Lower~panels}:
 Average shock velocity $v_{\mathrm{sh}}$ (black) and maximum velocity of the iso-surface
 of $X_{\nickel\mathord{+}\tracer}\,\mathord{=}\,0.03$ (blue) 
 in our 3D long-time simulations. The horizontal or vertical gray dashed and solid
 lines mark the location of the He/H interface and the surface of the progenitor star, respectively.
 The expansion of the forward shock in the ECSN-like progenitors proceeds with
 continuous deceleration nearly until the shock reaches the stellar surface.
 In contrast, the forward shock in model \snine accelerates strongly around the He/H interface.
 The reverse shocks in models \onemg and \znine form within $\mathord{\approx}1\,\s$ and 
 $\mathord{\approx}100\,\s$, respectively, whereas we witness the formation of a reverse shock 
 from the He/H interface in model \snine at $\mathord{\approx}2500\,\s$.
 Note that in model \snine the velocity of the fastest clump containing 
 $\nickel \mathord{+} \tracer$ is higher than the average shock velocity within the 
 hydrogen envelope of the star.
 }
 \label{fig:radii all times}
\end{figure*}

\begin{figure*}
 \centering
 \includegraphics[width=\textwidth, trim=0cm 0.0cm 0cm 0.0cm,clip]{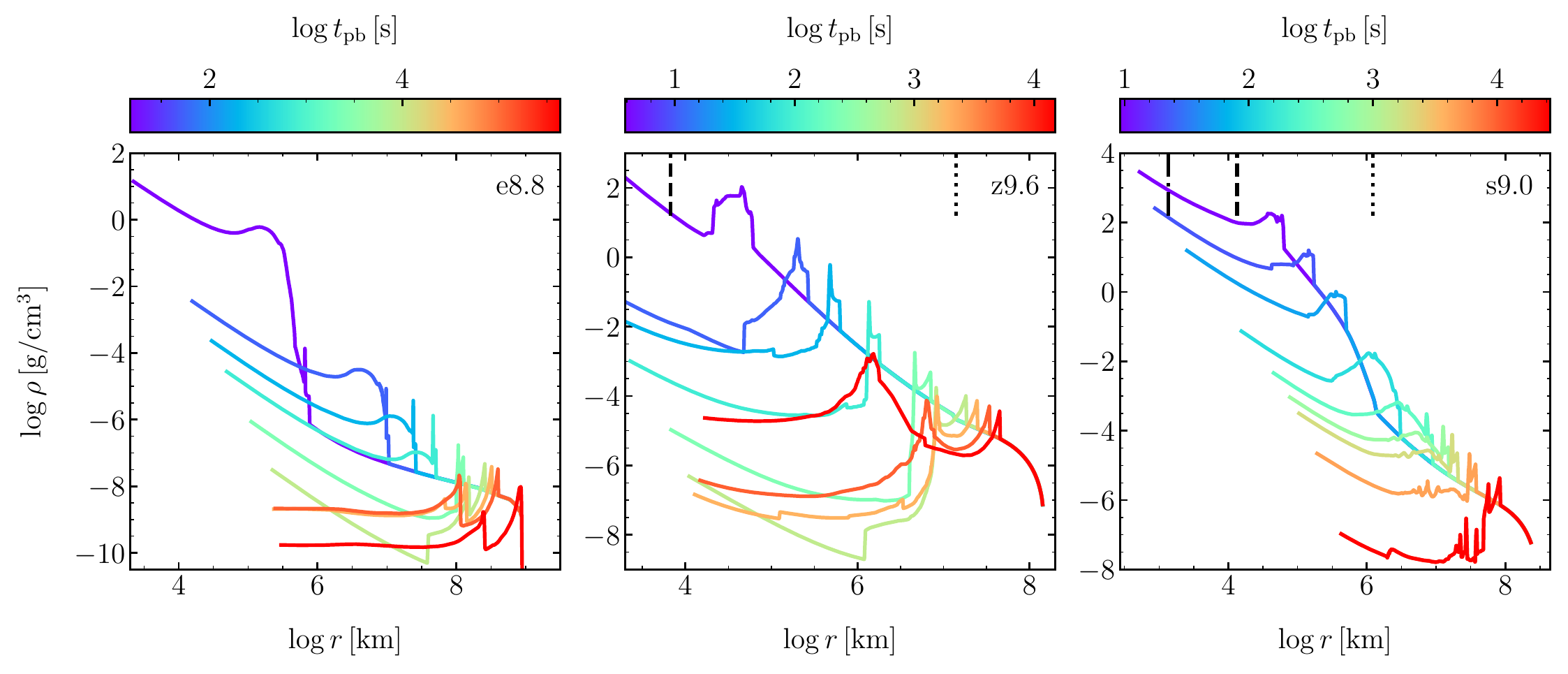}
 \caption{Radial profiles of the density of our 1D long-time simulations at representative times. 
 The dash-dotted, dashed and dotted vertical black lines represent the outer boundaries of the 
 degenerate core, CO, and He cores, respectively. See text for discussion.}
 \label{fig:density profiles all times}
\end{figure*}

\subsection{Propagation of the supernova shock}
\label{sec:Propagation of the forward shock}

In Figure~\ref{fig:radii all times} we show the angle-averaged shock radii $R_{\mathrm{sh}}$, 
reverse-shock radii $R_{\mathrm{rsh}}$ and shock velocities $v_{\mathrm{sh}}$ for our 
3D long-time simulations.

\subsubsection{3D model e8.8}
In model \onemg the forward shock has passed the He/H interface and is traveling through 
the H-envelope. It left the core with about $100,000\,\kms$ and is progressively 
decelerated due to the monotonically increasing $\rho r^3$ within the H-envelope 
(see also Figure~\ref{fig:eexp all}). 
Since the innermost neutrino-heated ejecta are not in sonic contact with the 
SN shock, they do not feel the deceleration and thus travel with constant velocity of 
$\mathord{\approx}35,000\,\kms$ until they catch up with the immediate post-shock matter at 
$\tpb\,\mathord{\approx}\,130\,\s$. The interaction of the fast neutrino-heated ejecta 
with the post-shock material keeps the shock at slightly 
higher velocities (see small kink at $r\,\mathord{\sim}\,10^7$\,km,
lower left panel in Figure~\ref{fig:radii all times}).
Nevertheless, the forward shock decelerates continuously (without relevant phases
of acceleration) until it reaches the surface of the star. Note the dramatic deceleration of the shock from its 
initial velocity of $75,000\,\kms$ at $\tpb\,\mathord{=}\,2.55\,\s$
down to $\mathord{<}1000\,\kms$ when it leaves the star. 
A large fraction ($\sim$75\%) of the kinetic energy of the shock wave is used up for
heating the hydrogen in the outer layers of the envelope, leaving only
around $2.5\mathord{\times}10^{49}\,\erg$ of kinetic energy in the
ejecta.

The early strong deceleration of the forward shock at the bottom of the 
H-envelope at $\tpb\,\mathord{\gtrsim}\,0.25\,\text{s}$ (see also Figure~\ref{fig:eexp all})
leads to the compression of the CO and He layers into a high density shell.
Less than 1\,s later, a strong reverse shock forms at the base of this shell 
as the forward shock is progressively slowed down. The dense shell separates the 
postshock material from the neutrino-heated wind ejecta, which can be seen
in the density profiles shown in Figure~\ref{fig:density profiles all times}. 
It is remarkable that in the ECSN the neutrino-heated ejecta are
denser than the postshock shell of the matter swept up by the shock in the 
H-envelope. The formation of the dense shell or ``wall'' \citep{Kifonidis2006} 
has important consequences for the evolution of the metal-rich ejecta as well as for the growth
of Rayleigh-Taylor instabilities. After $\tpb\,\mathord{\approx}\,1\,$h, continuous 
deceleration of the forward shock causes the reverse shock to travel back in mass coordinate
(see Figure~\ref{fig:density profiles all times}). It takes, however, a total of approximately
$\mathord{\sim}9\,\text{h}$ for the reverse shock to reach the inner boundary of our 
computational domain in model \onemg (Figure~\ref{fig:radii all times}). 

\subsubsection{3D model z9.6}
In model \znine the forward shock has already crossed the CO/He interface at $t_{\mathrm{map}}$
(see Table~\ref{tab:progenitors}) and is traveling at roughly $27,000\,\kms$. Behind the 
shock, a dense shell has formed due to the deceleration of the fast neutrino-driven 
wind in a wind-termination shock (Figure~\ref{fig:density profiles all times}). 
In between the shocks, the high density peak atop the bulk of the ejecta
is formed by the deceleration of the forward shock at the CO/He interface.
Outside of the CO/He interface, with similarity to model \onemg, 
a featureless density profile in the He-core and hydrogen envelope leads to an 
untroubled but gradually decelerated expansion of the forward shock,
which reaches the stellar surface at around $\tpb\mathord{\approx}\,1.3\,\mathrm{d})$. 
At the time of shock breakout the forward shock propagates only at $\mathord{\sim}2000\,\kms$ 
because of its strong deceleration in the hydrogen envelope of the star. 
Different from the ECSN progenitor, the wind-termination shock moves inward at 
$\tpb\,\mathord{\approx}\,10$--15\,s (see Figure~\ref{fig:density profiles all times}) 
to get reflected at the center before $\tpb\,\mathord{\approx}\,40\,\s$. A second
reverse shock forms within the He-core of the star and propagates back in radius
after $\sim$400\,s to reach the inner boundary
at $\tpb\,\mathord{\approx}\,800$\,s (Figure~\ref{fig:radii all times}), which is 
more than 8 hours earlier than in model \onemg.

\subsubsection{3D model s9.0}
For the 3D simulation of model \snine the trajectories of the forward and reverse shocks 
are shown in the right panels of Figure~\ref{fig:radii all times}. 
The long-time simulation is initiated at $\tpb\,\mathord{=}\,3.14\,\s$, when the 
forward shock has just crossed the CO/He interface
and is traveling at $v_{\mathrm{sh}}\,\mathord{\approx}\,11,000\,\kms$, only a fraction of the shock
velocity found in the ECSN-like models. Acceleration and deceleration
of the SN shock at the CO/He interface cause the formation of a dense shell in the 
post-shock region (see Figure~\ref{fig:density profiles all times}).
The density contrast between the shell and the ejecta is, however, around one 
order of magnitude smaller than found for the dense shells that formed at the core/envelope
boundary in model \onemg, and at the CO/He interface in model \znine. 
In the following, the shock slows down to 
$v_{\mathrm{sh}}\,\mathord{\approx}\,7,500\,\kms$  within the He-core of the 
star, before it accelerates again around the He/H composition
interface, reaching $v_{\mathrm{sh}}\,\mathord{\approx}11,500\,\kms$. This is due to 
the steep density gradient just below the He/H interface in the progenitor. 
Thereafter, the forward shock encounters the increasing $\rho r^3$ in the hydrogen 
envelope and is thereby strongly decelerated, causing a compression of the post-shock 
material into a double-peaked dense shell. At the bottom of the hydrogen shell,
a strong reverse shock forms shortly afterwards (see Figures~\ref{fig:radii all times} 
and \ref{fig:density profiles all times}). 
Note that the formation of this reverse shock from the He/H interface
occurs much later than witnessed in the ECSN-like progenitors. 
Eventually the reverse shock reaches the inner boundary of our numerical grid at 
$\tpb\,\mathord{\approx}\,16\,\mathrm{h}$. A first reverse shock from the 
CO/He interface had formed around $\tpb\,\mathord{\approx}\,30\,\s$ 
(Figure~\ref{fig:density profiles all times}), but was swept outward with the 
expanding ejecta. The main, spherically shaped SN shock encounters 
the stellar surface at $\tpb\,\mathord{\approx}\,2.8\,\mathrm{d}$, whereas the 
maximum radius of the shock, pushed by a giant, nickel-rich bubble, reaches the 
surface already at $\tpb\,\mathord{\approx}\,2.1\,\mathrm{d}$ 
(see discussion in Section~\ref{sec:Morphology of the ejecta} and 
Figures~\ref{fig:s9 nix cuts} and \ref{fig:s9 rho cuts}).

\begin{figure*}
 \centering
 \includegraphics[width=\textwidth,trim=0.2cm 0cm 0cm 0cm,clip]{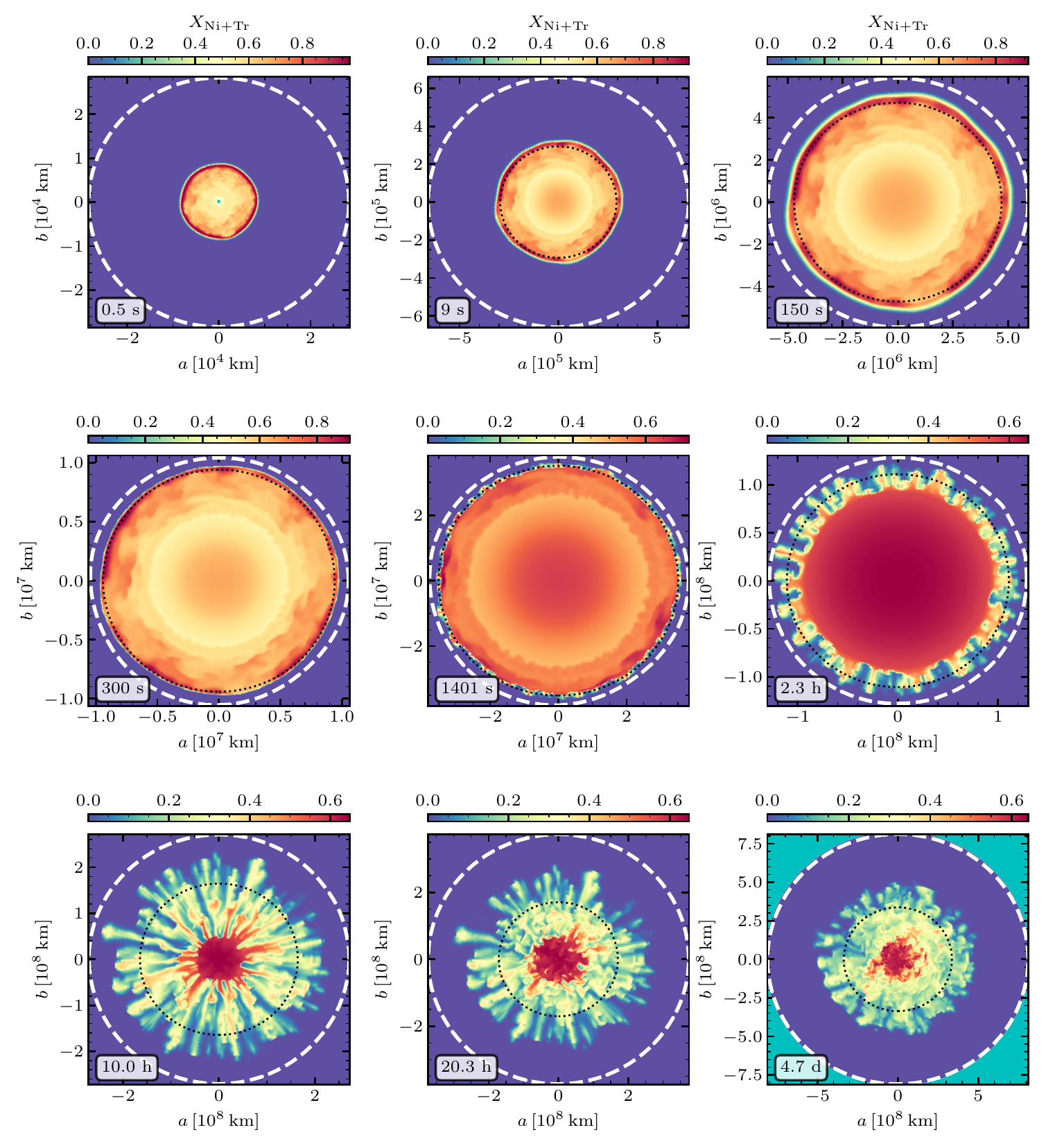} 
 \caption{Slices
    showing the $\nickel\mathord{+}\tracer$ mass fraction in the 3D simulation of model \onemg. 
    The dashed white line indicates 
    the position of the SN shock and the dotted black line
    marks the radial location where the enclosed mass is equal to the
    mass coordinate of the He/H shell interface of the progenitor. Cyan colored 
    regions in the bottom right panel represent the surrounding medium embedding the progenitor. 
    Until $\mathord{\approx}150\,\s$ the neutrino-heated 
    ejecta expand essentially self-similarly. This untroubled expansion ends at about
    $300\,\s$, when the material is decelerated in a dense postshock shell 
    (see Figure~\ref{fig:density profiles all times}). At $\mathord{\approx}1400\,\s$
    the growing RT instability at the He/H interface begins to affect the outer 
    layers of the neutrino-heated ejecta. From roughly $2\,\text{h}$ on the plumes 
    grow in size, while the reverse shock (visible at the base of the plumes)
    begins to propagate back in radius (see Figure~\ref{fig:radii all times}). 
    About $9\,\text{h}$ after bounce the reverse shock reaches the center, gets reflected
    there, and on the way compresses the metal-rich
    ejecta in the central region. Note that the plumes are almost evenly distributed in 
    angular direction and radial extent. }
 \label{fig:e8 nix cuts}
\end{figure*}

\begin{figure*}
\includegraphics[width=\textwidth]{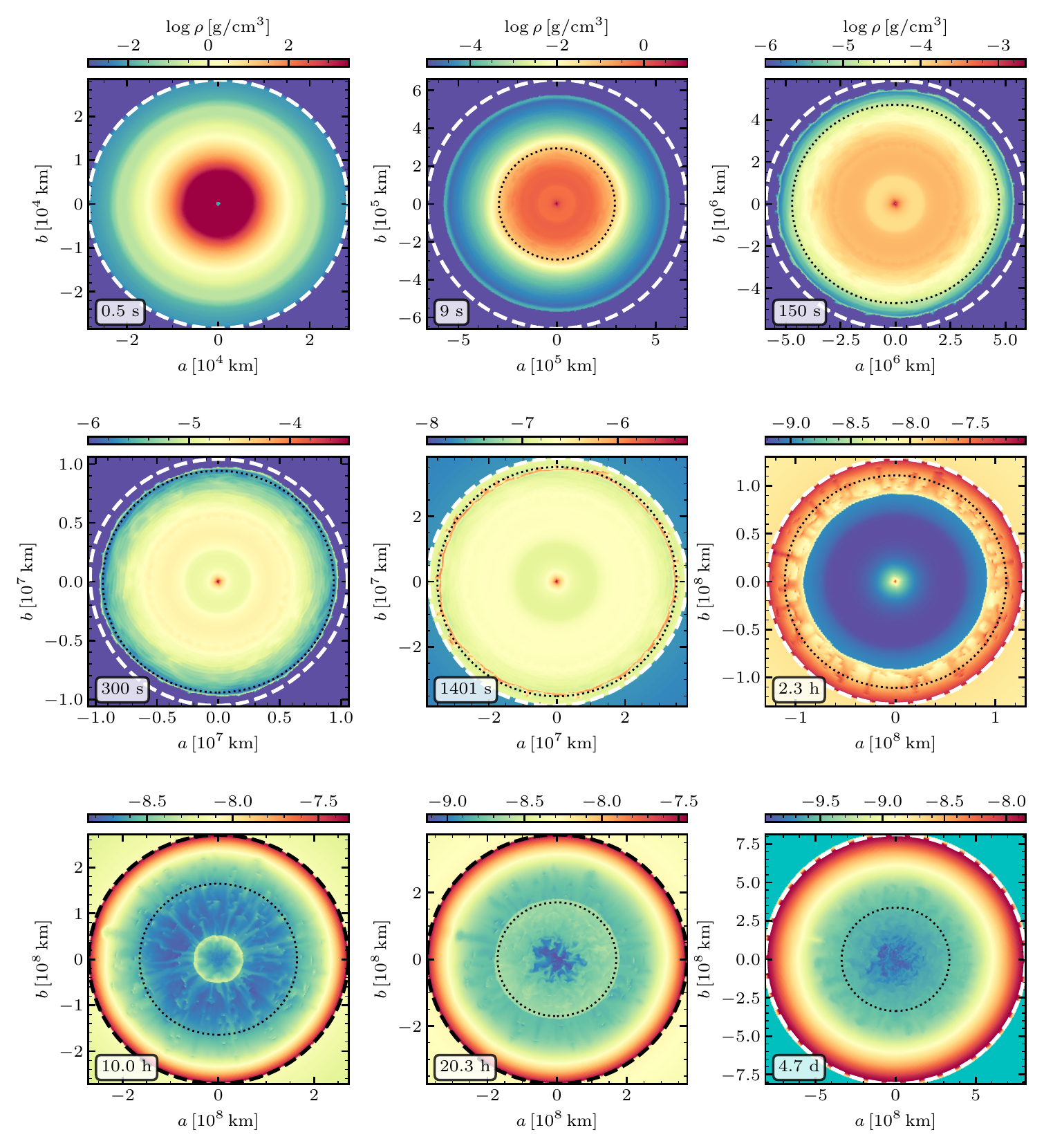}
    \caption{Slices showing the density distribution
    in the 3D simulation of model \onemg. The dashed white or black line indicates 
    the position of the SN shock and the dotted white or black line
    marks the radial location where the enclosed mass is equal to the
    mass coordinate of the He/H shell interface of the progenitor. Cyan colored 
    regions in the lower right panel represent the surrounding medium embedding the progenitor.  
    The dense shell at which the reverse shock will form (see also Figure~\ref{fig:e8 nix cuts}) 
    is first visible as the light yellow ring with a radius of about 
    $\mathord{\approx} 3\mathord{\times}10^5\,\mathrm{km}$ at $\tpb\,\mathord{=}\,9\,\s$.
    The panel at $\tpb\,\mathord{=}\,2.3\,\mathrm{h}$ shows the formation of small RT fingers.
    These fingers grow with time and partly take up the neutrino-heated material (see panels at
    $\tpb\,\mathord{\approx}\,2.3\,\mathrm{h}$ and $\tpb\,\mathord{\approx}\,10\,\mathrm{h}$).
    At $\tpb\,\mathord{\approx}\,4.7\,\mathrm{d}$ the innermost ejecta are characterized by an overall
    spherical shape, superimposed with the relics of the RT fingers.
    The reverse shock is very prominent at $\tpb\,\mathord{\approx}\,2.3\,\mathrm{h}$ 
    and travels inward from about 3\,h on (see Figure~\ref{fig:radii all times}). It reaches 
    the center at $\tpb\,\mathord{\approx}\,9\,\mathrm{h}$ 
    to be reflected outward again (see lower left panel).}
\label{fig:e8 rho cuts}
\end{figure*}

\begin{figure*}
\includegraphics[width=\textwidth]{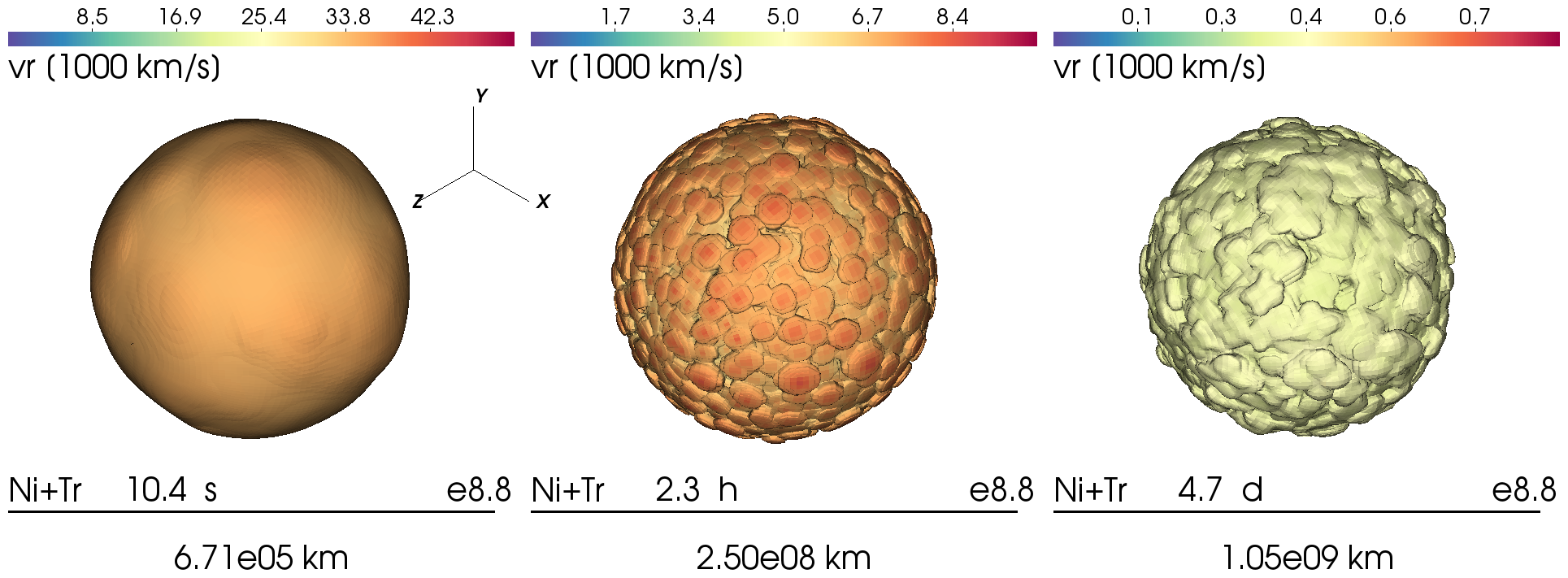}
    \caption{3D renderings of the $X_{\nickel\mathord{+}\tracer}\,\mathord{=}\,0.03$ 
             iso-surface of model \onemg at the indicated times. The black
             lines indicate the radial scales of the plots. The color-coding represents the radial velocity 
             of the material and the tripod in the left panel indicates the orientation
             of the global coordinate system. From the initially quasi-spherical distribution 
             of the neutrino-heated ejecta with only low-amplitude perturbations we observe the 
             growth of small-scale RT plumes. Because of the nearly spherical beginning of the 
             explosion and the growth of RT instabilities only on small angular scales, the final 
             distribution of the neutrino-heated ejecta is basically isotropic.}
\label{fig:e8 3d rendering}
\end{figure*}

\subsection{Morphology of neutrino-heated ejecta}
\label{sec:Morphology of the ejecta}

In the following we focus on the long-time development of the early-time asymmetries which we trace by 
the propagation of the neutrino-heated ejecta or more specifically the $\nickel\mathord{+}\tracer$-rich 
material. \cite{Kifonidis2003} already noted that \nickel is produced between the high-entropy bubbles
that expand due to strong neutrino-heating from below. Thus the distribution of \nickel traces the 
asymmetries that developed during the onset of the explosion. 

\subsubsection{3D model e8.8}
In Figure~\ref{fig:e8 nix cuts} we show slices of the $\nickel\mathord{+}\tracer$ mass fraction of the 
3D simulation of model \onemg. The dashed white line indicates the shock radius, whereas the thin 
dotted black line indicates the position where the enclosed mass
equals the mass interior to the He/H interface in the progenitor.

\begin{figure*}
 \centering
 \includegraphics[width=\textwidth,trim=0cm 0.25cm 0.2cm 0cm,clip]{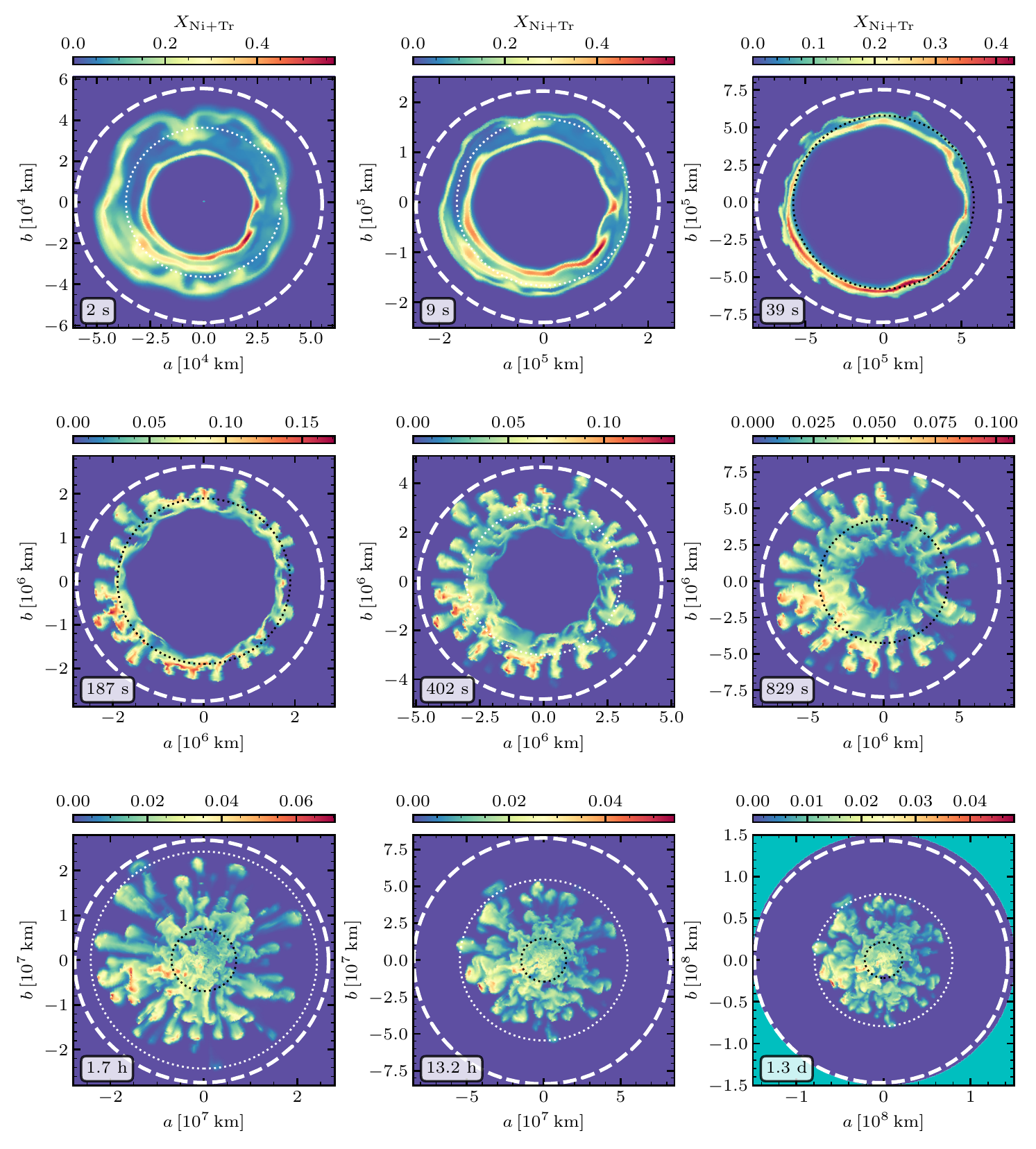}
 \caption{Slices of the $\nickel\mathord{+}\tracer$ mass fraction of the 3D simulation of model
   \znine at the indicated times. Cyan colored regions in the bottom right panel represent the 
   surrounding medium of the progenitor. The dashed white line marks the shock position, the dotted white or black lines indicate the radial locations where 
   the enclosed mass equals the 
   mass interior to the CO/He and He/H composition interfaces of the progenitor.
   The initial asymmetries that develope during the onset of the explosion are still visible at $\tpb\,\mathord{=}\,1.5\,\s$, but they are soon compressed to flat structures as they 
   collide with the RT unstable dense shell behind the CO/He interface
   (see times from $\tpb\,\mathord{=}\,9\,\s$ to $\tpb\,\mathord{=}\,39\,s$ and
   also Figure~\ref{fig:z9 rho cuts}). 
   Over the next, roughly, one hour, the growing RT instability
   mixes the outer \nickel-rich layers outwards in mass coordinate in numerous 
   small fingers. $1.7\,\mathrm{h}$ after bounce the RT instability has basically saturated 
   and the first iron-rich clumps reach the He/H interface of the
   progenitor, where in model \znine no secondary RT instability occurs. The final 
   morphology of the \nickel-rich ejecta has mostly lost any
   resemblance with the state at $t_{\mathrm{map}}$ and is dominated by
   small-scale asymmetries. However, the overall 
   distribution of \nickel and \tracer remains roughly spherical with
   many small-scale features and only a slight global deformation, which
   exhibits larger and stronger iron-rich plumes between the 8 o'clock and
   10 o'clock directions and weaker structures in the opposite directions. A
   corresponding hemispheric asymmetry is already visible at the beginning of the explosion, see top left panel for $\tpb\,\mathord{=}\,2\,\s$.
 }
 \label{fig:z9 nix cuts}
\end{figure*}

\begin{figure*}
\includegraphics[width=\textwidth]{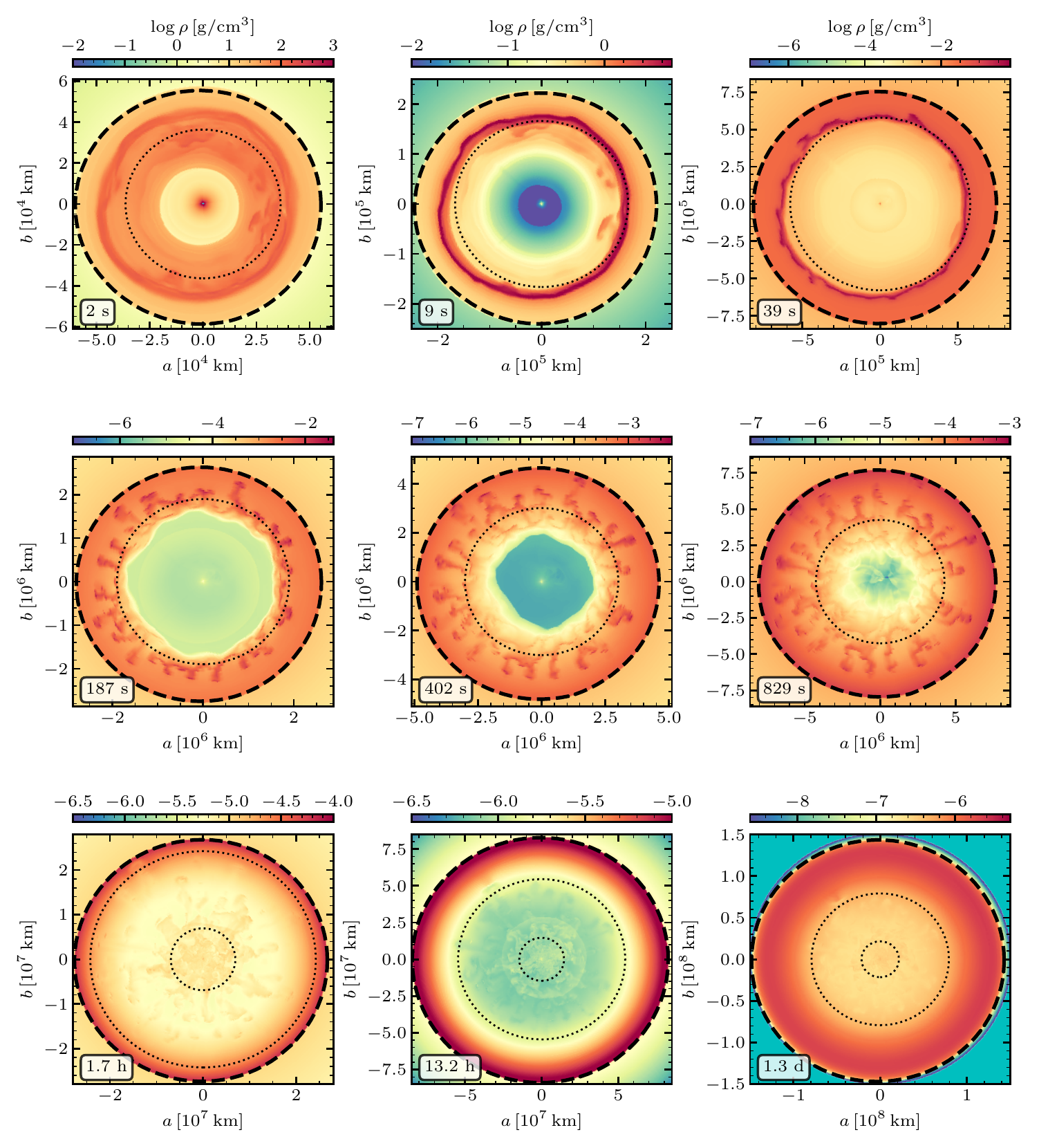}
    \caption{Slices showing the density in the 3D simulation of model \znine 
        at the indicated times. 
    Cyan colored regions in the bottom right panel represent the surrounding medium
    of the progenitor. The dashed black line marks the shock position, the dotted
    black lines indicate the radial locations where the enclosed mass equals the 
    mass interior to the CO/He and He/H composition interfaces of the progenitor.
    A termination shock of the neutrino-driven wind is
    visible as the yellow/orange discontinuity at $\tpb\,\mathord{\approx}\,2\,\s$.
    The shock deceleration in the He-layer leads to the formation of a dense shell that 
    gets extremely compressed by the outward shock from the reflection of the 
    wind termination shock at the center (see Figure~\ref{fig:density profiles all times}).
    Over the next $\mathord{\approx}100\,\s$, RT plumes
    start to grow within the unstable layer between the two shocks, thereby fragmenting
    the dense shell. A second reverse shock forms when the SN shock 
    propagates through the He-layer of the progenitor.
    As this reverse shock travels back in radius, similar to the 
    results presented for model \onemg, the plumes
    grow to their maximal radial extent 
    (see panels at $\tpb\,\mathord{=}\,187\,\s$--$829\,\s$).
    The final morphology of the ejecta at $\tpb\,\mathord{\approx}\,1.3\,\mathrm{d}$ resembles 
    the late-time morphology in model \onemg.}
\label{fig:z9 rho cuts}
\end{figure*}

\begin{figure*}
\includegraphics[width=\textwidth]{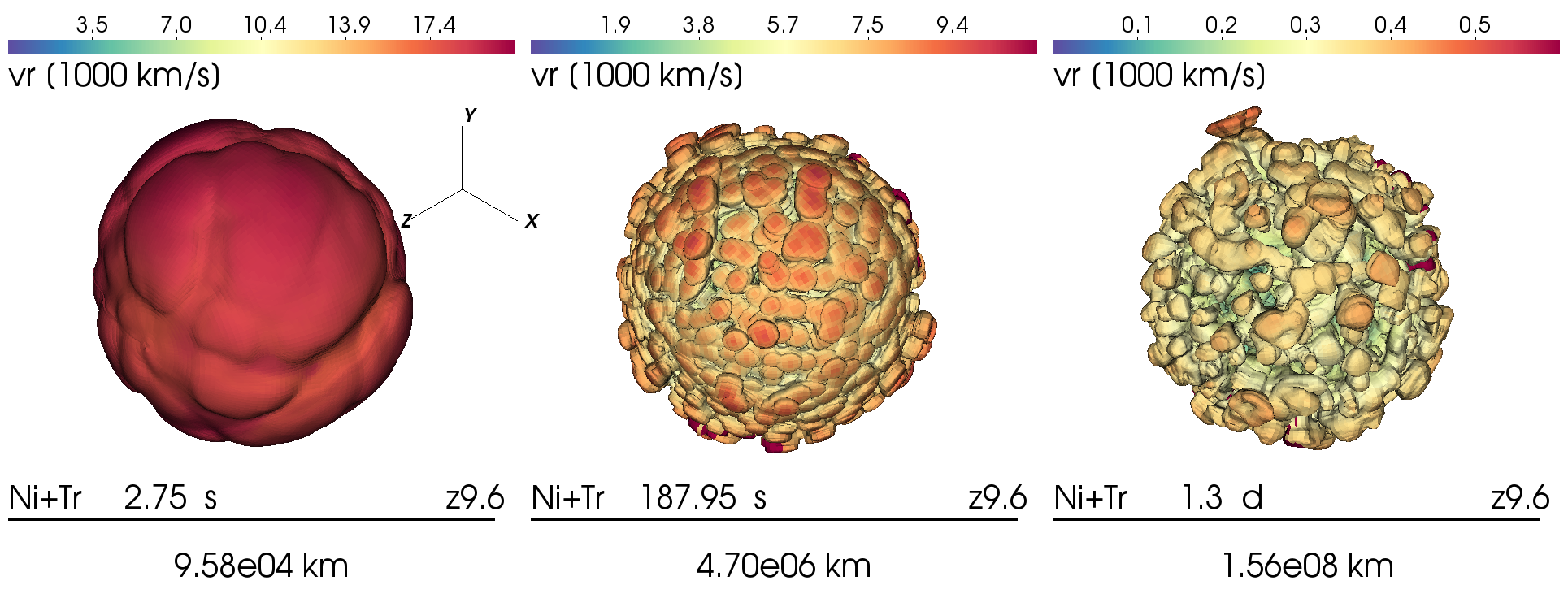}
    \caption{3D renderings of the $X_{\nickel\mathord{+}\tracer}\,\mathord{=}\,0.03$ 
             iso-surface of model \znine at the indicated times. The color-coding represents the radial 
             velocity of the material, and the tripod in the left panel indicates the orientation
             of the global coordinate system. We find the growth of small-scale (high spherical 
             harmonics mode numbers $\ell$) RT plumes on top of the initial asymmetries of the explosion. 
             The small amplitudes of the initial asymmetries and the narrow RT 
             unstable layer prevent the growth of large global asymmetries, 
             leaving the final state basically spherical. One initially
             bigger neutrino-heated bubble leads to a more extended
             plume near the 11 o'clock position in the right panel.}
\label{fig:z9 3d rendering}
\end{figure*}

As in the 1D simulation, the CO and He layers are compressed into a dense shell
just behind the SN shock at a radius of $\mathord{\approx}5\mathord{\times}10^5\,\rm km$
at $\tpb\,\mathord{=}\,9\,\s$ 
(see Figure~\ref{fig:density profiles all times} and Figure~\ref{fig:e8 rho cuts}).
Already at $\tpb\,\mathord{\approx}\,1.5\,\s$, a reverse shock begins to form at the bottom 
of this dense shell (see the sharp inner edge of the light-blue, narrow ring at
$5\mathord{\times}10^5$\,km in Figure~\ref{fig:e8 rho cuts} at $9\,\s$) .
By this time the bulk of the neutrino-heated ejecta has expanded in the 
volume 
inside $\mathord{\approx}3\mathord{\times}10^5\,\rm km$.
While this innermost metal-rich material seems to expand in a basically self-similar fashion
(see similarity of the snapshots until $\tpb\,\mathord{=}\,150\,\s$ in Figure~\ref{fig:e8 nix cuts}),
the growth of the RT instability in small protrusions at the unstable contact interface 
near the outer edge of the dense shell begins to corrugate the surface of 
the ($^{56}$Ni+Tr)-rich ejecta. At $\tpb\,\mathord{=}\,300\,\s$,
the neutrino-heated ejecta of \nickel+\tracer catch up with the expanding dense 
shell and are strongly decelerated. RT plumes become clearly 
visible near the He/H interface first at about $1400\,\s$ in Figure~\ref{fig:e8 nix cuts}.

As a consequence, the higher entropy/lower density bubbles are compressed to flat
structures, whereas the higher-density regions in between the bubbles 
are able 
to penetrate the dense shell, thereby inducing perturbations at the He/H interface. 
These RT plumes grow over the following hours
around the mass shell of the He/H interface, mixing clumps of 
\nickel from the central volume outward into the carbon and helium rich layers.

The reverse shock, which is visible at $\tpb\,\mathord{=}\,2.3\,\text{h}$ in 
Figure~\ref{fig:e8 rho cuts} as the yellow to blue discontinuity, 
begins to propagate back in radius, thereby 
compressing the innermost ejecta.
At $\tpb\,\mathrm{\approx }\,10\,\text{h}$ the inward passage of the reverse shock and 
the growth of the RT instability has erased the initial structures 
present at the onset of the explosion. 
However, the overall morphology of the neutrino-heated ejecta is still basically 
spherical with only small-scale asymmetries (see also Figure~\ref{fig:e8 3d rendering}).

\subsubsection{3D model z9.6}
In Figure~\ref{fig:z9 nix cuts}, we show slices of the $\nickel\mathord{+}\tracer$  
mass fraction in the 3D simulation of model \znine. 
The dashed white line marks the shock radius, while the dotted lines 
indicate the positions where the enclosed mass equals the 
mass interior to the CO/He and He/H interfaces of the progenitor.
The dense shell that forms after the forward shock has crossed 
the CO/He interface (see Figure~\ref{fig:density profiles all times}) 
is visible as the circular dark-red region at 
$\sim$1.5$\mathord{\times}10^5$\,km after 9\,s
in Figure \ref{fig:z9 rho cuts}. At that time the termination shock of
the neutrino-driven wind can be seen at about $5\mathord{\times}10^4$\,km,
moving inward.

Within the first $\mathord{\approx}9\,\s$ the fastest of the 
neutrino-heated ejecta encounter this dense CO-rich shell (see Figure~\ref{fig:z9 rho cuts})
and are compressed and squeezed to flat structures around $30\,\s$ later.
Similar to the results presented for the \onemg model, the slightly over-dense 
regions between the high-entropy plumes induce long-wavelength perturbations 
as they deform and try to penetrate the RT unstable dense shell. 
Over the next few minutes, RT fingers grow on top of these deformations and fragment the 
dense shell into numerous \nickel-rich shrapnels.

While the fingers grow progressively, the second reverse shock from the
shock propagation through the He layer (visible as the green to yellow discontinuity 
at $\tpb\,\mathord{=}\,187\,\s$ in Figure~\ref{fig:z9 rho cuts}) begins to 
propagate back in radius. At $\tpb\,\mathord{=}\,829\,\s$ the reverse shock 
has almost reached the center of our numerical grid, having 
compressed and decelerated the innermost ejecta material. 
At $\tpb\mathord{=}1.7\,\rm h$, the forward shock has crossed the He/H interface and the inner 
material has been fully shredded by the instability.

As the shock velocity does not change significantly at the He/H interface, we observe 
no additional growth of the RT instability nor the formation of another reverse shock.
Thus, the morphology of the innermost ejecta seems to be determined early on, already 
before the shock crosses the He/H interface.

Comparing the distribution of \nickel and \tracer of model \onemg 
(Figure~\ref{fig:e8 nix cuts}; $t_\mathrm{pb}\,\mathord{=}\,4.7$\,d) and 
model \znine (Figure~\ref{fig:z9 nix cuts}; $t_\mathrm{pb}\,\mathord{=}\,1.3$\,d) 
shortly before shock breakout, we find a slightly more clumped morphology in the 
$9.6\,\solm$ progenitor. The structure of the neutrino-heated ejecta in 
model \znine also remains fairly spherical with many small-scale clumps 
(see Figures~\ref{fig:z9 rho cuts} and \ref{fig:z9 3d rendering}). However, different from
model \onemg one can recognize a hemispheric asymmetry with bigger plumes between the 
8 o'clock and 10 o'clock positions in Figure~\ref{fig:z9 nix cuts} and weaker plumes in
the opposite hemisphere. These aspherical structures go back to asymmetries that existed
already in the first seconds of the explosion, visible by larger RT mushrooms and a 
slightly stronger shock expansion in the left hemisphere at 
$t_\mathrm{pb}\,\mathord{=}\,2$\,s.
The strongest plume sticks out near the 11 o'clock position 
in Figure~\ref{fig:z9 3d rendering}.

\subsubsection{3D model s9.0}
\begin{figure*}
\centering
\includegraphics[width=\textwidth]{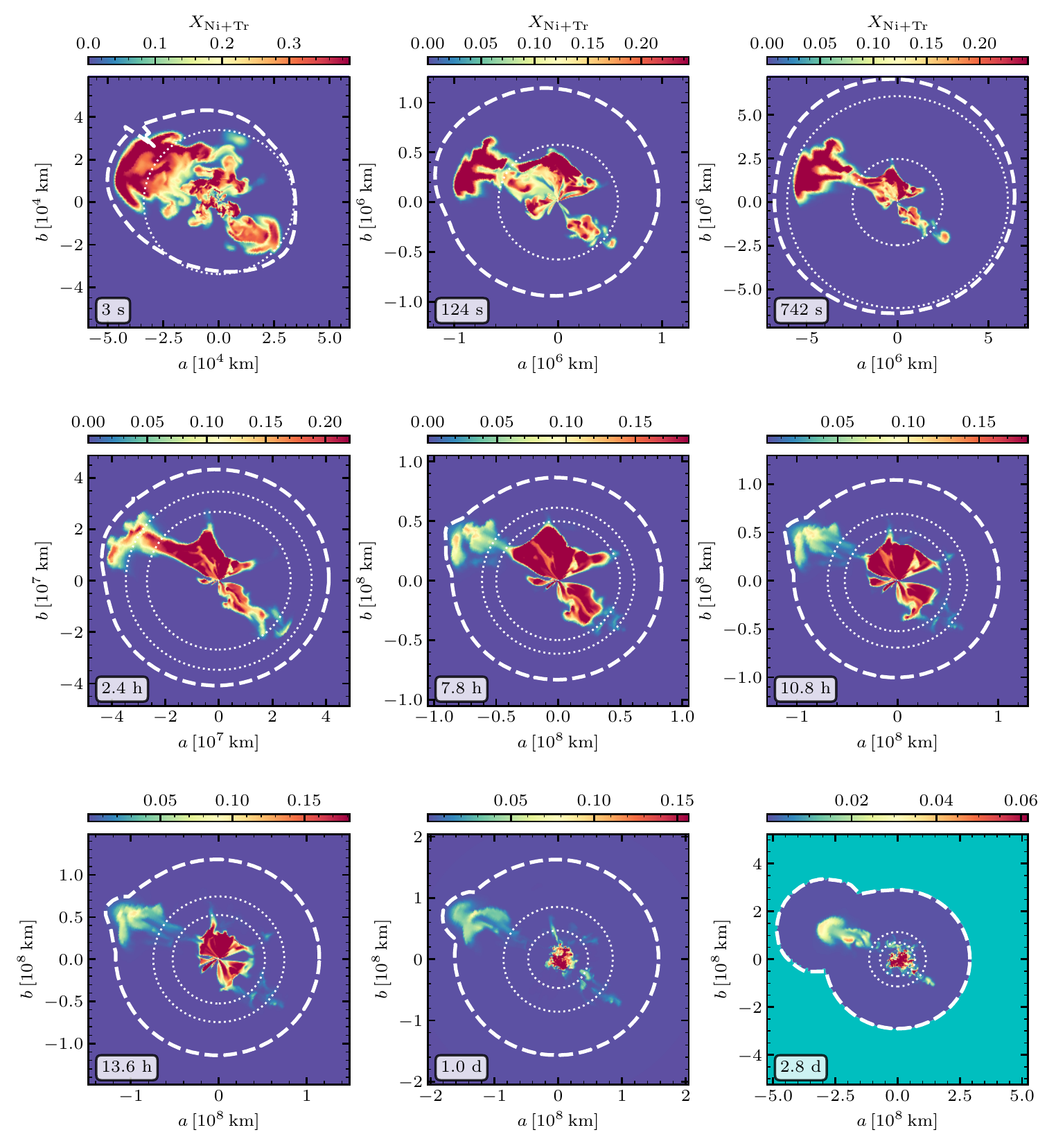}
    \caption{Slices of the $\nickel\mathord{+}\tracer$ mass fraction of model \snine at 
    the indicated times. The dashed white line marks the shock surface, and the dotted white 
    lines denote the positions where the enclosed mass 
    equals the mass interior to the CO/He and He/H composition 
    interfaces of the progenitor. The cyan colored region in the bottom 
    right panel represents the surrounding medium of the progenitor. 
    The $\nickel\mathord{+}\tracer$ distribution underlines the connection between the initial
    asymmetries, and the largest clump that overtakes the average shock even before shock breakout. 
    At $\tpb\,\mathord{=}\,3\,\s$ around 25\% of the \nickel-rich material are ejected in the 10 o'clock 
    direction. (The kink of the white dashed line for the shock surface
    at this time is a numerical artifact due to the shock-detection
    algorithm.) While the innermost ejecta are compressed and decelerated
    strongly by interaction with the dense shells 
    and reverse shocks that form at the CO/He and He/H interfaces, about half of the large clump experiences
    less dramatic deceleration and deforms the otherwise spherical shock wave well before shock breakout.
    }
\label{fig:s9 nix cuts}
\end{figure*}

\begin{figure*}
\includegraphics[width=\textwidth]{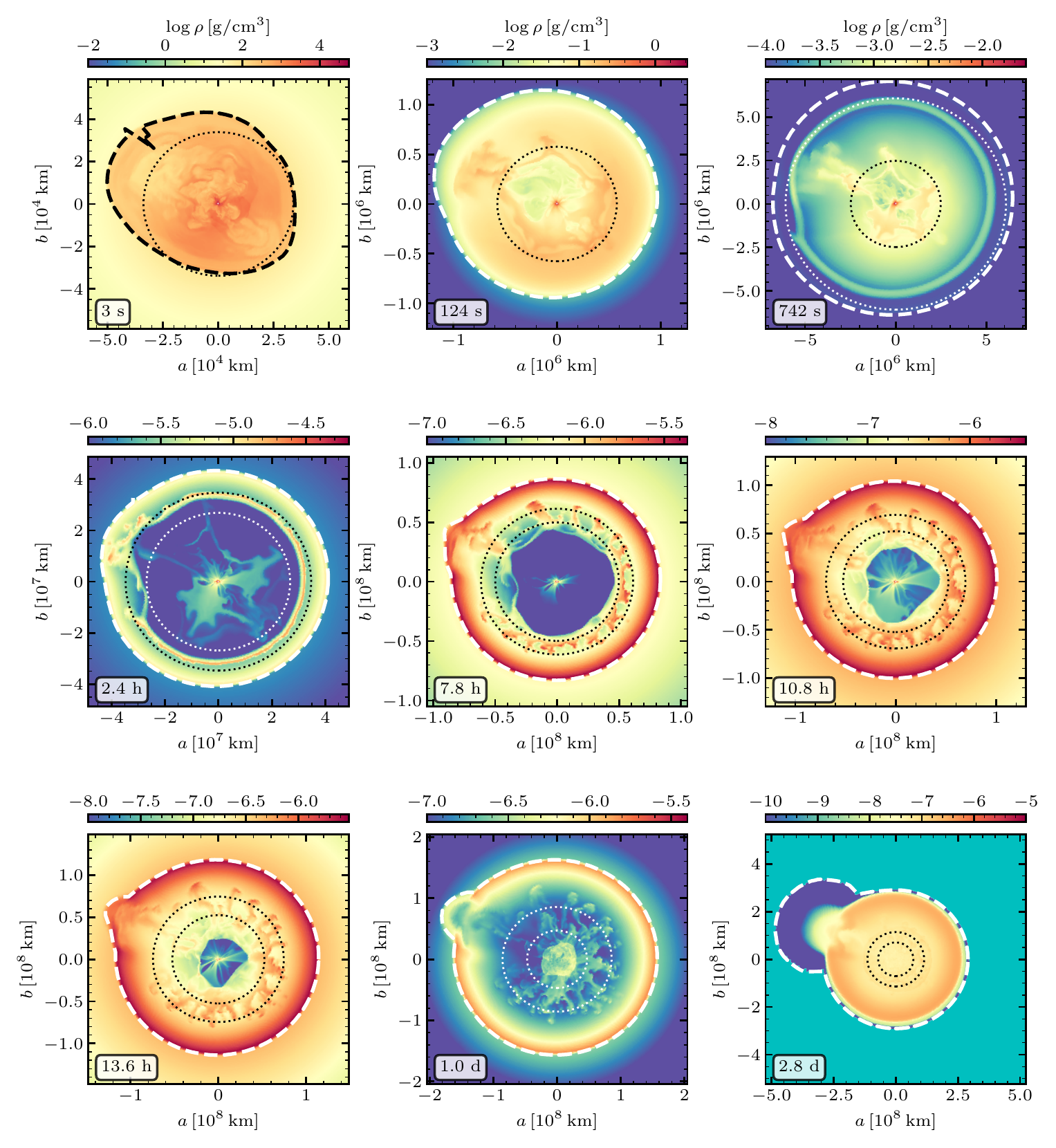}
    \caption{Slices of the density of model \snine at the indicated times. 
        The dashed white or black line indicates the shock surface, whereas the 
        dotted white or black lines mark the locations where the enclosed
        mass equals the mass interior to the CO/He and He/H composition interfaces of
        the progenitor. The cyan colored region in the bottom right panel represents
        the surrounding medium of the progenitor.
        Different from the ECSN-like progenitors, we observe here
        the growth of large-scale RT plumes caused by the great asymmetries at the onset of the explosion.
        The seed for the biggest later structure is set already at around $3\,s$ 
        after bounce when the shock passes the CO/He interface. (The kink 
        of the black dashed line for the shock surface at this time is a
        numerical artifact due to the shock-detection algorithm.) 
        We first observe the growth of the RT instability in the 
        direction perturbed by the largest and 
        fastest initial convective plume (see, e.g., $\tpb\,\mathord{=}\,124\,\s$, at the 10 o'clock position), 
        and there is less deceleration of this dense clump of $\nickel\mathord{+}\tracer$ when 
        it penetrates the CO/He interface. It therefore begins to move far ahead of the slower
        $\nickel\mathord{+}\tracer$. Shortly afterwards the plume arrives at the unstable He/H interface, 
        inducing a large-scale perturbation there.
        While the slower clumps of $\nickel\mathord{+}\tracer$ are further decelerated by the reverse shock that forms 
        at the He/H interface, the biggest clump begins to push the shock, thereby transporting a 
        significant amount of neutrino-heated ejecta to velocities
        larger than the average shock velocity (see also Figure~\ref{fig:radii all times}).
    }
\label{fig:s9 rho cuts}
\end{figure*}

\begin{figure*}
\includegraphics[width=\textwidth]{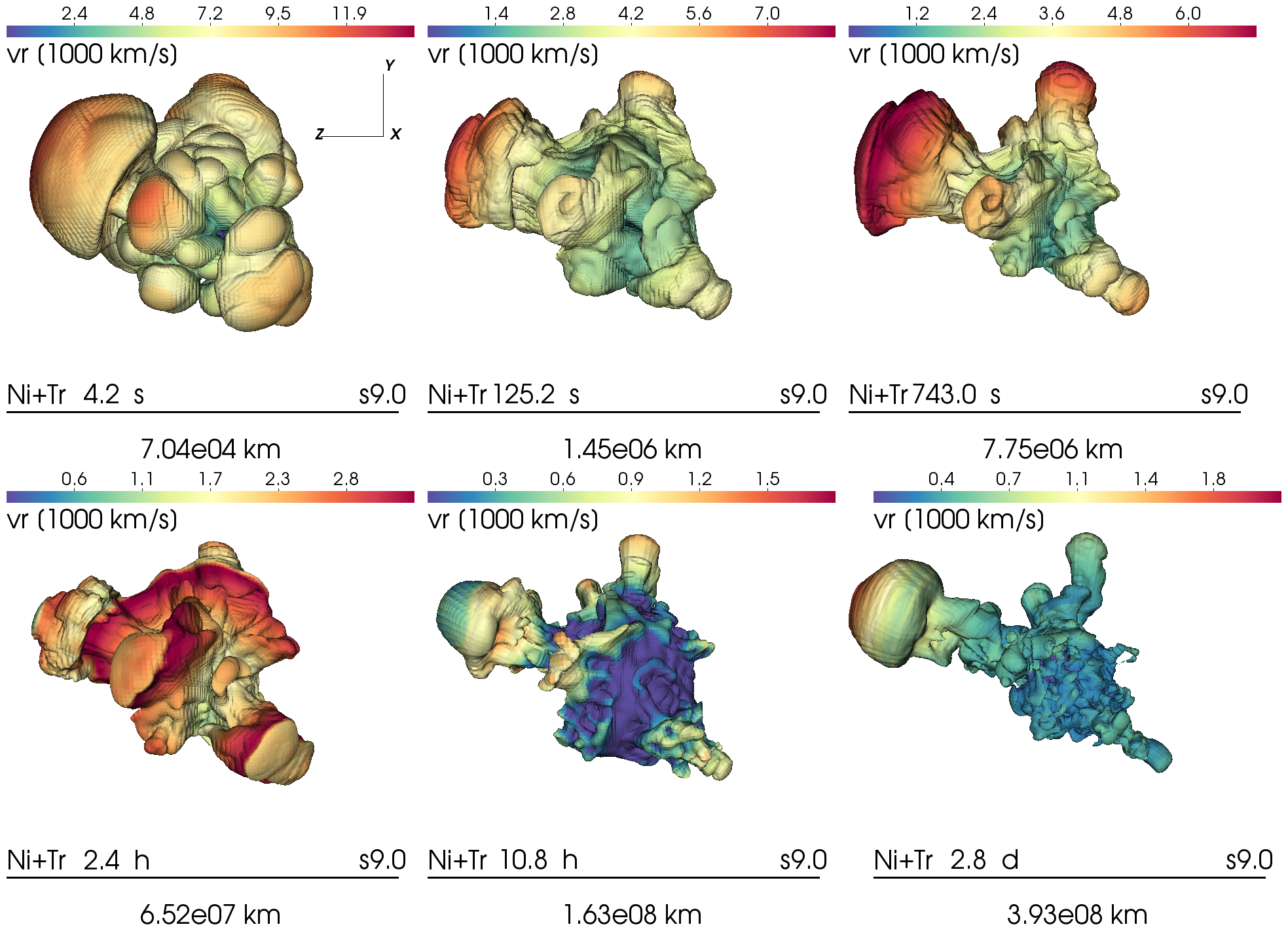}
    \caption{3D renderings of the $X_{\nickel\mathord{+}\tracer}\,\mathord{=}\,0.03$ 
             iso-surface of model \snine at the indicated times.  The color-coding represents the radial 
             velocity of the material, and the tripod in the left panel defines the orientation
             of the global coordinate system. During the first roughly $800\,\s$, the largest initial 
             asymmetries grow to extended metal-rich plumes. Slower iron-group matter in the
             interior is decelerated by the collision with the dense shell formed around the CO/He interface. 
             In the following, these fastest clumps encounter the dense shell behind the He/H and induce 
             large-amplitude perturbations in the RT unstable layer. Consequently, large metal rich RT fingers 
             begin to grow from the interface. At $\tpb\,\mathord{=}\,2.8\,\rm d$ we find one very big and two 
             smaller metal-rich plumes, of which the largest one penetrates the surface of the star even ahead
             of the average shock radius (see also Figures~\ref{fig:s9 nix cuts} and \ref{fig:s9 rho cuts}).}
\label{fig:s9 3d rendering}
\end{figure*}

The evolution of the neutrino-heated ejecta in model \snine proceeds
drastically differently from the ECSN-like progenitors 
(Figures~\ref{fig:s9 nix cuts}--\ref{fig:s9 3d rendering}). 
As discussed in Section~\ref{sec:Evolution during the first second}, 
the initial asymmetries and shock deformation seen in model \snine are 
considerably larger than found in the ECSN-like models. Strong convection leads 
to the formation of a large high-entropy plume, which is rich in iron-group material
and expands about two times 
faster than the surrounding material at the time of shock revival 
($\tpb\,\mathord{\approx}\,0.5\,\s$).  
It crosses the CO/He interface of the star at $\tpb\,\mathord{\approx}\,1.3\,\s$, 
shortly after the forward shock. In comparison, the slowest 
moving material reaches the interface around $ 0.65\,\s$ later. 

The crossing of the CO/He interface by the shock wave has several dynamical consequences. 
Due to the increasing $\rho r^3$ outside of the interface, the shock is 
decelerated and the postshock matter is swept up and compressed into a dense shell.
Note that the neutrino wind in this model is very weak and, different from model \znine,
there is no low-density central region and no wind termination shock. Instead, 
the central volume around the NS contains relics of (\nickel+Tr)-rich low-density
plumes and \nickel-poor, higher density downflows during the entire evolution.

The high-density shell behind the shock is aspherical in contrast to the 
shells found in the ECSN-like progenitors, which is also reflected by the still  
deformed SN shock (see first two panels in Figure~\ref{fig:s9 rho cuts} 
for  $\tpb\,\mathord{=}\,3$ and $124\,\s$).
Around the dense shell we observe the growth of large plumes, which stem from the initial 
asymmetries of the explosion. These can be seen in the 10 o'clock direction in
Figure~\ref{fig:s9 nix cuts} at $\tpb\,\mathord{=}\,124\,\s$ and in 
Figure~\ref{fig:s9 3d rendering}.
At the tops of these plumes small RT fingers grow, in line with the analysis of the amplification factors presented in Section~\ref{sec:Linear stability analysis}. 
Note that at this point in time the still deformed SN shock crosses the He/H interface 
(see Table~\ref{tab:progenitors} and Figure~\ref{fig:radii all times}).

Due to the varying $\rho r^3$ profile around this composition interface, the 
shock accelerates and decelerates, thereby forming a dense shell (see the yellow-green 
ring in Figure~\ref{fig:s9 rho cuts} at $\tpb\,\mathord{=}\,742\,\s$ and the density
spike in Figure~\ref{fig:density profiles all times}).
When the fast and dense metal-rich plume encounters the shell, it induces a high-amplitude perturbation in this RT unstable layer. 
From this perturbation, aided by the large initial momentum of the metal-rich plume, 
we observe the growth of a large RT structure, which assists the further
outward expansion of the $^{56}$Ni+Tr material of the initial, big plume. 
As the shock is decelerated in the H-envelope, 
the dense metal-rich plume retains higher velocities than the speed of the shock and thus 
the plume is able to deform the forward shock on its way 
(see Figure~\ref{fig:s9 rho cuts} at $\tpb\,\mathord{\geq}\,2.4\,\rm h$).

Concurrently, the reverse shock, which forms at the bottom of the He/H interface, 
propagates back into the ejecta and strongly decelerates and compresses the 
($\nickel\mathord{+}\tracer$)-rich material close to the center 
(see Figures~\ref{fig:s9 nix cuts} and \ref{fig:s9 rho cuts} at 
$\tpb\,\mathord{\approx}\,7.8\rm h$), whereas the fastest, biggest plume escapes the most dramatic deceleration, although its velocity also shrinks
with time (see Figure~\ref{fig:s9 3d rendering}). 
The growing RT instability around the CO/He 
interface (clearly visible in Figure~\ref{fig:s9 rho cuts} at
$\tpb\,\mathord{\approx}\,7.8\rm h$)
seems to only slightly affect the outer boundary of the central 
($\nickel\mathord{+}\tracer$)-rich material, as can be seen in 
Figure~\ref{fig:s9 nix cuts} at $\tpb\,\mathord{\geq}\,7.8\rm h$ (in
line with the small amplification factors found in this region).

Due to the strong deceleration of the innermost material, the fast plume almost fully detaches from the core material as it propagates through the hydrogen envelope of the star (see last two panels in Figures~\ref{fig:s9 nix cuts}, \ref{fig:s9 rho cuts}, and \ref{fig:s9 3d rendering}).
It encounters the surface of the star at around $\tpb\,\mathord{\approx}\,2.1\, \rm d$, 
so more than half a day earlier than the spherically shaped main shock front, which reaches the stellar surface at $\tpb\,\mathord{\approx}\,2.8\, \rm d$. Why is the large plume able to travel
with such high velocities, even deforming the forward shock, while the bulk of the $^{56}$Ni+Tr
mass travels at considerably slower speed?
First, the fastest $\nickel\mathord{+}\tracer$ material is, at all times, in close 
vicinity of the immediate postshock matter (see Figure~\ref{fig:radii all times}). 
Second, after the forward shock has crossed the He/H interface, the Ni-rich plume is decelerated less than the average shock (see Figure~\ref{fig:radii all times}), since this material is denser than its surroundings in the hydrogen layer.
Third, large growth rates at the He/H interface lead to an efficient outward mixing of
the dense plume within the unstable layers, and the plume can therefore also escape the strong deceleration by the reverse shock.
As a consequence, the ($\nickel\mathord{+}\tracer$)-rich plume catches up with the forward shock in the hydrogen envelope. Due to its large momentum it deforms the outgoing forward shock in its trajectory.
This is similar to a transient situation at about half an hour and 3 h in model \onemg, where the neutrino-heated ejecta in RT plumes catch up with the strongly decelerated immediate postshock material, thereby pushing the forward shock (see Figures~\ref{fig:radii all times} and 
\ref{fig:e8 rho cuts}).

\begin{figure*}
 \centering
 \includegraphics[width=\textwidth,trim=0cm 0.0cm 0cm 0cm,clip]{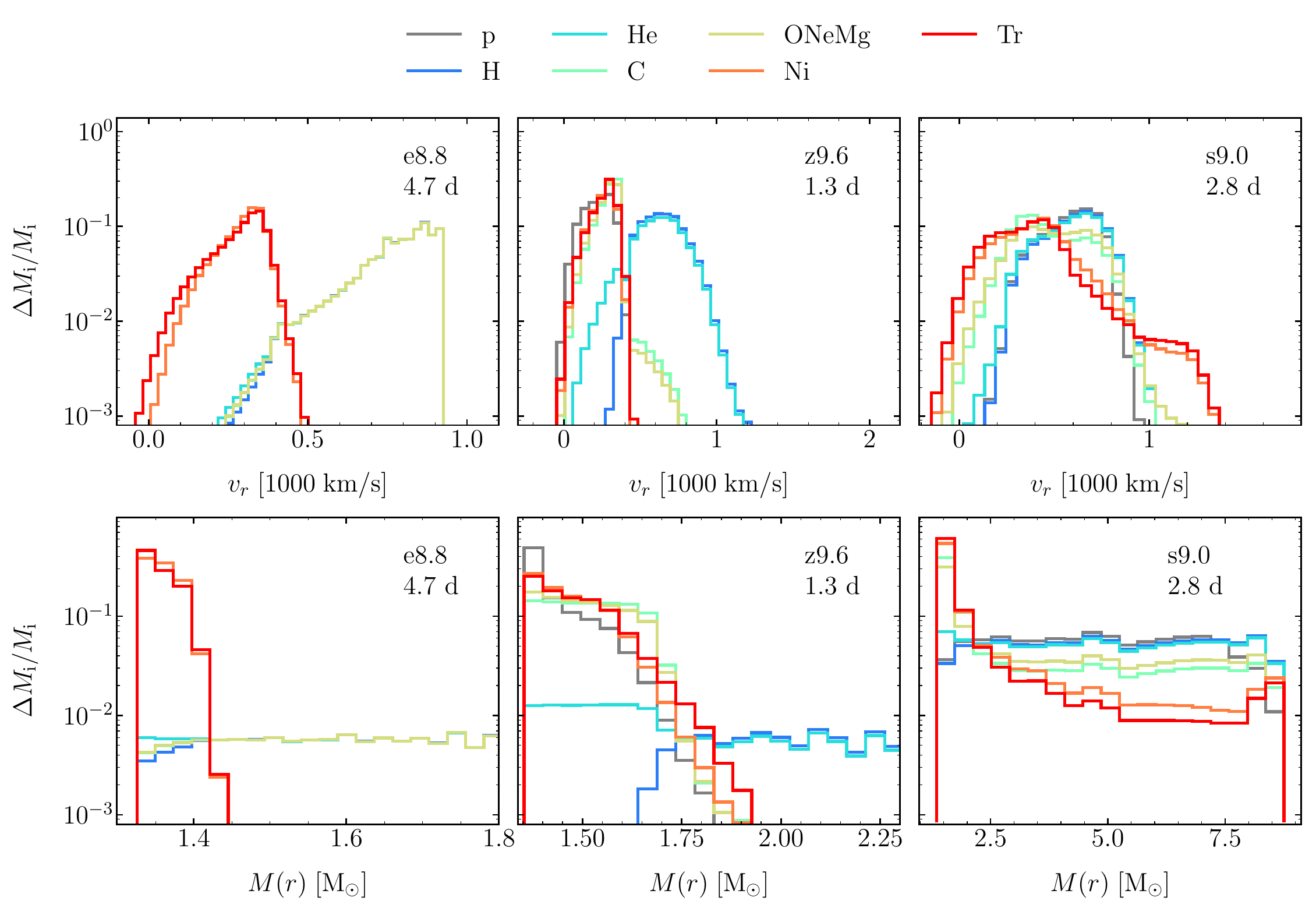}
 \caption{Normalized mass distributions of chemical elements versus radial 
 velocity (top row) and enclosed mass (bottom row) for the ejecta of 
 all of our 3D models at the time of shock breakout. We use 50 bins in 
 velocity space and 30 bins in mass space, starting outside of the compact
 remnant mass and only showing the distributions up to $1.8\,\solm$ and 
 $2.26\,\solm$ for models \onemg and \znine, 
 respectively, since mixing of the neutrino-heated ejecta beyond 
 these values of the enclosed mass is insignificant. For model 
 \snine we consider the whole stellar domain for the visualization.
 The ECSN-like models (\onemg and \znine) mix only a tiny amount of \nickel 
 into the bottom layers of the H-envelope, whereas efficient 
 mixing in model \snine transports a significant amount of 
 neutrino-heated ejecta to large mass coordinates up to the stellar surface.}
 \label{fig:mass distribution sbo}
\end{figure*}

\subsection{Extent of mixing}

In Figure~\ref{fig:mass distribution sbo}, we display the normalized
mass distributions of various nuclear species, including 
free protons from the freeze-out of NSE, hydrogen from the envelope, 
helium, carbon, oxygen-neon-magnesium, radioactive nickel, and the tracer 
nucleus of neutron-rich species for all of our models at the time of shock 
breakout as functions of radial velocity and mass coordinate (the latter
is defined as enclosed mass $M(r)$ at radius $r$). 

The distributions of the iron-group ejecta in the ECSN-like models (\onemg and \znine) have similar shapes in velocity and mass space 
(apart from the differences that result from the different initial progenitor composition). 
They are characterized by a maximum centred around $0.3\mathord{\times}10^3\,\kms$ 
and have a high-velocity tail which extends to 
$\mathord{\sim}0.5\mathord{\times}10^3\,\kms$ (using $\Delta M_{\mathrm{i}}/ M_{\mathrm{i}}\,\mathord{=}\,8\mathord{\times}10^{-4}$ as 
a threshold value). The mixing in velocity space corresponds to mixing in mass 
coordinate to a maximum of $\mathord{\sim}1.45\,\solm$ 
in model \onemg and $\mathord{\sim}1.95\,\solm$ in model \znine, which 
corresponds to the bottom of the respective hydrogen envelope. 

Thus, the mixing is more efficient in model \znine in comparison to model \onemg, although the explosion energy and the integrated 
growth factors are larger in the latter. Why is this the case? 
The answer can be found by inspecting Figures~\ref{fig:e8 nix cuts}, \ref{fig:e8 rho cuts} and 
\ref{fig:z9 nix cuts}, \ref{fig:z9 rho cuts}. 
Model \znine shows a larger asymmetry already at the onset of the explosion. 
Additionally, the density profile of the progenitor of 
model \znine exhibits less extreme declines
outside of the Si/CO and CO/He interfaces than the 
sharp drop of the density at the edge of the degenerate core in the ECSN progenitor. 
This permits a shock expansion that is not quite as rapid as in model e8.8
(Figures~\ref{fig:eexp all} and \ref{fig:radii all times}).
The initial asymmetry triggers more and faster growth of the RT instability
at the CO/He interface in model \znine (see Figures~\ref{fig:z9 nix cuts}, \ref{fig:z9 rho cuts}) on a time scale much shorter 
than the growth of the 
RT mushrooms in model \onemg. While in model \znine large RT plumes are visible
already at about 100 seconds after bounce, it takes a few hours for such structures
to develop at the He/H interface of model \onemg (see Figure~\ref{fig:e8 nix cuts}).
From this result we conclude that the extent of mixing during the SN blast 
is not only determined by the linear growth factors, but depends strongly
on the initial explosion asymmetry seeding the growth of the RT instability
at the unstable composition interfaces.

In contrast to the ECSN-like models, heavy elements are mixed to large mass coordinates and velocities in model \snine.
This is facilitated by the extreme initial asymmetries at the onset of the explosion.
Fast metal-rich plumes arrive quickly at the He/H interface and trigger the RT instability there, thereby transporting a 
significant amount of the total $\nickel\mathord{+}\tracer$ mass to large mass 
coordinates into the H-envelope.
Outward mixing of intermediate-mass elements is driven by the growth of the RT 
instability which causes a fragmentation of the dense shells that form after the 
forward shock crosses the CO/He and He/H interfaces.
Most importantly, the biggest plume is able to transport $\nickel\mathord{+}\tracer$-rich matter to large velocities and mass coordinates.
It contains about 25\% of the total $\nickel\mathord{+}\tracer$ mass at
$t_{\mathrm{map}}$ and carries about half of that to radii well ahead of the 
average radius of the shock when it reaches the stellar surface.
We find that about $4\%$ of the
$\nickel\mathord{+}\tracer$-rich material travels with more than $1,000\,\kms$ and thus far ahead of the bulk of the metal-rich matter.

\subsection{Long-time evolution of compact remnant properties}
\label{sec:Long-time evolution of compact remnant properties}

While the SN shock wave travels through the progenitor star, some material falls back 
onto the newly formed NS \citep[e.g.,][]{Chevalier1989,Woosley1989,Zhang2008,Fryer2009,Wong2014}. 
This fallback is needed to explain the observed broad range of compact remnant masses
\citep{Zhang2008,Wong2014,Ertl2020}.
A first episode of fallback occurs when the neutrino-driven wind abates and the wind termination
shock moves back towards the PNS \citep{Arcones2007}. This period happens at early times, 
roughly within $\sim$10\,s after bounce. At later times fallback is driven by the reverse shocks
that originate from the acceleration and deceleration phases of the SN shock passing the 
composition shell interfaces. As described in Sections~\ref{sec:Propagation of the forward shock} 
and \ref{sec:Morphology of the ejecta}, these reverse shocks
propagate backward into the central volume.

\begin{table*}
\caption{Overview of final NS properties in our 3D models at the time of shock breakout.}
\label{tab:neutron star final}
\setlength{\tabcolsep}{5pt}
\renewcommand{\arraystretch}{1.2}
\begin{tabular}{cccccccc||cccccccc}
    \hline 
            &
            $t_{\mathrm{map}}$ &
            $v_{\mathrm{NS}}^{\mathrm{tot,map}}$   &
            $J_{\mathrm{NS}}^{\mathrm{map}}/10^{45}$   &
            $\theta_{vJ}$    &
            $M_{\mathrm{map}}$ &
            $M_{\mathrm{g}}$   &
            $P_{\mathrm{NS}}^{\mathrm{map}}$ &
            $t_{\mathrm{fin}}$ &
            $v_{\mathrm{NS}}^{\mathrm{tot,fin}}$   &
            $J_{\mathrm{NS}}^{\mathrm{fin}}/10^{45}$   &
            $\theta_{vJ}^{\mathrm{fin}}$    &
            $M_{\mathrm{\mathrm{fb}}}$ &
            $M_{\mathrm{\mathrm{fin}}}$ &
            $M_{\mathrm{g}}^{\mathrm{fin}}$   &
            $P_{\mathrm{NS}}^{\mathrm{fin}}$
            \\
    Model &
    [s]   &
    $[\mathrm{km/s}]$ &
    $[\mathrm{cm^2 g/s}]$ &
    $[^{\circ}]$ &
    [\solm]    &
    [\solm]    &
    [s] &
    $10^{5}\,[\mathrm{s}]$   &
    $[\mathrm{km/s}]$ &
    $[\mathrm{cm^2 g/s}]$ &
    $[^{\circ}]$ &
    [$10^{-3}\,\solm$]    &
    [\solm]    &
    [\solm]    &
    [s]     \\
    
    \hline
    \onemg      & 0.47 &  0.44 & 0.70 &  90.0 & 1.326 &  1.210 &  10.58 & 4.5 &  0.46  &  1.77  &  114.3 &  0.316 & 1.326 &  1.210 & 4.16 \\
    \znine      & 1.44 & 34.90 & 2.55 &  45.4 & 1.353 &  1.231 &  2.96  & 5.0 & 31.48  &  40.2  &  163.6 &  0.065 & 1.353 &  1.231 & 0.19 \\
    \snine      & 3.14 & 40.87 & 8.05 &  31.3 & 1.351 &  1.230 &  0.94  & 4.1 & 41.32  &  253.2 &  101.7 &  4.999 & 1.356 &  1.234 & 0.030 \\
    \hline
\end{tabular}
\flushleft
\textit{Notes}: The left part of the table (columns 2--8) lists the properties of the PNS
at $t_{\mathrm{map}}$ (see also Table~\ref{tab:neutron star}).
The right part of the table (columns 10--16) gives the corresponding final NS properties
at the end of our long-time simulations, $t_\mathrm{fin}$. $v_{\mathrm{NS}}^{\mathrm{tot,fin}}$ is 
the final total kick velocity of the NS, $J_\mathrm{NS}^\mathrm{fin}$ its final angular
momentum, $\theta_{vJ}^\mathrm{fin}$ the angle between spin and (total) kick vectors, $M_{\mathrm{fb}}$ 
the total fallback mass, $M_\mathrm{fin}$ the final
baryonic mass, $M_\mathrm{g}^\mathrm{fin}$ the gravitational mass and $P_\mathrm{NS}^\mathrm{fin}$ the
spin period, adopting a NS radius of 12\,km. It is assumed that all fallback matter is accreted by the
NS.
\end{table*}

\begin{figure}
 \centering
 \includegraphics[width=0.48\textwidth,trim=0cm 0.25cm 0cm 0.3cm,clip]{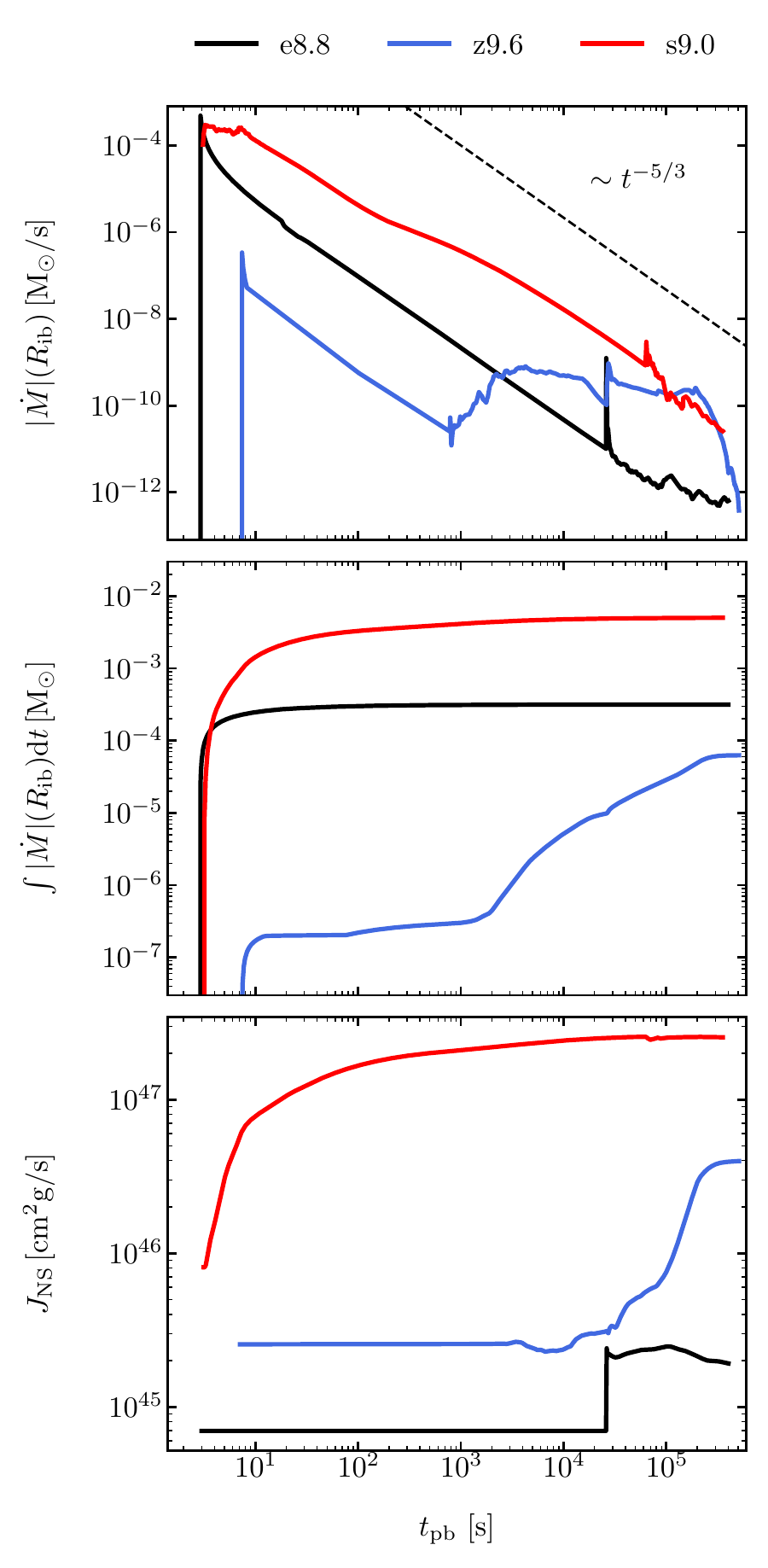}
 \caption{
Time evolution of the NS mass-accretion rate (top panel), cumulative accretion of mass
(middle panel), and NS angular momentum (bottom panel) associated with fallback
during the long-time simulations of all 3D models. After some initial transition
phase, which lasts longer in model \snine because of the 3D asymmetries of the 
slow ejecta in the vicinity of the NS, all models adjust to the well-known power-law
decline according to $\dot{M} \mathord{\propto}t^{-5/3}$ (black dashed line).
This continues until the reverse shock from the outer layers in each model has
propagated inward and reaches the inner boundary of the computational grid at
radius $R_{\mathrm{ib}}$. In models \onemg and \snine early fallback clearly dominates, 
whereas in model \znine the reverse shock contributes significantly to the total 
fallback mass. Angular momentum carried through $R_{\mathrm{ib}}$ by the infalling 
matter changes the initial angular momentum of the NS under the assumption that all
fallback matter is accreted by the NS. 
In models \onemg and \znine late fallback associated with the reverse shock 
causes the main effect on $J_\mathrm{NS}$, whereas in model \snine the total angular
momentum rises continuously and most steeply at early times, when some of the highly 
asymmetric ejecta fall back onto the NS.}
 \label{fig:fallback 3d}
\end{figure}

In Figure~\ref{fig:fallback 3d} we show, starting from $t_{\mathrm{map}}$, the corresponding time-dependent mass accretion rate through 
the inner boundary
of our computational grid, $|\dot{M}(t)|(R_{\mathrm{ib}})$ (top panel), the associated time-integrated
accreted mass, $\int_{t_\mathrm{map}}^t\mathrm{d}t'\,|\dot{M}(t')|(R_{\mathrm{ib}})$ (middle panel), and
the evolution of the angular momentum of the NS, $J_\mathrm{NS}(t)$, driven by the angular 
momentum that is carried by fallback material into the central volume and that is 
assumed to be accreted into the compact remnant (bottom panel) for all of our 3D long-time
simulations. For comparison, Table~\ref{tab:neutron star final} 
lists the values of the NS masses (baryonic and gravitational), total kick velocities,
$v_{\mathrm{NS}}^{\mathrm{tot}}$, total angular momenta, $J_\mathrm{NS}$,
and spin periods, $P_\mathrm{NS}$, as well as the angle between NS spin and kick vectors, 
$\theta_{vJ}$, at $t_\mathrm{map}$ and at the end of the 3D simulations, $t_\mathrm{fin}$.

During the first fallback episode, i.e., during the first tens of seconds,
all three models behave differently because of their different
evolution during the neutrino-wind phase. In model \onemg the forward shock is very fast and the
slower neutrino-driven wind (adopted from a parametric 1D simulation; see Fig.~\ref{fig:wind})
expands freely behind the shock without deceleration and without
developing a reverse shock. After the termination of the neutrino-driven wind at $\sim$2.5\,s
post bounce, fallback sets in, but the mass accretion onto the NS declines steeply,
because the ejecta move rapidly outward, evacuating the surroundings of the compact object.

In model \znine a similar situation applies, but the faster and more long-lasting
neutrino-driven wind (adopted from a self-consistently computed NS cooling model; 
see Fig.~\ref{fig:wind}) is stronger and evacuates the 
neighbourhood of the NS even more extremely than in model \onemg. However, a reverse shock forms
when the forward shock passes the CO/He interface and decelerates, while at the same time
the neutrino-driven wind pushes from behind and
compresses the postshock matter into a dense shell. This reverse shock moves inward
within a few seconds and creates the short accretion spike at $\sim$8\,s
(Figure~\ref{fig:fallback 3d}). The accretion peak
decays within only a second when a reflected wave sweeps through the medium surrounding the PNS
outward again. Model \znine displays the lowest mass accretion of all three models.
In contrast, model \snine has by far the highest mass accretion through the inner grid boundary 
since its neutrino-driven wind is very weak. Therefore the NS is not surrounded by a large
low-density wind bubble, but instead accretion downflows and rising plumes of neutrino-heated 
matter continue to coexist in the vicinity of the NS for many seconds of postbounce evolution.
The mass accretion rate during the early fallback phase exhibits a correspondingly high plateau
between $\sim$3\,s and $\sim$10\,s.

Models \onemg and \znine reach the asymptotic scaling $\dot{M}\mathord{\propto}\,t^{-5/3}$ 
\citep{Chevalier1978,Zhang2008,Dexter2013,Wong2014} already after about 10\,s, and this
persists until
the late fallback associated with the reverse shocks sets in at about 1000\,s in \znine 
and at about 9\,h in \onemg. In contrast, model \snine displays clear deviations from the
$-$5/3 power-law until roughly 1\,h, and only gradually approaches the 
$-$5/3 power-law behavior later, because of large-scale asymmetries in 
the fallback material. Late accretion due to the reverse shock from the 
He/H interface is triggered only after $\sim$16\,h. 

With an integral value around $5\times 10^{-3}$\,M$_\odot$ model \snine has the highest
total fallback mass (middle panel of Fig.~\ref{fig:fallback 3d}), the other two models 
possess at least 10 times lower values. In all of the three cases the fallback mass is
too low to have any significant impact on the NS mass or kick (see 
Table~\ref{tab:neutron star final}). In principle, the asymmetric fallback of matter
can change the net momentum carried by the ejecta with a corresponding increase of
the momentum of the NS in the opposite direction. In the considered models this modifies
the NS velocity by at most a few kilometers per second (certainly 
$\lesssim$10\,km\,s$^{-1}$), with the biggest effect in model \znine because
of the early onset of the reverse-shock induced accretion in this case.

However, the asymmetric fallback of matter also carries angular momentum through the 
inner grid boundary. Although very little mass is accreted, matter that falls back from
large radii can transport appreciable amounts of angular momentum, accounting for the 
dominant contribution to the total NS angular momentum in all of our simulations.
In model \snine the angular momentum of the NS increases by more than a factor
of 10 to $10^{47}$\,g\,cm$^2$s$^{-1}$ within the first 10\,s of fallback accretion.
It further grows continuously by another factor 2.5 until the 3D simulation was stopped 
over 4\,d later. The final angular momentum of the NS is more than 30 times bigger
than at 3\,s after bounce. In models \onemg and \znine the NS angular momentum begins to
change only when the late reverse shocks reach the center and anisotopic structures begin
to be accreted. This amplifies the initial angular momentum of the NS in \znine still by 
a factor of 15 (see bottom panel in Figure~\ref{fig:fallback 3d}). 
In contrast, in model \onemg the total effect is much more modest because of the low mass 
associated with the late fallback and the weak asymmetry of the explosion in this case.

Nevertheless, also in model \onemg the spin period estimated for a NS with 12\,km
radius decreases from $\sim$11\,s at 0.5\,s after bounce to a final value of nearly 4\,s
(Table~\ref{tab:neutron star final}).
For \znine the spin period shrinks from 3\,s at $t_\mathrm{map} = 1.44$\,s to finally
0.19\,s, and for \snine the final period is as low as 0.030\,s, whereas it had been 
$\sim$1\,s before the fallback.

The NS spins seem to be randomly oriented relative to the NS kick directions, i.e.,
the angles between NS spin vector and total kick vector, $\theta_{vJ}$, do not show any
preference for spin-kick alignment, neither at early times nor days later when the
fallback is complete and the simulations are terminated. This is not unexpected
and it is in line with previous findings \citep{Wongwathanarat2013,Mueller2019,Chan2020,Powell2020}.
To date there is no suggestion based on well accepted physics (i.e., without invoking
uncertain ingredients or extreme physical assumptions) for a convincing 
mechanism that could provide spin-kick alignment. Even if such an alignment were 
achieved during the first seconds of
the explosion, when ejecta and NS are still in contact through hydrodynamical and
gravitational forces and the NS is kicked by anisotropic neutrino radiation, it 
is very hard to imagine how this initial alignment could not be overruled by the
stochastic effects of the later fallback and its dominant influence on the NS 
spin. It might require
very rapid progenitor rotation, possibly very strong magnetic fields, to impose a 
preferred direction for the explosion, correlating the NS recoil acceleration and 
the spin-up of the NS either by angular momentum inherited during the collapse
from the rotating progenitor or later through accretion of stellar angular momentum 
by fallback \citep[for a suggestion of such a mechanism, see][]{Janka2017}.

Recently, \citet{Chan2020} presented results of an explosion simulation for a non-rotating,
zero-metallicity (Pop~III) 12\,M$_\odot$ progenitor, including the effects of anisotropic
fallback. The overall behavior of this model is similar to our model s9.0, however
much more extreme, because the model has a considerably higher fallback mass (nearly
0.2\,M$_\odot$), and the corresponding change of the NS kick is several 10\,km\,s$^{-1}$.
The accretion of angular momentum associated with the fallback spins the NS up from an
early period around 100\,ms to a final period of only a few milliseconds. Interestingly,
in this model as well as in the other two cases considered by \citet{Chan2020}, in which 
black holes are formed by fallback, 
the final angles between spin and kick vectors are close to 90 degrees,
very similar to our result for model s9.0. \citet{Chan2020} explain such a perpendicular
orientation by the fact that the directions of the accretion streams remain relatively
constant in time, thereby corresponding to a single, off-center momentum impulse onto the 
remnant. Our models e8.8 and s9.0 comply with this pattern, but model z9.6 deviates from
this behavior, showing spin-kick anti-alignment.

\begin{figure*}
 \centering
 \includegraphics[width=0.95\textwidth]{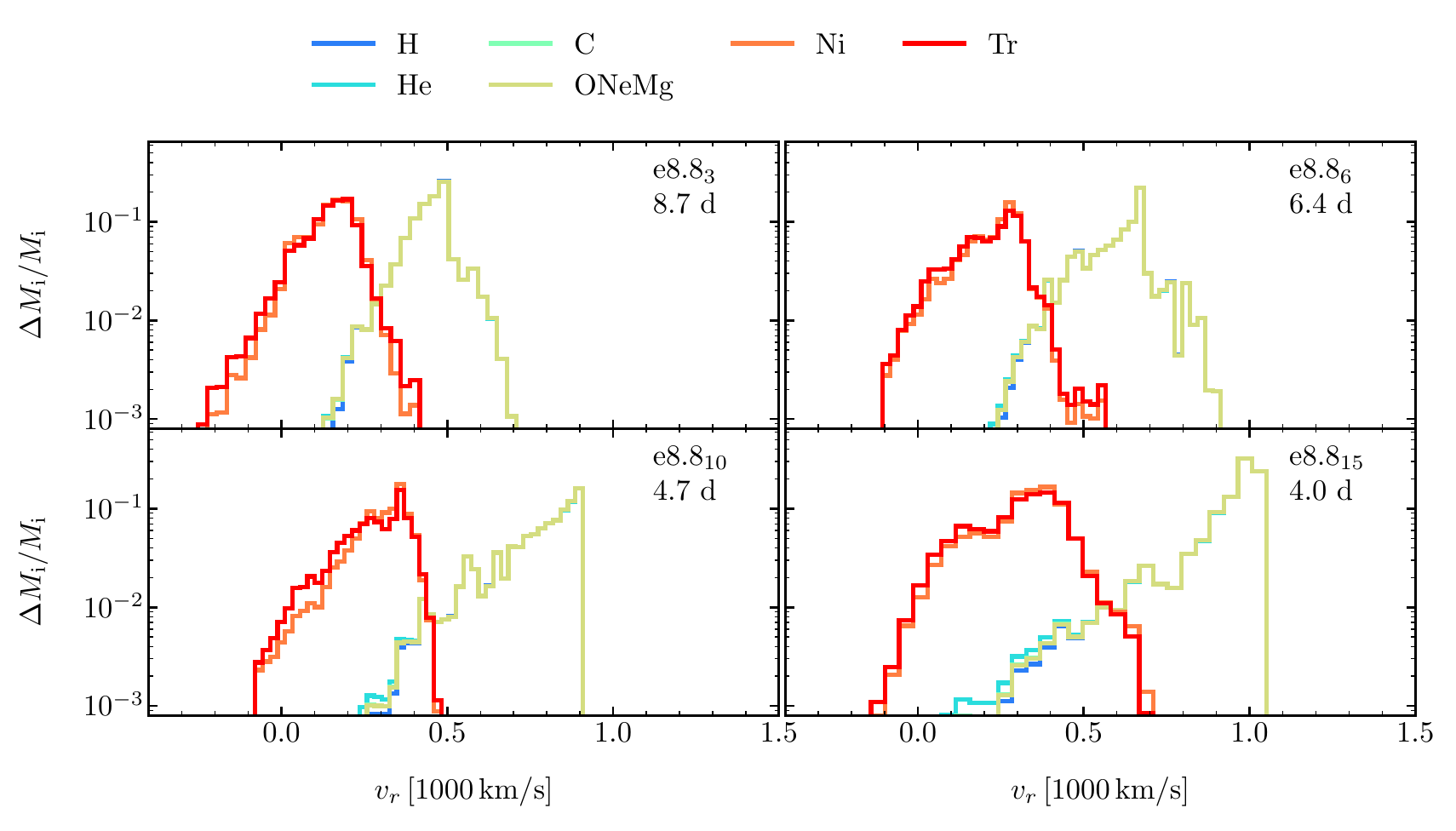}
 \caption{Normalized mass distributions of chemical elements as functions of
 radial velocity for the long-time 2D simulations of model \onemg (based on the 
 calibrations listed in Table~\ref{table:e8param}) 
 at the time of shock breakout. 
 Maximum velocities and the velocities of the bulk of 
 \nickel and \tracer scale roughly with $\sqrt{E_{\mathrm{exp}}}$, 
 implying slightly enhanced mixing for more energetic models 
 (see also Figure~\ref{fig:e8 2d massDis sbo mass}). Note that the
 line corresponding to carbon is hardly visible because it coincides with
 the line for ONeMg.}
 \label{fig:e8 2d massDis sbo}
\end{figure*}

\begin{figure*}
 \centering
 \includegraphics[width=0.95\textwidth]{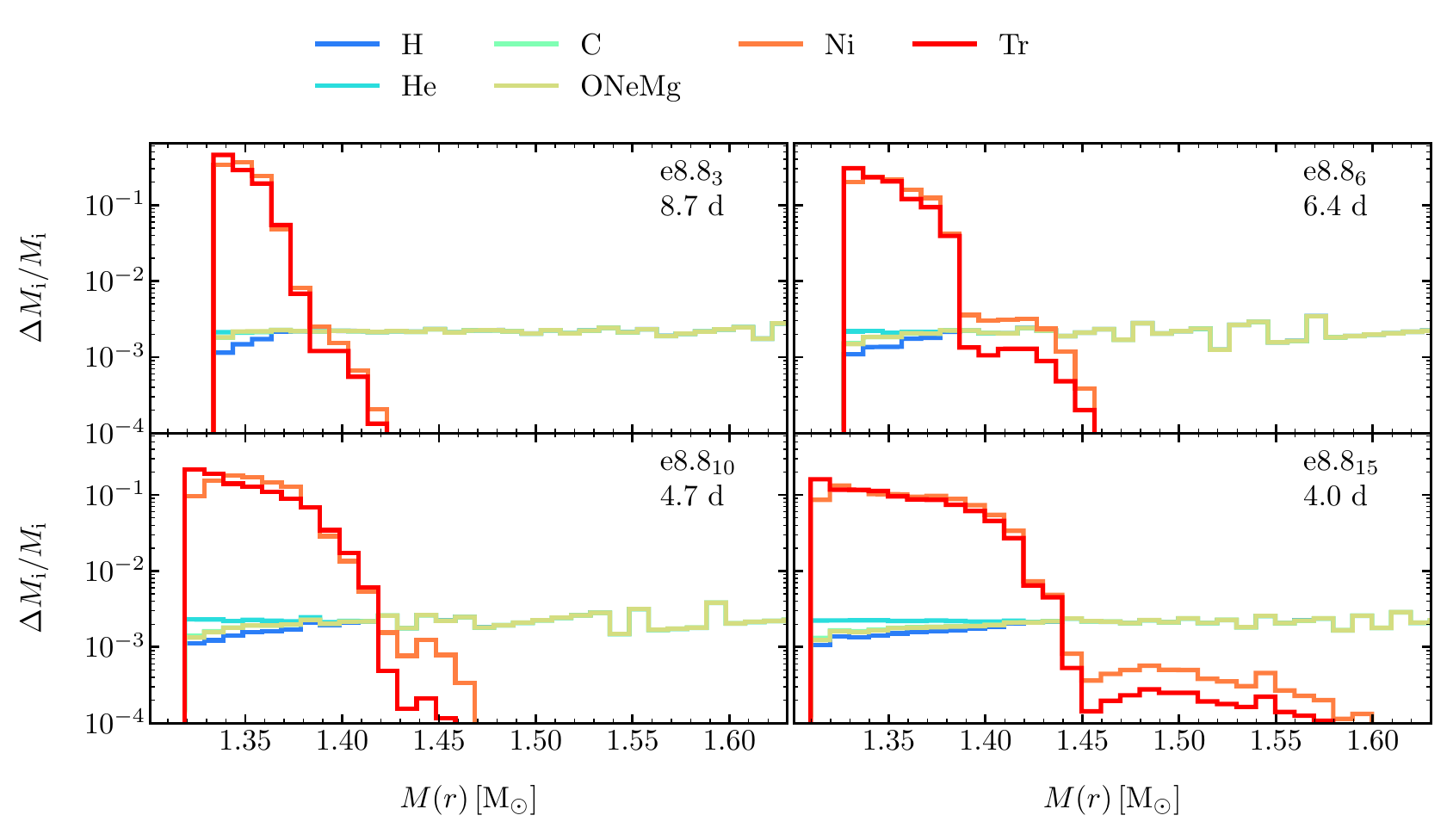}
 \caption{Normalized mass distributions of chemical elements as functions of
 enclosed mass for the 2D simulations of model \onemg  
 (based on the calibrations listed in Table~\ref{table:e8param}) at the 
 time of shock breakout. Models with higher
 $E_{\mathrm{exp}}$ show slightly more efficient mixing of iron-group material.
 Note that the line corresponding to carbon is hardly visible because it 
 coincides with the line for ONeMg.}
 \label{fig:e8 2d massDis sbo mass}
\end{figure*}

\subsection{Dependence on the explosion energy}
\label{sec:Dependence on the explosion energy}

In Figure~\ref{fig:e8 2d massDis sbo} we present the normalized 
mass distributions of chemical elements as functions of
radial velocity for the 2D simulations of model \onemg, using the explosion
calibrations as listed in Table~\ref{table:e8param}, at the time of shock breakout. 

Comparing the distributions of model e8.8$ _{10}^{\mathrm{2D}}$ with the 3D 
simulation, the extent of mixing in model \onemg does not seem to be 
significantly dependent on the chosen dimensionality. This allows us to 
investigate the influence of the explosion energy on the efficiency of mixing 
in our 2D simulations of model \onemg.
 
Inspecting Figure~\ref{fig:e8 2d massDis sbo}, we find that the bulk of \nickel 
resides at low velocities of $\mathord{\sim}(0.2\mathord{-}0.4)\mathord{\times}10^3\,\kms$, 
increasing with explosion energy roughly as $v_{r}\,\mathord{\sim}\,\sqrt{E_{\mathrm{exp}}}$. 
For larger explosion energies the downward mixing (to smaller velocities) of 
lighter elements also seems to be more efficient, but the effect is small and 
affects only $\mathord{\lesssim}1\%$ of the respective masses. For lower explosion
energies the amount of $\nickel\mathord{+}\tracer$ experiencing fallback 
($v_{r}\,\mathord{<}\,0$) grows.

The mixing in velocity space corresponds to a distribution of \nickel and 
\tracer in mass space to a maximum mass coordinate of about 
$1.42\mathord{-}1.60\,\solm$ (see 
Figure~\ref{fig:e8 2d massDis sbo mass}), also increasing with explosion 
energy, using $\Delta M/M\,\mathord{=}\,1\mathord{\times}10^{-4}$ as a threshold 
value for plotting the distributions. 
Note that the 4.49\,\solm of the hydrogen envelope of the progenitor 
extended initially from 1.34\,\solm to 5.83\,\solm, thus mixing only affects 
the innermost part of the envelope.

As the iron-core progenitors are exploded self-consistently, their explosion 
energies are fixed within our framework. 
Drawing direct connections between the amount of mixing and the explosion energy can 
therefore not be done. Comparing the \znine model 
with the electron-capture model, which both have similar explosion energies, however, suggests
that the influence is secondary. More decisive for the amount of nickel mixing are the progenitor
structure and initial asymmetries right after shock revival. This view is supported by the 
strong mixing exhibited by model \snine, which has a comparable explosion energy as well. 
The velocity and density perturbations that are present at the onset of the explosion, 
combined with the strong acceleration and deceleration of the forward shock in the
envelope of the progenitor, yield high growth rates of RT instability 
over a larger range in the mass coordinate and thus dominate over effects 
caused by different explosion energies in a given progenitor.

\section{Comparison to previous studies}
\label{sec:Comparison to previous studies}

Some previous works also explored the explosion properties
of CCSNe of low-mass iron and ONeMg-core progenitors.
The study by \cite{Radice2017} simulated the onset of the explosion
of models $\mathrm{e8.8_n}$, \znine, and \snine in 2D with varied 
microphysics. They found for model $\mathrm{e8.8_n}$ an explosion 
energy of up to $1.8\mathord{\times}10^{50}\,\erg$, which is about 
a factor of two higher than obtained for the same $8.8\,\solm$ star 
by \cite{Kitaura2006,Janka2008,Huedepohl2010,Fischer2010}, and \cite{Groote2014}.
Consistently, their 2D simulations yielded explosion energies for 
models \znine and \snine that were $\mathord{\sim}50\%$ higher than in
our simulations, namely $1.2\mathord{\times}10^{50}\,\erg$ and 
$0.7\mathord{\times}10^{50}\,\erg$, respectively; their 3D calculations
of model \snine yielded even $1.0\mathord{\times}10^{50}\,\erg$ 
\citep{Burrows2020}.

\cite{Mueller2019} focused on the onset of the explosion in helium core
progenitors from binary evolution and low-mass single stars 
also including model \znine. 
They found a slightly higher explosion energy of model \znine of around
$1.3\mathord{\times}10^{50}\,\erg$ at the end of their simulation. 
Moreover, their explosion energy still seems to grow fairly 
steeply when they stopped their simulation.

However, \cite{Mueller2019} used an approximative neutrino transport
treatment (``fast multi-group transport'', FMT) with simplified neutrino
interactions, and therefore their explosion energies cannot be 
considered as quantitatively absolutely reliable.\footnote{In
contrast to \textsc{Vertex}, the FMT treatment neglects fluid-velocity
effects and neutrino energy-bin coupling, computes the flux factor from
a two-stream approximation (i.e. only for ingoing and outgoing directions,
while \textsc{Vertex} solves the Boltzmann equation for many directions), 
and uses a subset of additionally simplified neutrino opacities. Therefore, 
naturally, FMT is not on equal level in accuracy with the \textsc{Vertex} neutrino 
transport. Correspondingly, comparisons of FMT and \textsc{Vertex} results 
in \citet{Mueller2015b} showed ``reasonably good overall agreement'', but 
differences in the neutrino luminosities and mean energies can still be
considerable \citep[see also][for a comparison of FMT with Boltzmann 
transport]{Chan2020b}.} 
The reason why \cite{Radice2017} and \cite{Burrows2020} 
obtained consistently more energetic explosions 
than in our \vertexprom simulations is unclear to us, in particular 
because their \textsc{FORNAX} code is claimed to possess neutrino
physics that is compatible with that of \vertexprom.

We note that, due to the higher explosion energies found in the
studies of other groups, the final PNS masses there are smaller than ours
in general. A more energetic explosion
drives a stronger wind from the PNS, which carries away mass from its
surface or reduces further accretion. For example 
\cite{Mueller2019} found the PNS baryonic mass of model \znine 
to be 1.35\,\solm, close to ours, and \cite{Burrows2019} determined
the PNS mass of model \snine to be 1.342\,\solm, whereas we get 
1.351\,\solm before fallback accretion and 1.356\,\solm afterwards. 
The explosion energies and PNS masses of \cite{Radice2017} and
\cite{Burrows2020} are not only in conflict with ours but also with
other previous studies as for example \cite{Kitaura2006,Janka2008,Huedepohl2010,Groote2014} for model $\mathrm{e8.8}$ and \cite{Glas2019} for model \snine. These studies 
consistently attain lower explosion energies with around $10^{50}\,\erg$
for model \onemg and around $0.5\mathord{\times}10^{50}\,\erg$ for model \snine, 
respectively.

Concerning the long-time evolution of the explosion of low-mass CCSNe
progenitors, \cite{Mueller2018} followed the expansion of the SN shock from 
its initiation by neutrino heating until shock breakout in an ultra-stripped helium
star (he2.8) with a helium-core mass of 1.49\,\solm, which is structurally 
similar to our model \znine. The SN runs (s2.8) of 
\cite{Mueller2018} and \cite{Mueller2019} also explode with energies
only slightly higher than our simulation of model \znine, and 
comparing the maximum mass coordinate of the neutrino-heated ejecta, 
they exhibit a similar extent of mixing. In their simulation a small 
fraction of the total iron-group material is mixed to the edge of the 
helium core of the progenitor star, quite analogously to the case of 
our model \znine, where we find a small 
fraction of the total neutrino-heated ejecta to be mixed out to $M(r)\,\mathord{\sim}\,1.8\,\solm$.

Other studies such as those of \citet{Kifonidis2006}, \cite{Hammer2010}, 
\citet{Wongwathanarat2015}, \citet{Chan2018}, \citet{Chan2020}, and 
\cite{Ono2020} and \citet{Orlando2019} also performed
simulations of the long-time evolution of CCSNe after shock revival, based on 2D/3D 
initial data of the beginning explosion. 
These studies, however, focused on more massive RSG progenitors or BSG 
progenitors that stand as a proxy for Sanduleak -69 202, the progenitor of SN1987A. \cite{Kifonidis2006}
and \citet{Hammer2010} used a 15\,\solm BSG, while \citet{Wongwathanarat2015} 
explored various BSG and RSG models from 15--20\,\solm, and \citet{Chan2018} and 
\citet{Chan2020} investigated zero-metallicity (Pop III) stars, 
i.e., a 12\,M$_\odot$ progenitor and a 
black-hole forming 40\,M$_\odot$ model. \cite{Ono2020} employed two BSG 
models with 16.3\,\solm and 18.3\,\solm and two RSG stars with ZAMS 
masses of 18\,\solm and 19.8\,\solm. 
While the studies of \cite{Kifonidis2006,Hammer2010,Wongwathanarat2015,Chan2018,Chan2020} 
started with initial data from neutrino-driven explosion simulations, \cite{Ono2020} 
and \citet{Orlando2019} initiated the explosions by injecting energy near the IG/Si 
interface and parameterized the deformation of the outgoing shock wave in order
to mimic the effect of non-radial instabilities at the onset of the explosion and 
to reproduce the observed morphology of SN1987A.
Long-time 3D SN simulations for studying mixing
and explosion asymmetries by \cite{Ellinger2012} and 
\cite{Joggerst2009,Joggerst2010} started their runs of zero metallicity, 
low-metallicity, and solar metallicity 15\,\solm and 25\,\solm 
progenitors from spherically symmetric explosions.

These more massive progenitors differ strongly in their $\rho r^3$-profiles when compared to our 
ECSN-like models. Only model \snine exhibits structural features (e.g., a significant 
variation of the density gradient at the He/H interface) similar to the RSG models of 
\citet{Wongwathanarat2015}. Consequently, all models of the mentioned
studies evolve considerably
differently from the ECSN-like progenitors presented here. 
Larger amplification factors at the composition interfaces and a more
extended region of instability lead to the growth of large RT plumes, 
which are absent in models \onemg and \znine. 
Model \snine, however, behaves in a more similar way, despite the smaller ZAMS 
mass and considerably lower explosion energy. Although the
previous studies are tuned to give around $10^{51}\,\erg$ for the explosion energy 
(e.g., to be compatible with observations of SN1978A), we find a similar 
efficiency of mixing in s9.0, reflected by the distributions of the 
chemical elements at the end of our simulation. This efficient mixing 
is facilitated by the strongly asymmetric onset of the explosion and
the growth of strong secondary RT instability at the composition interfaces 
triggered by the initial ejecta asymmetries.

\section{Summary and Conclusions}
\label{sec:summary}
In this study we presented results of 1D, 2D, and 3D SN simulations
for three non-rotating low-mass progenitors, two of which were RSGs
of 9.6\,M$_\odot$ (z9.6) and 9\,M$_\odot$ (s9.0), respectively,
which had formed $\sim$1.30\,M$_\odot$ iron cores at the end of
their lives, and the third one was a newly constructed 8.8\,M$_\odot$
super-AGB star as
ECSN progenitor (e8.8) with a highly degenerate $\sim$1.34\,M$_\odot$
ONeMg core\footnote{But note the remark in Sect.~\ref{sec:onemgcoreprog}
that the true mass of the ONeMg core should have been 1.39\,M$_\odot$.} 
and a pre-collapse mass of 5.83\,M$_\odot$.
Our aim was a comparison of observable features between the models,
including the remnant properties, ejecta composition and asymmetries,
as well as the radial mixing of chemical species during the SN blast.
To this end our simulations covered continuously all evolutionary 
phases, from the onset of stellar core collapse to core bounce,
shock formation, shock stagnation, delayed shock revival 
by neutrino heating, shock propagation through the stellar mantle 
and envelope, to shock breakout from the surface of the star, and beyond
this moment until fallback of matter to the central compact remnant
was complete.

Our investigation by means of neutrino-hydrodynamical simulations 
was focused on neutrino-driven explosions, because all of the 
considered progenitors explode self-consistently and fairly quickly
after core bounce by the delayed neutrino-driven mechanism. We 
therefore did not invoke any additional effects such as ``jittering jets''
(e.g., \citealt{Soker2010}), which have been suggested
as alternative or additional mechanism to revive the stalled SN
shock (e.g., \citealt{Soker2019}), and whose effects have recently been claimed
to play a role in the low-energy explosions of low-mass SN progenitors \citep{Gofman2019}.

\subsection{Explosion energies and NS properties}

Collapse and explosion of ONeMg-core progenitors have been
investigated extensively before by fully self-consistent simulations
\citep{Kitaura2006,Janka2008,Huedepohl2010,Fischer2010,Mueller2013,Groote2014,Radice2017}, yielding
quite a spread of blast-wave energies, ranging from several $10^{49}$\,erg
to nearly $2\mathord{\times} 10^{50}$\,erg, depending on details such as the
treatment of neutrino transport, general relativity, and the EoS
of supranuclear matter in the PNS. Therefore, we simulated the explosions 
of model e8.8 in 1D, 2D, and 3D with the \prom code by imposing suitable
neutrino luminosities to tune the SN energies to values between
$3\mathord{\times} 10^{49}$\,erg and $1.5\mathord{\times} 10^{50}$\,erg. Our models reproduce 
the generic behavior seen previously when the shock wave expands extremely
rapidly at running down the steep density gradient that marks the edge of
the ONeMg core.

We picked our 3D ECSN model of e8.8 to have an explosion energy
of $1.0\mathord{\times} 10^{50}$\,erg, which is in the ballpark of the previous
self-consistent runs by \citet{Kitaura2006} and \citet{Huedepohl2010},
but it is only about half of the energy obtained by \cite{Radice2017}. The
hydrodynamic kick of the NS in this 3D model was less than
0.5\,km\,s$^{-1}$, and in all of our 2D realizations 
it stayed below about 1.5\,km\,s$^{-1}$,
in agreement with the low kick values obtained by \citet{Gessner2018}.

Models z9.6 and s9.0 were exploded fully
self-consistently in 3D by applying the \vertexprom code with 
sophisticated neutrino transport. First results of the initial
$\sim$0.5\,s after bounce for these two simulations were presented by
\citet{Melson2015a} and \citet{Melson2019}. 
The explosion energies saturate at values below $10^{50}$\,erg,
in the case of z9.6 at $0.85\mathord{\times} 10^{50}$\,erg, and for s9.0 at
about $0.5\mathord{\times} 10^{50}$\,erg (compatible with \citealt{Glas2019}). 
This is again roughly a factor of 
two lower than the energies found in 2D and 3D simulations for s9.0
by \cite{Radice2017}, \cite{Burrows2019}, and \cite{Burrows2020},
and in 3D for z9.6 by \cite{Mueller2019}. 
However, all of our 3D models explode with energies
consistent with the $5\mathord{\times} 10^{49}$--$10^{50}$\,erg of the Crab SN \citep{Yang2015},
which has been interpreted as ECSN or as
CCSN of a low-mass iron-core progenitor (e.g., \citealt{Smith2013,Tominaga2013}).
Moreover, all models produce small ejecta masses of radioactive $^{56}$Ni,
between about $4\mathord{\times}10^{-3}$\,M$_\odot$ and roughly 
$6\mathord{\times}10^{-3}$\,M$_\odot$.

Both iron-core SN models develop a pronounced dipole mode of the
lepton-number emission by neutrinos, which goes back to the 
LESA (Lepton-Emission Self-sustained Asymmetry) phenomenon.
The NS kicks induced by the corresponding dipole component
of the total neutrino luminosity are around 25\,km\,s$^{-1}$,
which is subdominant compared to the hydrodynamical NS kick 
($\sim$30\,km\,s$^{-1}$) in s9.0 but higher than the hydrodynamical
kick ($\sim$10\,km\,s$^{-1}$) in z9.6. The neutrino-induced kicks
might further increase after the end of the evolution that we
simulated with detailed neutrino transport (about 0.5\,s after bounce),
although in model z9.6 a quadrupole emission mode begins to dominate
the dipole mode at that time, and the NS acceleration by 
anisotropic neutrino radiation becomes correspondingly small.

All of our SN simulations of the low-mass progenitors
produce NSs with baryonic masses between $\sim$1.30\,M$_\odot$ and 
$\sim$1.35\,M$_\odot$, corresponding to gravitational masses between
$\sim$1.20\,M$_\odot$ and $\sim$1.23\,M$_\odot$.
Right after their formation, during the first seconds of their
lives, the new-born NSs spin slowly with periods in the range of 
seconds, corresponding to angular momenta of the order of 
some $10^{45}$\,g\,cm$^2$s$^{-1}$. However, this initial spin
is dwarfed by later fallback effects. Despite the small fallback
masses of at most a few $10^{-3}$\,M$_\odot$, which neither
change the NS masses nor NS kicks to any relevant extent, the fallback
transports large amounts of angular momentum to the compact remnant.
The angular momentum received from fallback material outruns the previous
angular momentum by up to a factor of 30 and can shrink the NS spin period
from seconds to tens of milliseconds.

For example, the compact remnant in model s9.0 thus attains a spin period 
of nearly 1\,s before fallback due to asymmetric mass ejection during
the early stages of the explosion that transferred an angular momentum
of only $8\mathord{\times} 10^{45}$\,g\,cm$^2$s$^{-1}$ to the NS. After 
fallback the spin period of the new-born NS is 30\,ms only. This is 
very close to
the 17--19\,ms estimated for the birth period of the Crab pulsar 
in SN~1054 (\citealt{Manchester1977,Bejger2003,Lyne2015}). The NS in this model 
received a kick velocity of 41\,km\,s$^{-1}$. However, the velocity is
still rising when we stopped monitoring the neutrino- (LESA-) induced 
component of the NS kick (reaching $\sim$25\,km\,s$^{-1}$ in our models),
and the final value of the total kick velocity might well be higher. 
Therefore, we consider this result to be
in the ballpark of the spatial velocity of the Crab pulsar, which 
is inferred to be around 160\,km\,s$^{-1}$ with rather big uncertainties
(\citealt{Hester2008,Kaplan2008}).
A possible spin-kick alignment of the Crab pulsar (see the detailed discussion
in \citealt{Kaplan2008}), however, cannot be explained by our model;
s9.0 yields a final angle of $\sim$100$^\circ$ between NS spin and kick.
Since the dominant mechanism for spinning up the NS is fallback,
it is very difficult to understand how such a late-time effect,
whose angular momentum is connected to stochastic asymmetries of the
fallback matter, could correlate with the NS kick direction, which is 
determined in the first seconds after the onset of the explosion
\citep[see also][for similar conclusions for explosions of more massive
progenitors]{Chan2020}. It is also not easy to imagine a scenario where either
rotation of the progenitor star (which is not taken into account in our
pre-collapse models) or the inclusion of the NS motion in 
the modelling (which we do not follow because of our use of an inner 
grid boundary for the long-time runs)
could lead to spin-kick alignment as a deterministic consequence of 
physical effects. We therefore hypothesize that the spin-kick alignment
of the Crab pulsar, if true and a relic of the SN explosion and not just 
a projection effect, is a purely incidental outcome. 

From all of the three progenitors, only s9.0 seems to have favorable 
properties to explain the explosion energy of SN~1054 as well as the
magnitude of the spin and kick of the Crab pulsar. Models e8.8 and z9.6
explode too symmetrically and possess too little fallback to yield
the short spin period. Moreover, in such symmetric explosions
the NS kick is strongly
dominated by a component associated with anisotropic neutrino 
emission (which our simulations tracked only in models z9.6 and s9.0).
Model s9.0 also demonstrates that very low explosion energies 
($\mathord{\sim}0.5\mathord{\mathord{\times}}10^{50}$\,erg in this case)
do not exclude sizable NS kick velocities when the explosion
occurs highly asymmetrically; the momentum asymmetry of the ejecta is 
$\alpha_\mathrm{ej}\mathord{\sim}10$\% after $\sim$3\,s in this model.

Therefore we reason that the observed kick and spin of the Crab pulsar 
may be most naturally explained by considerable asymmetries of the
explosion and fallback in a CCSN of a low-mass Fe-core progenitor.
It is also important to note that the considered progenitors of 
\znine and \snine are just two samples of this class of SN progenitors,
and there is a large variety of them, filling the gaps in between, and 
beyond, including more extreme hAGB-like models beyond \znine.
Therefore our study is just a starting point in exploring this most
interesting regime, which is not small in terms of its weight by the
stellar initial mass function (e.g., for the z-series it spans the 
range of $\sim$9.6--10.3\,M$_\odot$).

Nevertheless, alternative possibilities cannot be excluded on grounds
of our results, because our conclusions apply for the considered low-mass 
progenitor properties, which do not include rotation at the onset of core 
collapse. Rotation of the progenitor core can be relevant for the spin of 
the NS, in particular in explosions with little fallback or little angular 
momentum connected to the fallback \citep[for a recent discussion
of the many facets of angular momentum transport in massive-star evolution 
and possible implications for NS rotation, see][]{Ma+Fuller2019}. 
For example, the degenerate 
pre-collapse core may spin up considerably during contraction, and the 
compact ONeMg core near the Chandrasekhar mass limit might rotate rapidly,
if the initial angular momentum is (partially) conserved. This latter
requirement may be enabled by the fact that the core evolves very 
much independently from the envelope due to the very steep density gradient 
around the core-envelope interface of a super-AGB star. Thus, the angular 
momentum transport from the core to the extended envelope could be small.
Rotation could even increase the mass of the degenerate ONeMg core 
compared to the non-rotating case (``super-Chandra'' cores), with a mass
excess that depends on the degree of differential rotation~\citep{Uenishi+2003,Benvenuto+2015,Hachisu+2012}.

Another possibility could be a close binary progenitor scenario,
in which, for example, the formation and collapse of an ONeMg core 
might occur as a result of the merging of two white dwarfs in a 
certain common-envelope configuration formed during the close binary 
evolution~\citep{Nomoto1985}.  In such a case, the
disrupted white dwarf material would form a relatively dense envelope
around the collapsing ONeMg core and affect the properties of the
resulting NS. This scenario as well as the previous one might
account for the spin of the Crab pulsar, but both of them are likely
to share the problem with our explosion models of 
\onemg and \znine that the NS kicks stay too low.

If the NS progenitor was a member of a close binary 
system, as was speculated for the Crab pulsar~\citep{Tsygan1975},
the NS properties may have been affected by the binary nature.
In this case the observed NS velocity might originate from the breakup of the
binary when SN~1054 exploded~\citep{Blaauw1961} instead of an 
intrinsic SN kick. However, spin-kick alignment of the Crab pulsar
in such a scenario would require extreme fine tuning of the binary
evolution~\citep{Horvat+2018}.

\subsection{Ejecta composition, asymmetries, and mixing}

Our three 3D models exhibit considerable differences in their ejecta
morphology and long-time evolution. The degree of asymmetry and 
extent of radial mixing show a clear dependence on the steepness
of the density profile around the degenerate core. If the density
drops steeply, as in the ECSN-like models of e8.8 and z9.6, 
quick shock revival and
fast shock expansion favor buoyant plumes and bubbles from postshock 
convection to freeze in on relatively small scales when the shock
starts to accelerate outwards. Consequently, asymmetries in the ejecta
possess small scales, corresponding to higher-order spherical 
harmonics modes.
In contrast, in model s9.0, which has a significantly flatter 
density decline around the iron core and higher mass-accretion 
rate during the shock stagnation phase, the $\sim$300\,ms delay 
for the onset of the explosion permits the development of 
large-scale asymmetries with dominant dipolar and quadrupolar 
deformation modes.

Despite these differences, both iron-core models resemble each other
in the directional $Y_e$ variations imposed on the neutrino-heated
material by the neutrino-emission dipole of the LESA:
neutron-rich material is predominantly ejected in one hemisphere,
whereas an excess of proton-rich matter is expelled on the opposite
side. The width of the mass distribution, however, is broader
in the case of z9.6, with $Y_e$ reaching down to nearly 0.39 and 
up to $\sim$0.63, in contrast to $0.46\lesssim Y_e \lesssim 0.58$
in s9.0. 

In both the e8.8 and z9.6 progenitors, the SN shock expands very 
rapidly at the beginning, but is dramatically decelerated when it
travels through the hydrogen plus helium shells with their 
positive gradients of $\rho r^3$. This triggers the formation
of a strong reverse shock propagating inward as well as the 
creation of conditions for the growth of RT instability, which
distributes the neutrino-processed material including freshly 
nucleosynthesized radioactive species like $^{56}$Ni in stretched 
fingers and plumes within an extended spatial volume. Nevertheless,
very little of this material gets mixed into the 
hydrogen envelope, and the corresponding distribution remains
narrow in mass space, stretched out only over the innermost 
$\sim$0.1\,M$_\odot$ of the ejecta in the case of e8.8 and
$\sim$0.6\,M$_\odot$ in z9.6. In both cases the maximum 
$^{56}$Ni velocities are around 500\,km\,s$^{-1}$. 

In contrast, in model s9.0 the SN shock propagates rather steadily,
yet less rapidly (though still with $\mathord{\sim}10^4$\,km\,s$^{-1}$),
during the first 150\,s, but it is also strongly decelerated after
entering the hydrogen envelope. Because the shock is highly deformed
in this explosion and the postshock ejecta are extremely asymmetric from
the beginning, these initial ejecta asymmetries trigger the rapid and
powerful growth of Richtmyer-Meshkov instability and RT instability 
at the He/H interface. Big plumes of nickel-rich matter, originating
from the biggest bubbles at the time of shock revival, shape the 
large-scale asymmetry of the SN blast by
penetrating deep into the hydrogen envelope. In fact,
the most extended of these plumes is pushed by buoyancy forces so
strongly that it catches up with the continuously decelerated 
SN shock and creates a massive deformation of the shock front, thus
overtaking the average shock radius. This effect persists until the
shock breaks out from the stellar surface, which therefore happens
highly asymmetrically. The first, biggest plume pushes the shock
through the surface of the progenitor after roughly 2.1 days, 
whereas the main sphere of the shock reaches the stellar surface 
considerably later after 2.8 days.
Since radioactive nickel is mixed through the entire hydrogen 
envelope in this model, it expands with velocities up to 
1400\,km\,s$^{-1}$ after the breakout from the star (and might
be even further accelerated due to radioactive decay heating).

The bubble-driven, asymmetric breakout of the SN shock in our
9\,M$_\odot$ model will have ramifications for theoretical
studies of this evolution phase and for the interpretation
of corresponding observations (e.g., \citealt{Nakar2010,Kozyreva2020}). Since at
that time the giant plume contains 
about 10--20\% of the radioactive material produced
in the deepest layers of the SN, the one-sided expansion of such
a feature will cause a strongly direction-dependent emission of
gamma-rays and X-rays, which will occur particularly early in the 
hemisphere of the bubble breakout \citep[for recent investigations
of high-energy radiation emission based on asymmetric 3D explosion 
models, see][]{Alp2018,Alp2019,Orlando2019,Jerkstrand2020}.
The deep mixing of iron-group nuclei and radioactive species from
the core through the whole hydrogen shell also plays
an important role in shaping the Type-II SN light 
curve during the luminosity peak \citep{Utrobin2017}.
Moreover, the large-scale deformation of the stellar envelope by 
extended, wide-angle plumes enriched with heavy elements might 
also offer an explanation why some Type-IIP SNe show an unusually
early rise of the polarization {\it before} the tail phase is entered,
and thus {\it before} the helium core is exposed by the transparency of
the hydrogen envelope \citep{Nagao2019}. Last but not least,
an outward pushing RT plume with its long-stretched
stem might be a mechanism to create the faint protrusion
extending out from the northern rim of the visible Crab Nebula,
which is often called northern ejecta `jet' 
(e.g., \citealt{Gull1982,Blandford1983,Davidson1985,Fesen1993,Black2015}). 
Such an origin would naturally
allow for understanding the fact that the jet's sharp western limb
and  its blueshifted and redshifted sides seem to be radially aligned
with the center of expansion of the remnant \citep{Black2015}.
The hollow appearance of the jet in [OIII] line emission,
its remarkably empty, elliptical shape, and its growing diameter 
with larger distance from the remnant center, all of which define a 
funnel-like structure, could be naturally explained in such a picture.
The jet walls would be expected to contain oxygen swept up when the
RT plume penetrates the oxygen shell of the progenitor, whereas the 
jet's interior should contain a reduced oxygen fraction but enhanced
content of iron-group elements and possibly also silicon.

An alternative possibility to create Crab's northern ejecta funnel might
be a jet produced by a fallback disk. The specific angular momentum
associated with the $5\,\mathord{\times}\,10^{-3}$\,M$_\odot$ of fallback matter (about $2.5\,\mathord{\times}\,10^{16}$\,cm$^2$\,s$^{-1}$ for a total
of $\sim$\,$2.5\,\mathord{\times}\,10^{47}$\,g\,cm$^2$\,s$^{-1}$) is large enough
to keep the matter on Keplerian orbits at a radius of 30--35\,km around
the NS (Noam Soker, 2020, private communication). If an accretion 
disk forms, a part of the disk matter might be expelled again in 
jets. With an escape velocity around $10^{10}$\,cm\,s$^{-1}$, even only
10\% of the fallback mass would carry a considerable amount of energy
\citep{Soker2020}
with sufficient excess energy to punch a funnel into the SN ejecta. It
is interesting to note that the disk and jet would be expected at a 
large angle relative to the NS kick direction, because the angular 
momentum and kick vectors are nearly perpendicular (see Table~\ref{tab:neutron star final}). This resembles the geometry in the
Crab case, where the northern jet has a large angle relative to the 
NS's projected direction of motion.

A more detailed analysis of the evolving plume structure, based 
preferably also on higher-resolution 3D simulations, 
and more detailed studies of the
fallback accretion are needed to consolidate such
appealing scenarios. Another interesting extension of our work concerns
the nucleosynthetic post-processing of the LESA-affected neutron-rich
and proton-rich components of the neutrino-heated ejecta. Since our
2D and 3D results of the mass distributions as functions of $Y_e$ 
exhibit, overall, fairly close similarities, however, we expect basic 
confirmation of the trends already seen in a recent study of element 
formation in the ejecta of 2D simulations for low-mass stars including
the e8.8 and z9.6 progenitors \citep{Wanajo2018}. Finally, it would 
be desirable to repeat our 3D explosion modeling with 3D pre-collapse 
conditions originating from the latest phases of convective oxygen and 
silicon shell burning (e.g., 
\citealt{Arnett2011,Couch2015,Mueller2016,Mueller2016b,Yoshida2019,Yadav2019}).
The large-scale density and velocity perturbations created by the convective
shell burning prior to collapse are more realistic seeds for the growth of
postshock instabilities than the artificial, small-amplitude,
stochastic cell-by-cell seed perturbations that we imposed on the 
spherical progenitor models in our present calculations. 
The larger physical perturbations of
3D initial models might lead to earlier explosions, in particular of
the s9.0 progenitor, and might also affect the shock and explosion
asymmetry right after shock revival 
(see \citealt{Couch2013,Mueller2015b,Mueller2017,Kazeroni2019}).

\section*{Acknowledgements}
Valuable comments by Alexandra Kozyreva on the manuscript and by
Noam Soker and Thomas Tauris on the arXiv posting are acknowledged.
G.S.\ and M.G.\ thank Margarita Petkova of the Computational Center for
Particle and Astrophysics (C2PAP) for assistance in developing a new 
parallel version of the \prom code. G.S.\ also wants to thank Naveen Yadav 
and Ninoy Rahman for fruitful discussions 
and comments on the paper and also thanks Anders Jerkstrand, Oliver Just,
and Ricard Ardevol-Pulpillo for useful hints and guidance at an early stage 
of the project.
H.-T.J.\ is very grateful to Rob Fesen for sharing his deep insights into the 
observational properties of the Crab SN remnant.
K.N.\ would like to thank Sam Jones and Raphael Hirschi for the
collaborative work on the evolution of super-AGB stars.
At Garching, funding by the
European Research Council through Grant ERC-AdG No.~341157-COCO2CASA
and by the Deutsche Forschungsgemeinschaft (DFG, German Research Foundation)
through Sonderforschungsbereich (Collaborative Research Centre)
SFB-1258 ``Neutrinos and Dark Matter in Astro- and Particle Physics
(NDM)'' and under Germany's Excellence Strategy through
Cluster of Excellence ORIGINS (EXC-2094)---390783311 is acknowledged. 
Computer resources for this project have been provided by the Max Planck 
Computing and Data Facility (MPCDF) on the HPC systems Cobra and Draco, 
and by the Leibniz Supercomputing Centre (LRZ) under LRZ project ID: pn69ho,
GAUSS Call 13 project ID: pr48ra, and GAUSS Call 15 project ID: pr74de.
S.-C.L.\ acknowledges support from HST-AR-15021.001-A.
K.N.\ received support by the World Premier International
Research Center Initiative (WPI Initiative), MEXT, Japan, and JSPS
KAKENHI Grant Numbers JP17K05382 and JP20K04024.
A.H.\ was supported, in part, by JINA-CEE through US NSF grant PHY-1430152; by
the Australian Research Council Centre of Excellence for Gravitational Wave
Discovery (OzGrav), through project number CE170100004; by the Australian
Research Council Centre of Excellence for All Sky Astrophysics in 3
Dimensions (ASTRO~3D), through project number CE170100013; by a grant from
Science and Technology Commission of Shanghai Municipality (Grants
No.~16DZ2260200); and by the National Natural Science Foundation of China
(Grants No.~11655002).

\textit{Software:} \vertexprom \citep{Fryxell1989,Rampp2002,Buras2006}, \prom \citep{Kifonidis2003,Scheck2006,Arcones2007,Ertl2016}, Numpy and SciPy \citep{Jones2001}, IPython \citep{Perez2007}, Matplotlib \citep{Hunter2007}, VisIt \citep{Childs2012}.

\textit{Data availability:} The data underlying this article will be shared on reasonable request to the corresponding author.

\label{lastpage}
\bibliographystyle{mnras}
\bibliography{bibtex}       

\newpage
\appendix
\section{PNS cooling model and inner boundary condition in Prometheus-HotB}
\label{Appendix:prom inner boundary}

As stated in Section~\ref{sec:Collapse and post-bounce setup in prom}, we use the modeling approach of \cite{Ugliano2012}, \cite{Sukhbold2016}, and \cite{Ertl2020} in simulations with \prom. 
The central 1.1\,\solm of the PNS are excised from the computational domain and replaced by an inner grid boundary at $R_{\mathrm{ib}}$. The
shrinking of the cooling and deleptonizing PNS is mimicked by the contraction of the inner grid boundary,
whose time dependence (see also \citealt{Arcones2007}) is given by
\begin{equation}
    R_{\text{ib}}(t) = R_{\text{ib,f}} + (R_{\text{ib,i}} - R_{\text{ib,f}})\, \exp\left(-\frac{t}{t_0}\right)\,,
\end{equation}
where $R_{\text{ib,f}}$ is the final radius, $R_{\text{ib,i}}$ the initial radius and $t_0$ the contraction timescale. $R_{\text{ib,f}}$ and $t_0$ are two representatives of our set of free parameters and are chosen to mimic the behavior of the PNS contraction found in more sophisticated simulations of PNS cooling \citep[see][for comparisons with such results]{Scheck2006,Sukhbold2016}.

As detailed in \cite{Ugliano2012}, the PNS core of mass $M_{\mathrm{c}}\,\mathord{=}\,1.1\,\solm$ is described by an analytic one-zone model under the constraints of energy conservation and the virial theorem including the effects associated with the growing pressure of the accretion layer. The accumulation of mass around the PNS core is followed by the hydrodynamic simulations. The one-zone model provides the time-dependent total neutrino luminosity that leaves the excised core and is imposed as boundary condition (split up into contributions of each of the neutrino species) at the bottom of the computational domain at $R_{\mathrm{ib}}$. It is given by 
\begin{equation}
\begin{split}
\label{eqn:lib}
    L_{\mathrm{\nu,tot}} = & - \frac{2}{5} \frac{3\Gamma - 4}{3(\Gamma - 1)} \frac{GM^2_{\mathrm{c}} \dot{R}_{\mathrm{c}}}{R_{\mathrm{c}}^2} \\
            &  - \frac{3\Gamma - 4} {3(\Gamma - 1)} \frac{a\,GM_{\mathrm{c}} m_{\mathrm{acc}}\dot{R}_\mathrm{c}}{R^2_{\mathrm{c}}}  -\frac{a\,GM_{\mathrm{c}} \dot{m}_{\mathrm{acc}}}{3(\Gamma - 1)R_{\mathrm{c}}}.
\end{split}
\end{equation}
Here, $\Gamma\,\mathord{=}\,3$ is the adiabatic index of the PNS core (assumed to be homogeneous), $G$ is the
gravitational constant, $M_{\mathrm{c}}$ is the core mass, $a$ is a parameter which characterizes the
accretion luminosity, and $m_{\mathrm{acc}}(t)$ is the mass contained between the radius of the inner grid boundary, $R_\mathrm{ib}(t)$, and
the radius $r_0$ at which the density falls below $\rho_0\,\mathord{=}\,10^{10}\gcc$. We define
$\dot{m}_{\mathrm{acc}}(t)\,\mathord{=}\,-4\pi r_0^2 v_0 \rho_0$, where $v_0$ is the fluid velocity at the position $r_0$.
In multi-dimensional simulations, $\dot{m}_{\mathrm{acc}}$ is determined from angle-averaged values.
The time dependence of the core radius $R_\mathrm{c}$ in Equation~(\ref{eqn:lib}) is prescribed by
\begin{equation}
    R_{\mathrm{c}}(t) =R_{\text{c,f}} + (R_{\mathrm{c,i}} - R_{\mathrm{c,f}}) \left(1+\frac{t}{t_{\text{L}}}  \right)^p,
\end{equation}
where $p\,\mathord{<}\,0$ is another parameter. We always set the characteristic time scale  $t_\mathrm{L}\,\mathord{=}\,1$\,s and use initially $R_{\mathrm{c,i}}\,\mathord{=}\,R_{\mathrm{ib,i}}$. Note that in general the PNS core radius and the radius of the inner grid boundary can differ during parts of the evolution.
The quintuple of $p$, $R_{\mathrm{ib,f}}$, $a$, $R_{\mathrm{c,f}}$, and $t_0$ constitutes our set of five parameters to approximate the physics of the time evolution of the PNS and to enable SN explosions with chosen energy.
The calibration of these parameters was done using the method described in \citet{Ertl2016}.

\section{Correction to Neutrino-Nucleon Scattering in Prometheus-HotB}
\label{Appendix:Neutrino}

Because of numerical issues in the neutrino transport module of the previous version of \prom, long-time simulations ($\tpb\,\mathord{>}\,3\,\s$) including our neutrino transport approximation showed spurious oscillations in the energy source terms $Q_{\nu}$. In particular, cases where no or only a very late explosion was observed were affected. 
These undesired effects were caused by an improper treatment of the energy source terms connected to our non-conservative description of neutrino-nucleon scattering.
\cite{Scheck2006} coined the net energy exchange rate through neutrino-nucleon scattering, following \citet{Tubbs1979} and \citet{Janka1991PhDT}, in a closed form:
\begin{equation}\label{equ:nns}
\begin{aligned}
    Q_{\mathrm{\nu N}} = \, & \frac{1}{4} \frac{\sigma_0 c}{(m_{\mathrm{e}}c^2)^2} \mathcal{C}_N \mathcal{E}_{\mathrm{N}} \frac{n_{\mathrm{N}}}{m_{\mathrm{N}}c^2}
    \{\langle \epsilon^4 \rangle - 6 T\langle \epsilon^3 \rangle  \} \\
    & \times  \frac{L_{e,\nu}}{4\pi r^2 f_{\nu}c\langle \epsilon \rangle}, \\
\end{aligned}
\end{equation}
which is their equation (D.68). Here, the $\langle \epsilon^n \rangle$ 
($n$ being a natural number) are spectal averages of powers of the neutrino energy,
and $T$ is the gas temperature in MeV.
When the temperature exceeds $\langle \epsilon^4 \rangle / 6 \langle \epsilon^3 \rangle $, the net rate $Q_{\mathrm{\nu N}}$ is negative and neutrinos {\it receive} energy from the stellar medium
through scattering reactions with nucleons. 
Since the scattering term was implemented in this closed form and could change its sign, it appeared either as an opacity producing
contribution to the neutrino-energy absorption rate $Q^-$ (when $Q_{\mathrm{\nu N}}$ was positive)
or as an energy source rate for neutrinos, $Q^+$ (when $Q_{\mathrm{\nu N}}$ was negative), in the analytic integral of the transport equation employed by \citet{Scheck2006}. Because of the associated big changes and alternating signs, the tight coupling of fluxes and source terms caused large-amplitude oscillations in the transport solution for the neutrino fluxes, induced by large variations of the heating and cooling source terms $Q^+$ and $Q^-$. This led to unphysical heating of the matter, creating additional luminosity and preventing further cooling of the PNS.

A simple solution to this problem is to split Equation~\eqref{equ:nns} into two separate source-/sink terms,
a neutrino-energy emission rate represented by the temperature-dependent term,
\begin{equation}\label{equ:nns1}
    Q^{\mathrm{em}}_{\nu \mathrm{N}} = \frac{3}{2}\frac{\sigma_0 c}{(m_{\mathrm{e}}c^2)^2} \mathcal{C}_{\mathrm{N}} \mathcal{E}_{\mathrm{N}} \frac{n_{\mathrm{N}}}{m_{\mathrm{N}}c^2}
    T \langle \epsilon^3 \rangle \frac{L_{e,\nu}}{4\pi r^2 f_{\nu}c\langle \epsilon \rangle}\,,
\end{equation}
and a neutrino-energy absorption rate,
\begin{equation}\label{equ:nns2}
    Q^{\mathrm{abs}}_{\nu \mathrm{N}} = \frac{1}{4} \frac{\sigma_0 c}{(m_{\mathrm{e}}c^2)^2} \mathcal{C}_{\mathrm{N}} \mathcal{E}_{\mathrm{N}} \frac{n_{\mathrm{N}}}{m_{\mathrm{N}}c^2}
    \langle \epsilon^4 \rangle \frac{L_{e,\nu}}{4\pi r^2 f_{\nu}c\langle \epsilon \rangle}\,.
\end{equation}
The latter rate corresponds to an exponential attenuation factor of the luminosity connected with the
neutrino absorption opacity. This factor also accounts for the re-absorption of neutrinos that are locally 
produced in each computational cell, and thus it can damp source-rate variations very efficiently.

\section{Multipole decomposition of the neutrino lepton-number flux}
\label{app:neutrinomultipoles}

For our discussion, we decompose the electron-neutrino 
lepton-number flux density $F^\mathrm{n}_{\nu_e}\,\mathord{-}\,F^\mathrm{n}_{\bar{\nu}_e}$ into spherical harmonics. 
$F^\mathrm{n}_{\nu_e}$ and $F^\mathrm{n}_{\bar{\nu}_e}$
denote the individual radial number flux densities of $\nu_e$ 
and $\bar{\nu}_e$,
respectively.  The real spherical harmonics are defined as
\begin{equation}
    Y_\ell^m(\theta,\phi) = \begin{cases}
    \sqrt{2} N_\ell^m \, P_\ell^m(\cos\theta) \, \cos (m\phi) & \quad m>0\,,\\
    N_\ell^0 \, P_\ell^0(\cos\theta) & \quad m=0\,,\\
    \sqrt{2} N_\ell^{|m|} \, P_\ell^{|m|}(\cos\theta) \, \sin (|m|\phi) & \quad m<0\,,
\end{cases}
\end{equation}
with normalization factors
\begin{equation}
    N_\ell^m = \sqrt{\frac{2\ell+1}{4 \pi} \, \frac{(\ell-m)!}{(\ell+m)!}}
\end{equation}
and associated Legendre polynomials $P_\ell^m(\cos\theta)$.  
The coefficients for the multipole analysis are
\begin{equation}
    c_\ell^m = \sqrt{\frac{4 \pi}{2\ell + 1}} \int \mathrm{d} \Omega \, r^2
    \left[ F^\mathrm{n}_{\nu_e} (\theta,\phi) - F^\mathrm{n}_{\bar{\nu}_e}  (\theta,\phi) \right]
    Y_\ell^m(\theta,\phi)\,.
\end{equation}
In our chosen normalization, the reconstruction reads
\begin{equation}
    F^\mathrm{n}_{\nu_e}(\theta,\phi)-F^\mathrm{n}_{\bar{\nu}_e}(\theta,\phi) =
    \frac{1}{r^2} \sum_{\ell=0}^\infty \sqrt{\frac{2\ell + 1}{4 \pi}}
    \sum_{m=-\ell}^\ell c_\ell^m Y_\ell^m(\theta,\phi)\,.
\label{eq:reconstruction}
\end{equation}
From the coefficients $c_\ell^m$, the multipole moments can be calculated 
according to
\begin{equation}
    A_\ell = \sqrt{\sum_{m=-\ell}^\ell \left( c_\ell^m \right)^2}\,.
\label{eq:LESAmultipole}
\end{equation}

\cite{Tamborra2014} used a different definition for the monopole and
dipole components of the lepton-number flux density. Their $A_\mathrm{Monopole}$
is equal to our $A_0$, whereas $A_\mathrm{Dipole}\,\mathord{=}\,3 A_1$. This can be easily
seen if we consider the lepton-number flux density to consist only of monopol
and dipole components. If we also align the coordinate system into the dipole
direction, the reconstruction of Equation~(\ref{eq:reconstruction}) reads
\begin{equation}
    F^\mathrm{n}_{\nu_e}(\theta,\phi)-F^\mathrm{n}_{\bar{\nu}_e}(\theta,\phi) 
    = \frac{1}{4\pi	r^2} c^0_0 + \frac{3}{4 \pi r^2} c^0_1 \cos \theta\,.
\end{equation}
Note that $c^{-1}_1$ and $c^1_1$ vanish in this orientation of the coordinate
system. Expressed in terms of the multipole moments, the latter expression
becomes
\begin{equation}
    F^\mathrm{n}_{\nu_e}(\theta,\phi)-F^\mathrm{n}_{\bar{\nu}_e}(\theta,\phi) 
    = \frac{1}{4\pi r^2} A_0 + \frac{3}{4 \pi r^2} A_1 \cos \theta\,.
\end{equation}
According to \cite{Tamborra2014}, $F^\mathrm{n}_{\nu_e} (\theta,\phi) 
- F^\mathrm{n}_{\bar{\nu}_e} (\theta,\phi)$ is proportional to 
$A_\mathrm{Monopole}\,\mathord{+}\,A_\mathrm{Dipole}
\cos \theta$, if higher-order multipoles do not contribute. We can directly
infer $A_\mathrm{Dipole}\,\mathord{=}\,3 A_1$.

\section{Neutron star kick and spin}
\label{Appendix:Neutron Star Properties}

Two mechanisms are considered that can lead to a recoil kick of 
the newly formed NS during the SN blast. 
First, asphericities developing during the explosion exert hydrodynamic and gravitational forces on the PNS. Both accelerate the PNS in the direction opposite to the strongest direction of the explosion, compatible with global momentum conservation. Since the gravitational effects become dominant during the long-time evolution, this kick mechanism was termed ``gravitational tug-boat mechanism'' \citep{Wongwathanarat2013}. For a thorough discussion of its physics details, the reader is referred to \citet{Scheck2006}, \citet{Wongwathanarat2013}, \citet{Janka2017}, \citet{Gessner2018}, and \citet{Mueller2019}.
Second, anisotropic emission of neutrinos, specifically the anisotropic neutrino energy flux density, $F_{\mathrm{\nu}}^{\mathrm{e}}(\theta,\phi)$, radiated from the surface of the PNS ($\theta$ and $\phi$ are the direction angles in a polar grid of the star), exerts a force onto the PNS in the direction opposite to the most intense neutrino emission.

Using the momentum conservation equation, the hydrodynamic PNS kick can be simply estimated as
\begin{equation}
    \pmb{P}_{\mathrm{NS}}^{\mathrm{hyd}} = \pmb{v}_{\mathrm{NS}}^{\mathrm{hyd}}\mns = -\pmb{P}_{\mathrm{gas}}\,,
\end{equation}
where $\pmb{v}_{\mathrm{NS}}^{\mathrm{hyd}}$ is the hydrodynamic kick velocity, $M_\mathrm{{b}}$ is the baryonic (PNS) mass contained inside the radius $R_{\mathrm{NS}}$ where the angle-averaged density drops below $10^{11}\, \gcc$, and $\pmb{P}_{\mathrm{gas}}\,\mathord{=}\int_{_{R_{\mathrm{gain}}}}^{R_\mathrm{sh}} \rho\pmb{v}\ud V$ is the total linear momentum of the ejecta between the gain radius, $R_\mathrm{gain}$, and the SN shock, $R_\mathrm{sh}$.
The momentum transfer by escaping neutrinos is given by
\begin{equation}
\label{equ:nukick}
    \dot{\pmb{P}}_{\nu} (t) = \oint_{r = R_\mathrm{free}}
    \frac{F_{\nu}^{\mathrm{e}}}{c}\ \textit{\textbf{e}}_r \, \ud S
    = - \dot{\pmb{P}}_\mathrm{NS}^{\,\nu} (t) \, ,
\end{equation}
where $F_{\nu}^{\mathrm{e}}$ is the neutrino energy flux summed over all species, $c$ the speed of light, $\pmb{e}_r$ the unit vector in radial
direction, and $R_\mathrm{free}$ the radius of evaluation (typically about 400\,km), exterior to which neutrinos stream essentially freely and a tiny fraction of still interacting neutrinos can be ignored. Using $\textit{\textbf{P}}_\nu (t)\,\mathord{=}\int_0^{\, t} \dot{\textit{\textbf{P}}}_{\nu} 
(t')\mathrm{d}t'$, the total kick velocity of the PNS at any time $t$ can be calculated as
\begin{equation}
    \pmb{v}_{\mathrm{NS}}^{\mathrm{tot}}(t) = \frac{\pmb{P}_{\mathrm{NS}}^{\mathrm{hyd}}(t) + \pmb{P}_\mathrm{NS}^{\,\nu}(t)}{\mns(t)} = - \frac{\pmb{P}_{\mathrm{gas}}(t) + \pmb{P}_{\mathrm{\nu}}(t)}{\mns(t)}\,.
\label{equ:momentum_kick}
\end{equation}

One can characterize the asymmetry of the ejecta and neutrino emission by means of anisotropy parameters. The hydrodynamic parameter reads 
\begin{equation}
    \alpha_{\mathrm{ej}} = \frac{|\pmb{P}_{\mathrm{gas}}|}{P_{\mathrm{ej}}}\,,
\end{equation}
where 
\begin{equation}
    P_{\mathrm{ej}}=\int_{R_{\mathrm{gain}}}^{R_{\mathrm{sh}}}\rho |\pmb{v}| \ud V
\end{equation}
is the total momentum stored in the ejecta, which becomes equal to the total radial momentum when the ejecta expand essentially radially.
For the neutrino anisotropy parameter we use 
the total energy loss rate in neutrinos, which is given by
\begin{equation}
\label{nuerg}
    \dot{E}_\nu (t) = \oint_{r = R_\mathrm{free}} F_{\nu}^{\mathrm{e}}\, \ud S\,.
\end{equation}
The time-dependent total flux of neutrino momentum through the sphere at 
$R_\mathrm{free}$ is given by $c^{-1}\dot{E}_\nu$
allowing us to define the
instantaneous neutrino emission anisotropy parameter as
\begin{equation}
\label{nuanis}
    {\widetilde\alpha}_\nu (t) = 
    c\,\frac{|\dot{\textit{\textbf{P}}}_{\nu}(t)|}{\dot{E}_\nu(t)}\,.
\end{equation}
In analogy to the linear ejecta momentum at a time $t$, the momentum radiated by neutrinos until time $t$ is
\begin{equation}
\label{pnu}
    \frac{1}{c}\, E_\nu (t)
    = \frac{1}{c}\int_0^{\, t} \oint_{r = R_\mathrm{free}}F_{\nu}^{\mathrm{e}}(t')\, \ud S \,\mathrm{d}t'\,
\end{equation}
so that the time-integrated neutrino emission asymmetry at time $t$ becomes
\begin{equation}
\label{anu}
    \alpha_\nu (t) = c\,\frac{|{\textit{\textbf{P}}}_{\nu}(t)|}{E_\nu(t)}\,.
\end{equation}

In addition, we compute the PNS spin by integrating the flux of angular momentum through a sphere of radius $r_0$ around the origin,
\begin{equation}
\label{equ:jns}
    \frac{\mathrm{d}\pmb{J}_{\mathrm{NS}}}{\mathrm{dt}} = - r_0^2\int_{4\pi} 
    \rho v_r \,\pmb{r}\times\pmb{v} \,\mathrm{d}\Omega\,,
\end{equation}
where $\rho$ is the matter density, $v_r$ the radial velocity, and $\pmb{r}$ and $\pmb{v}$ 
are the position and velocity vectors, respectively. During the early post-bounce
evolution with neutrino treatment, $r_0\,\mathord{=}\,100\,\km$, whereas we use 
$r_0\,\mathord{=}\,R_\mathrm{ib}$ during the long-time simulations.

In order to estimate the spin period of the PNS $P_{\mathrm{NS}}\,\mathord{=}\,2\pi I_{\mathrm{NS}}/|J_{\mathrm{NS}}|$, with $I_{\mathrm{NS}}$ being the moment of inertia of the PNS, we use the approximation by \citet{Lattimer2005},
\begin{eqnarray}
    I_{\mathrm{NS}} &=& 0.237M_{\mathrm{g}}R_{\text{NS}}^2\left[1 + 4.2 A  + 90 A^4\right],\\
    A &=&  M_{\text{g},\solm}R_{\text{NS},\km}^{-1}\,,\nonumber
\label{equ: ins}
\end{eqnarray}
where $M_{\text{g},\solm}$ is the gravitational mass of the PNS in units of $\solm$, and $R_{\text{NS},\km}$ is the radius of the PNS in units of $\km$.
In the above equation, the gravitational mass $M_{\mathrm{g}}$ can be estimated from the baryonic mass $M_{\mathrm{b}}$  as \citep{Lattimer2000}
\begin{equation}
    M_{\mathrm{g}} = \mns - \frac{0.6 \beta}{1-0.5\beta}  M_{\mathrm{g}}\,, 
\end{equation}
where $\beta\,\mathord{=}\,GM_{\mathrm{g}} / R_{\mathrm{NS}}c^2 $.

\section{Simplified neutrino treatment in Vertex-Prometheus}
\label{appendix:scaling relations}

The computational demands of the full-fledged neutrino transport in our \vertexprom code are a severe obstacle for continuing simulations well beyond post-bounce times of $\mathord{\sim}\,0.5\,\mathrm{s}$. In order to follow explosions to much later times including neutrino effects, we newly implemented a simplified neutrino treatment based on a light-bulb-like scheme for neutrino emission and absorption. Within this framework, we do not solve the neutrino transport equations (i.e., we switch off the transport module \textsc{Vertex}). Instead, we obtain local neutrino source terms for application in the hydrodynamics module \textsc{Prometheus} from an analytical scaling and transformation of the neutrino source terms that are present in a model at the end of the \vertexprom simulation. This moment in time is chosen to be in an evolutionary phase where the model is well on the way to a successful explosion. 

Doing so, we aim at capturing the most crucial neutrino effects after the onset of the explosion in a computationally efficient way, while keeping numerical transients at a minimum and thus ensuring a seamless continuation of our simulations after switching off the \textsc{Vertex} transport.
To further increase the time step, we remap the 3D (hydrodynamical) PNS data within 10~km to 1D and slightly reduce the radial resolution in the PNS interior at the beginning of our simulations with simplified neutrino treatment.
Moreover, adding more radial zones, the outer grid boundary is shifted from initially 10.000~km to $\sim$\,70.000~km to be able to follow the outward propagation of the shock through the exploding star at later times ($t\,\mathord{\gtrsim}\,1\,\s$ after bounce).
In the following, we will elaborate on our new approximate neutrino treatment.

\subsection{Source terms for energy and lepton number}

In our new simplified approach, we apply the following expressions for the net neutrino cooling and heating rates per volume (i.e., the energy source terms):
\begin{align}
    Q_\mathrm{erg}^{-}(\vect{r}) &= Q_\mathrm{erg}^0(x) \cdot \left(\frac{\rho(\vect{r})}{\rho_0(r_0)}\right)^{\! a}\left(\frac{T(\vect{r})}{T_0(r_0)}\right)^{\! 6}\,,\label{eq:q_erg-}\\
    Q_\mathrm{erg}^{+}(\vect{r}) &= Q_\mathrm{erg}^0(x) \cdot \left(\frac{\rho(\vect{r})}{\rho_0(r_0)}\right)
    \left(\frac{R_\mathrm{gain}(t_0)}{R_\mathrm{gain}(t)}\right)^{\! 2}
    \cdot f_L\,f_E\,,\label{eq:q_erg+}
\end{align}
where Equation~(\ref{eq:q_erg-}) is employed in regions where $Q_\mathrm{erg}^0(x)\,\mathord{<}\,0$ (i.e., PNS cooling), while Equation~(\ref{eq:q_erg+}) is applied where $Q_\mathrm{erg}^0(x)\,\mathord{>}\,0$ (i.e., gain-layer heating).\footnote{\label{fn:heating}We use Equation~(\ref{eq:q_erg+}) only outside of the PNS, i.e. at radii $r\,\mathord{>}\,R_\mathrm{NS}$, while we do not allow for heating inside the PNS. Analogously, we do not allow for an increase of the electron fraction by neutrino sources via Equation~(\ref{eq:q_lep+}) in the interior of the PNS.} The quantities with superscript or subscript ``0'' are the angle-averaged net heating and cooling rates, $Q_\mathrm{erg}^0$, the angle-averaged density, $\rho_0$, and the angle-averaged temperature, $T_0$, at time $t_0$ when we switch from the transport calculation with \textsc{Vertex} at $t\,\mathord{\leqslant}\,t_0$ to our simplified neutrino treatment at $t\,\mathord{>}\,t_0$.\footnote{The radial profiles of $Q_\mathrm{erg}^0$, $\rho_0$, $T_0$, and $Y_e^0$ of Equation~(\ref{eq:q_lep-}) are smoothed by time-averaging over the last few ms of the calculations with detailed transport.}
To adjust the radial profile of heating and cooling to the contraction of the gain radius, which roughly follows the contraction of the PNS radius, we define the variable $x\,\mathord{\equiv}\,r_0/R_\mathrm{gain}(t_0)\,\mathord{=}\,r(t)/R_\mathrm{gain}(t)$, which connects the radial coordinate $r\,\mathord{=}\,r(t)$ with $r_0\,\mathord{=}\,r(t_0)$. The factor $[R_\mathrm{gain}(t_0)/R_\mathrm{gain}(t)]^2$ results from the transformation responsible for the inward shift of the heating profile and ensures that the net heating rate drops like $r^{-2}$ at large radii.

The functional ansatz of Equations~(\ref{eq:q_erg-}) and (\ref{eq:q_erg+}) fulfills the requirement of a continuous and smooth transition of cooling
and heating before and after $t_0$.
The scaling of the heating and cooling rates is motivated by the
rough scaling of the $\nu_e$ and $\bar\nu_e$ absorption rates with the density of a gas of free nucleons, and the scaling of
(nondegenerate) electron and positron capture rates on nucleons with the temperature \citep[see, e.g.,][]{Janka2001}. The parameter $a$ in Equation~(\ref{eq:q_erg-}) is chosen to be of order unity, depending on the model under consideration, and adjusted to best reproduce the PNS contraction behaviour obtained in a corresponding 1D PNS-cooling simulation with mixing-length convection. For the cases considered in this work, we found $a\,\mathord{=}\,2$ to be a reasonable choice. The factors $f_L$ and $f_E$ in Equation~(\ref{eq:q_erg+}) contain the dependencies of the neutrino heating rate on the radiated
luminosities $L_{\nu_e}$ and $L_{\bar\nu_e}$ (for the energy transferred by $\nu_e$ and $\bar\nu_e$ absorption per unit of time) and on the mean squared neutrino
energy (for the basic energy dependence of the absorption cross section), which we replace by the squared arithmetic average of the mean energies $E_{\nu_e}$ and $E_{\bar\nu_e}$ of $\nu_e$ and $\bar\nu_e$ leaving the PNS. In the spirit of our approach explained above, we therefore use the following functional scaling prescriptions:
\begin{equation}
    f_L = \frac{L_{\nu_e}(t) + L_{\bar{\nu}_e}(t)}{L_{\nu_e}(t_0) + L_{\bar{\nu}_e}(t_0)}\,, 
\label{eq:f_L}
\end{equation}
which holds for core-luminosity-dominated conditions, and
\begin{equation}
    f_E = \left(\frac{E_{\nu_e}(t) + E_{\bar{\nu}_e}(t)}{E_{\nu_e}(t_0) +
	E_{\bar{\nu}_e}(t_0)}\right)^{\! 2} \left(\frac{M_\mathrm{b}(t)}{M_\mathrm{b}(t_0)}\right)^{\! 2}\,.
\label{eq:f_E}
\end{equation}
Guided by the findings of \cite{Mueller2014}, $f_E$ is assumed to scale with the square of the (baryonic) PNS mass, $M_\mathrm{b}$. The approximate time dependence of the $\nu_e$ and $\bar{\nu}_e$ luminosities and of the corresponding mean energies is adopted from the PNS-cooling behavior in a 1D explosion simulation of the \znine progenitor with \vertexprom, which results in a NS of similar mass ($M_\mathrm{b}\,\mathord{=}\,1.36\,\mathrm{M_\odot}$, $M_\mathrm{g}\,\mathord{=}\,1.26\,\mathrm{M_\odot}$) as obtained in our 3D simulations of the \znine and \snine models (see Table~\ref{tab:neutron star}). The mass scaling in Equation~(\ref{eq:f_E}) shall account for a possible difference in the evolution of the PNS mass between the 1D and 3D models.

In a manner analogous to Equations~(\ref{eq:q_erg-}) and (\ref{eq:q_erg+}), we apply the following source terms for the net rates of change (per volume) of the electron number density by neutrino emission and absorption:
\begin{align}
    Q_\mathrm{lep}^{-}(\vect{r}) &= Q_\mathrm{lep}^0(x) \cdot \left(\frac{\rho(\vect{r})}{\rho_0(r_0)}\right)^{\! a}\left(\frac{T(\vect{r})}{T_0(r_0)}\right)^{\! 5}\max\left\{0\,, \frac{Y_e(\vect{r}) - 0.01}{Y_e^0(r_0) - 0.01}\right\}\,,\label{eq:q_lep-}\\
    Q_\mathrm{lep}^{+}(\vect{r}) &= Q_\mathrm{lep}^0(x) \cdot \left(\frac{\rho(\vect{r})}{\rho_0(r_0)}\right) \left(\frac{R_\mathrm{gain}(t_0)}{R_\mathrm{gain}(t)}\right)^{\! 2} \cdot f_L\,f_E\,f_{Y_e}\,,\label{eq:q_lep+}
\end{align}
where $Q_\mathrm{lep}^0$ denotes the angle-averaged net change-rate of the electron density and $Y_e^0$ the angle-averaged electron fraction at time $t_0$, when we switch from the calculation with detailed neutrino transport to our simplified neutrino treatment.
Deleptonization, as per Equation~(\ref{eq:q_lep-}), applies in regions where $Q_\mathrm{lep}^0(x)\,\mathord{<}\,0$ (and $Y_e\,\mathord{>}\,0.01$), whereas for $Q_\mathrm{lep}^0(x)\,\mathord{>}\,0$ (and outside of the PNS, cf.\ footnote~\ref{fn:heating}), the electron number density is changed according to Equation~(\ref{eq:q_lep+}). Since outflowing ejecta in the neutrino-heating layer gain electron number by neutrino absorption while accretion flows deleptonize, which our angle-averaged source terms cannot capture in detail in multi-dimensional conditions, we include a factor $f_{Y_e}$ in Equation~(\ref{eq:q_lep+}). It is defined in terms of the electron fraction ($Y_e$) as
\begin{subnumcases}{f_{Y_e}=}
    \mkern-8mu \max\left\{0\,, \min\left\{1\,, \frac{1.1\,Y_e^\mathrm{eq} - Y_e(\vect{r})}{0.1\,Y_e^\mathrm{eq}}\right\}\right\}\,, & \kern-1.5em if $v_r(\vect{r}) \geqslant 0$\,,\label{eq:f_Ye+}\\
    \mkern-8mu (-1) \cdot \Theta\{Y_e(\vect{r}) - 0.1\}\,, & \kern-1.5em if $v_r(\vect{r})<0$\,,\label{eq:f_Ye-}
\end{subnumcases}
with $v_r$ denoting the radial velocity and
$\Theta$ being the Heaviside step function.
The negative sign of Equation~(\ref{eq:f_Ye-}) is supposed to approximately account for the fact that accretion downflows settling onto the PNS lose electron number. Applying Equation~(\ref{eq:f_Ye+}) shall ensure that $Y_e$ in the ejecta is limited by its kinetic equilibrium value of $Y_e^\mathrm{eq} = \left[1 + \lambda_{\bar\nu_e}/\lambda_{\nu_e}\right]^{-1}$,
which is approached when $\nu_e$ absorption on neutrons is balanced by $\bar\nu_e$ absorption on protons \citep{Qian1996}.
For the capture rates of $\nu_e$ and $\bar\nu_e$, $\lambda_{\nu_e}$ and $\lambda_{\bar\nu_e}$, respectively, we employ equations~(5)--(8) of \citet{Pllumbi2015}, which provide the detailed expressions including corrections associated with the neutron-proton mass difference and with weak magnetism~\citep{Horowitz2002}. The time-dependent neutrino luminosities and energy moments needed to evaluate the neutrino absorption rates are again
taken from a corresponding 1D simulation of PNS cooling.\footnote{Constraining $Y_e$ from the high side in Equation~(\ref{eq:f_Ye+}) is relevant only when our approach is applied in spherically symmetric explosion models. In our 3D simulations we found that $Y_e$ always remains fairly close to 0.5 or the equilibrium value because of the effects of coexisting outflows and downflows. Artificial regulation towards the equilibrium value by the factor $f_{Y_e}$ in Equation~(\ref{eq:f_Ye+}) did therefore not become active in the 3D cases.}

The gain radius in Equations~(\ref{eq:q_erg-}), (\ref{eq:q_erg+}), (\ref{eq:q_lep-}) and (\ref{eq:q_lep+}), which governs the radial migration of the heating and cooling profile, also needs to be prescribed as a function of time, because its behavior can be tracked only with self-consistent neutrino transport. In our simple heating and cooling treatment we couple the evolution of $R_\mathrm{gain}(t)$ to the contraction of the PNS radius $R_\mathrm{NS}(t)$ in the following way:
\begin{equation}
    R_\mathrm{gain}(t) = \left[ \left( C_0 - b \right)
    \frac{\dot{M}(t)}{\dot{M}(t_0)} + b \right] \cdot R_\mathrm{NS}(t)\,,
\label{eq:gain_radius}
\end{equation}
where $C_0\,\mathord{=}\,R_\mathrm{gain}(t_0)/R_\mathrm{NS}(t_0)$ denotes the ratio of the gain radius to the PNS radius at time $t_0$, and $b\,\mathord{=}\,\min\{1.01, C_0\}$ determines the assumed asymptotic value of $R_\mathrm{gain}(t)$ at late times when the mass accretion rate $\dot{M}(t)$ has declined to an insignificant level. This prescription accounts for an inflated PNS mantle and therefore increased gain radius due to ongoing accretion, and it ties in continuously with the evolution of $R_\mathrm{gain}(t)$ at times before we switch our neutrino treatment. $\dot{M}$ is evaluated at a fixed radius of 100\,km and for downflows ($v_r\,\mathord{<}\,0$) only.

A note of caution is indicated here. We noticed that our cooling prescription according to Equation~(\ref{eq:q_erg-}) can lead to runaway cooling, because a built-in mechanism of self-regulation is missing in the cooling rate. This can affect regions in the deep interior of the PNS at late evolution times ($t\,\mathord{\gtrsim}\,1$\,s after bounce), but it can also cause artifacts already early on in progenitors that are more massive than the ones considered in the present work. In massive stars it can happen that powerful accretion flows reach down to, or even below, the gain radius, in course of which cooling in the narrow layer between the neutrinosphere and the gain radius is overestimated, causing accelerated PNS contraction. An optical-depth-dependent exponential damping term turns out to be ineffective in such a situation, because the artificial effect occurs in a region of rather low optical depth. Devising a cure of these problems is currently work in progress. In the simulations of \znine and \snine, the cooling implementation described above worked well and ensured a smooth and continuous contraction of the PNS with a gradient nicely extrapolating the behavior of the \vertexprom simulations.      

Nevertheless, in the case of the \snine progenitor we encountered the mentioned issues with runaway cooling in a thin shell in the deep interior of the PNS. Instead of fixing them by a hard cut-off of the local cooling, we globally switched off the anyway low energy and lepton-number loss rates of Equations~(\ref{eq:q_erg-}) and (\ref{eq:q_lep-}), respectively, at $\sim$\,1\,s after bounce and continued our simulation with gain-layer heating only. Since the PNS radius at that time had decreased to 20\,km already and its further contraction was only slow, we could not witness any transient or discontinuous behavior as a consequence of our measure. In order to account for the subsequent shallow PNS contraction seen in the guiding 1D PNS-cooling simulation, which would shift $R_\mathrm{gain}$ in the 3D run closer in (see Equation~(\ref{eq:gain_radius})) and thus would enhance the heating of the PNS surroundings, we scale Equations~(\ref{eq:q_erg+}) and (\ref{eq:q_lep+}) with another factor
\begin{equation}
    f_\mathrm{NS} = \max\left\{1\,, \left(\frac{R_\mathrm{NS}(t)}{R_\mathrm{NS}^\mathrm{1D}(t)}\right)^{\! 2}\right\}\,.
\end{equation}
Here, $R_\mathrm{NS}^\mathrm{1D}$ is adopted from the corresponding 1D model that also provides the time-dependent neutrino luminosities and mean energies used in the scaling factors of $f_L$ and $f_E$ in Equations~(\ref{eq:f_L}) and (\ref{eq:f_E}), as well as in $Y_e^\mathrm{eq}$. This scaling increases the source rates $Q_\mathrm{erg}^{+}$ and $Q_\mathrm{lep}^{+}$ when the PNS radius in the 1D simulation with \vertexprom becomes smaller than in our 3D simulation with simplified neutrino treatment, thus ensuring that neutrino heating at the bottom of the SN ejecta is not underestimated.

\subsection{Neutrino pressure correction}

Switching off the neutrino transport would also lead to a sudden drop of pressure support in the high-density regime, if the contributions of neutrinos to the total pressure were ignored. To avoid unphysical artifacts (such as PNS oscillations), we replace the neutrino-momentum source term in the hydrodynamics equations by an adequate neutrino-pressure contribution that is added to the gas pressure. 

Assuming that neutrinos are in local chemical equilibrium with matter at sufficiently high densities, we can employ the following analytic expression \citep[using the sums of relativistic fermi integrals from][]{Bludman1978}:
\begin{equation}
    p_{\nu}^\mathrm{eq} = \frac{4 \pi (k_{\mathrm{b}} T)^4}{3 (h c)^3} \left[ \frac{21 \pi^4}{60} + \frac{1}{2} \eta_{\nu_e}^2 \left( \pi^2 + \frac{1}{2} \eta_{\nu_e}^2 \right) \right]\,, \label{eq:p_nu_eq}
\end{equation}
where $T$ is the local temperature, $k_\mathrm{b}$ Boltzmann's constant, and $\eta_{\nu_e}\,\mathord{=}\,\mu_{\nu_e}/(k_\mathrm{b}T)$ the degeneracy parameter of electron neutrinos, with $\mu_{\nu_e}$ denoting the $\nu_e$ chemical potential. Equation~(\ref{eq:p_nu_eq}) includes the pressure contributions from neutrinos and antineutrinos of all three flavors with $\mu_\nu = 0$ for both muon and tau neutrinos.

Since the assumption of local chemical equilibrium does not hold in regions of low matter densities, i.e., close to and outside of the neutrinosphere, we describe the neutrino pressure in the entire computational domain according to
\begin{equation}
    p_{\nu} = p_{\nu}^\mathrm{eq} \cdot f_p(\rho) \cdot \mathrm{min}\left\{ 1, \frac{\rho}{10^{13}\,\mathrm{g/cm^3}} \right\}\,,
\end{equation}
where the expression in the curly braces reduces the neutrino pressure gradually with decreasing density, and the density dependent
factor $f_p(\rho)$ is defined as
\begin{equation}
    f_p(\rho) = \frac{p_\mathrm{\textsc{vertex}}(\rho,t_0)}{p_{\nu}^\mathrm{eq}(\rho,t_0)}\,.
\label{eq:fpress}
\end{equation}
%
It is computed as the ratio of the neutrino pressure from the \textsc{Vertex} transport module, $p_\mathrm{\textsc{vertex}}$, to the analytical equilibrium-neutrino pressure according to Equation~(\ref{eq:p_nu_eq}). Both the numerator and the denominator of Equation~(\ref{eq:fpress}) are evaluated at time $t_0$ (when we change our neutrino treatment) and averaged over all directions. The ratio is then tabulated as a function of the matter density, $\rho$. At high densities the factor of Equation~(\ref{eq:fpress}) is applied to correct for a possible small mismatch of the analytic equilibrium pressure of Equation~(\ref{eq:p_nu_eq}) and the numerical value from the \vertex transport, connected to resolution and discretization effects. With the same prescriptions we also include the neutrino energy and pressure contributions to the general relativistic corrections in the effective gravitational potential (case A) of \cite{Marek2006}. Again, this treatment ensures a minimum of numerical noise and, as well as possible, it allows for a transient-free transition when the \textsc{Vertex} transport is replaced by our approximate heating and cooling description.

\end{document}